\pdfoutput=1
\documentclass[url,11pt,fleqn]{book} 

\usepackage[top=3cm,bottom=3cm,left=3.2cm,right=3.2cm,headsep=10pt,letterpaper]{geometry} 
\usepackage{xcolor} 
\definecolor{ocre}{RGB}{52,177,201} 

\usepackage{float} 

\usepackage{avant} 
\usepackage{mathptmx} 

\usepackage{microtype} 
\usepackage[utf8]{inputenc} 
\usepackage[T1]{fontenc} 
\usepackage[official]{eurosym} 

\usepackage{titlesec} 

\usepackage{graphicx} 
\graphicspath{{Pictures/}{figures/cosmo/}} 
\usepackage{wrapfig}
\usepackage{textcomp}
\usepackage{lipsum} 

\usepackage{tikz} 

\usepackage[english]{babel} 

\usepackage{enumitem} 
\setlist{nolistsep} 

\usepackage{booktabs} 

\usepackage{eso-pic} 

\usepackage{setspace}

\usepackage{titletoc} 

\contentsmargin{0cm} 
\titlecontents{chapter}[1.25cm] 
{\addvspace{15pt}\large\sffamily\bfseries} 
{\color{ocre!60}\contentslabel[\Large\thecontentslabel]{1.25cm}\color{ocre}} 
{}
{\color{ocre!60}\normalsize\sffamily\bfseries\;\titlerule*[.5pc]{.}\;\thecontentspage} 
\titlecontents{section}[1.25cm] 
{\addvspace{5pt}\sffamily\bfseries} 
{\contentslabel[\thecontentslabel]{1.25cm}} 
{}
{\sffamily\hfill\color{black}\thecontentspage} 
[]
\titlecontents{subsection}[1.25cm] 
{\addvspace{1pt}\sffamily\small} 
{\contentslabel[\thecontentslabel]{1.25cm}} 
{}
{\sffamily\;\titlerule*[.5pc]{.}\;\thecontentspage} 
[]

\titlecontents{lsection}[0em] 
{\footnotesize\sffamily} 
{}
{}
{}

\titlecontents{lsubsection}[.5em] 
{\normalfont\footnotesize\sffamily} 
{}
{}
{}

\usepackage{fancyhdr} 

\fancypagestyle{plain}{\fancyhead{}} 

\makeatletter
\renewcommand{\cleardoublepage}{
\clearpage\ifodd\c@page\else
\hbox{}
\vspace*{\fill}
\thispagestyle{empty}
\newpage
\fi}

\usepackage{amsmath,amsfonts,amssymb,amsthm} 

\newtheoremstyle{ocrenumbox}
{0pt}
{0pt}
{\normalfont}
{}
{\small\bf\sffamily\color{ocre}}
{\;}
{0.25em}
{\small\sffamily\color{ocre}\thmname{#1}\nobreakspace\thmnumber{\@ifnotempty{#1}{}\@upn{#2}}
\thmnote{\nobreakspace\the\thm@notefont\sffamily\bfseries\color{black}---\nobreakspace#3.}} 

\newtheoremstyle{blacknumex}
{5pt}
{5pt}
{\normalfont}
{} 
{\small\bf\sffamily}
{\;}
{0.25em}
{\small\sffamily{\tiny\ensuremath{\blacksquare}}\nobreakspace\thmname{#1}\nobreakspace\thmnumber{\@ifnotempty{#1}{}\@upn{#2}}
\thmnote{\nobreakspace\the\thm@notefont\sffamily\bfseries---\nobreakspace#3.}}

\newtheoremstyle{blacknumbox} 
{0pt}
{0pt}
{\normalfont}
{}
{\small\bf\sffamily}
{\;}
{0.25em}
{\small\sffamily\thmname{#1}\nobreakspace\thmnumber{\@ifnotempty{#1}{}\@upn{#2}}
\thmnote{\nobreakspace\the\thm@notefont\sffamily\bfseries---\nobreakspace#3.}}

\newtheoremstyle{ocrenum}
{5pt}
{5pt}
{\normalfont}
{}
{\small\bf\sffamily\color{ocre}}
{\;}
{0.25em}
{\small\sffamily\color{ocre}\thmname{#1}\nobreakspace\thmnumber{\@ifnotempty{#1}{}\@upn{#2}}
\thmnote{\nobreakspace\the\thm@notefont\sffamily\bfseries\color{black}---\nobreakspace#3.}} 
\makeatother

\newcounter{dummy}
\numberwithin{dummy}{section}
\theoremstyle{ocrenumbox}
\newtheorem{theoremeT}[dummy]{Theorem}

\newtheorem{exerciseT}{Exercise}[chapter]
\theoremstyle{blacknumex}
\newtheorem{exampleT}{Example}[chapter]
\theoremstyle{blacknumbox}

\newtheorem{definitionT}{Definition}[section]
\newtheorem{corollaryT}[dummy]{Corollary}
\theoremstyle{ocrenum}

\RequirePackage[framemethod=default]{mdframed} 

\newmdenv[skipabove=7pt,
skipbelow=7pt,
backgroundcolor=black!5,
linecolor=ocre,
innerleftmargin=5pt,
innerrightmargin=5pt,
innertopmargin=5pt,
leftmargin=0cm,
rightmargin=0cm,
innerbottommargin=5pt]{tBox}

\newmdenv[skipabove=7pt,
skipbelow=7pt,
rightline=false,
leftline=true,
topline=false,
bottomline=false,
backgroundcolor=ocre!10,
linecolor=ocre,
innerleftmargin=5pt,
innerrightmargin=5pt,
innertopmargin=5pt,
innerbottommargin=5pt,
leftmargin=0cm,
rightmargin=0cm,
linewidth=4pt]{eBox}	

\newmdenv[skipabove=7pt,
skipbelow=7pt,
rightline=false,
leftline=true,
topline=false,
bottomline=false,
linecolor=ocre,
innerleftmargin=5pt,
innerrightmargin=5pt,
innertopmargin=0pt,
leftmargin=0cm,
rightmargin=0cm,
linewidth=4pt,
innerbottommargin=0pt]{dBox}	

\newmdenv[skipabove=7pt,
skipbelow=7pt,
rightline=false,
leftline=true,
topline=false,
bottomline=false,
linecolor=gray,
backgroundcolor=black!5,
innerleftmargin=5pt,
innerrightmargin=5pt,
innertopmargin=5pt,
leftmargin=0cm,
rightmargin=0cm,
linewidth=4pt,
innerbottommargin=5pt]{cBox}

\makeatletter
\renewcommand{\@seccntformat}[1]{\llap{\textcolor{ocre}{\csname the#1\endcsname}\hspace{1em}}}
\renewcommand{\section}{\@startsection{section}{1}{\z@}
{-4ex \@plus -1ex \@minus -.4ex}
{1ex \@plus.2ex }
{\normalfont\large\sffamily\bfseries}}
\renewcommand{\subsection}{\@startsection {subsection}{2}{\z@}
{-3ex \@plus -0.1ex \@minus -.4ex}
{0.5ex \@plus.2ex }
{\normalfont\sffamily\bfseries}}
\renewcommand{\subsubsection}{\@startsection {subsubsection}{3}{\z@}
{-2ex \@plus -0.1ex \@minus -.2ex}
{.2ex \@plus.2ex }
{\normalfont\small\sffamily\bfseries}}
\renewcommand\paragraph{\@startsection{paragraph}{4}{\z@}
{-2ex \@plus-.2ex \@minus .2ex}
{.1ex}
{\normalfont\small\sffamily\bfseries}}

\usepackage{hyperref}
 \hypersetup{hidelinks,
            breaklinks=true,
            bookmarksopen=false,
            pdftitle={Title},
            pdfauthor={Author}
            colorlinks=true,
            pdfstartview=FitV,
            linkcolor=Blue,
            citecolor=MediumBlue,
            urlcolor=IndianRed}

\newcommand{\thechapterimage}{}
\newcommand{\chapterimage}[1]{\renewcommand{\thechapterimage}{#1}}

\def\thechapter{\arabic{chapter}}
\def\@makechapterhead#1{
\thispagestyle{empty}
{\centering \normalfont\sffamily
\ifnum \c@secnumdepth >\m@ne
\if@mainmatter
\startcontents
\begin{tikzpicture}[remember picture,overlay]
\node at (current page.north west)
{\begin{tikzpicture}[remember picture,overlay]
\node[anchor=north west,inner sep=0pt] at (0,0) {\includegraphics[width=\paperwidth]{\thechapterimage}};
\draw[anchor=west] (5cm,-9cm) node [rounded corners=20pt,fill=ocre!10!white,text opacity=1,draw=ocre,draw opacity=1,line width=1.5pt,fill opacity=.6,inner sep=12pt]{\huge\sffamily\bfseries\textcolor{black}{\thechapter. #1\strut\makebox[22cm]{}}};
\end{tikzpicture}};
\end{tikzpicture}}
\par\vspace*{230\p@}
\fi
\fi}

\def\@makeschapterhead#1{
\thispagestyle{empty}
{\centering \normalfont\sffamily
\ifnum \c@secnumdepth >\m@ne
\if@mainmatter
\begin{tikzpicture}[remember picture,overlay]
\node at (current page.north west)
{\begin{tikzpicture}[remember picture,overlay]
\node[anchor=north west,inner sep=0pt] at (0,0) {\includegraphics[width=\paperwidth]{\thechapterimage}};
\draw[anchor=west] (5cm,-9cm) node [rounded corners=20pt,fill=ocre!10!white,fill opacity=.6,inner sep=12pt,text opacity=1,draw=ocre,draw opacity=1,line width=1.5pt]{\huge\sffamily\bfseries\textcolor{black}{#1\strut\makebox[22cm]{}}};
\end{tikzpicture}};
\end{tikzpicture}}
\par\vspace*{230\p@}
\fi
\fi
}
\makeatother

\usepackage{graphicx}
\graphicspath{{./figures/}}
\usepackage{appendix}
\usepackage{latexsym,amsmath,amsfonts,amssymb,booktabs}
\usepackage[font=small]{caption}
\usepackage{slashed,upgreek,amscd,cancel,tensor,color}
\usepackage{adjustbox}
\usepackage[numbers,compress,square]{natbib}
\usepackage{epsfig,latexsym}
\usepackage{url}
\numberwithin{equation}{section}
\usepackage{doi}
\usepackage{subcaption}
\usepackage{mathtools}
\usepackage{upgreek}
\usepackage{stfloats}
\usepackage{afterpage}
\usepackage{multirow}

\usepackage{pdfpages}
\usepackage[separate-uncertainty=true]{siunitx}
\usepackage{acronym}
\usepackage{tcolorbox}
\usepackage{etoolbox}
\usepackage{multicol}
 \usepackage{vwcol}
\usepackage{placeins}
\usepackage[absolute]{textpos}
\usepackage{ulem} 

\newcommand{\etboxwidth}{0.975\textwidth}
\newcommand{\etboxskip}{0.0125\textwidth} 
\newcommand{\etboxmargin}{0.5cm}
\usepackage{float}
\floatstyle{ruled} 
\newfloat{ETbox}{t}{lop}[section] 

\newenvironment{ETboxenvironment}[1][\hsize]{%
\MakeFramed{\hsize#1\advance\hsize-\width\FrameRestore}}%
{\endMakeFramed}

\def\longetbox#1#2#3#4{ 
\definecolor{framecolor}{cmyk}{1,1,1,1}%
\ifthenelse{\equal{#1}{r}}{%
\definecolor{darkred}{rgb}{0.6,0,0}
\definecolor{semidarkred}{rgb}{0.8,0,0}
\floatname{ETbox}{\color{darkred}\large Box}%
\renewcommand{\titlecolor}{darkred}%
\renewcommand{\captioncolor}{darkred}%
\renewcommand{\contentcolor}{semidarkred}%
\renewcommand{\ccolor}{\definecolor{citecolor}{rgb}{0,0,0.8}}%
\renewcommand{\lcolor}{\definecolor{linkcolor}{rgb}{.0,.8,0}}%
\renewcommand{\ucolor}{\definecolor{urlcolor}{rgb}{0,0.8,0.8}}%
\definecolor{shadecolor}{rgb}{1,0.9,0.7}}{%
\ifthenelse{\equal{#1}{i}}{%
\floatname{ETbox}{\color{blue}\large Box}%
\renewcommand{\titlecolor}{blue}%
\renewcommand{\captioncolor}{blue}%
\renewcommand{\contentcolor}{blue}%
\renewcommand{\ccolor}{\definecolor{citecolor}{rgb}{0,.5,0}}
\renewcommand{\lcolor}{\definecolor{linkcolor}{rgb}{.8,0,0}}
\renewcommand{\ucolor}{\definecolor{urlcolor}{rgb}{0.9,0.,0.9}}
\definecolor{shadecolor}{cmyk}{.07,.01,0,0}}{%
\ifthenelse{\equal{#1}{h}}{%
\floatname{ETbox}{\color{blue}\large Box}
\renewcommand{\titlecolor}{blue}%
\renewcommand{\captioncolor}{blue}%
\renewcommand{\contentcolor}{black}%
\renewcommand{\ccolor}{\definecolor{citecolor}{rgb}{0,.5,0}}%
\renewcommand{\lcolor}{\definecolor{linkcolor}{rgb}{.8,0,0}}%
\renewcommand{\ucolor}{\definecolor{urlcolor}{rgb}{0,0,.7}}%
\definecolor{shadecolor}{cmyk}{0,0,.3,0}}{}{}}}%
\vspace{\parskip} 
\color{\contentcolor}
\fboxrule 0.75pt
\setlength{\fboxsep}{\etboxmargin}
\begin{ETboxenvironment}[\etboxwidth] 
\arrayrulecolor{\contentcolor}
\ccolor
\lcolor
\ucolor
\captionsetup{labelfont={color=\contentcolor,bf}, textfont={color=\contentcolor}, labelsep=colon}
\captionof{ETbox}[#3]{\textcolor{\captioncolor}{\large\bf#3}} 
\captionsetup{labelfont={color=\contentcolor,bf}, textfont={color=\contentcolor},labelsep=colon}
#4
\label{#2} 
\end{ETboxenvironment} 
\arrayrulecolor{black} 
\color{black} 

}

\def\etbox#1#2#3#4{ 
\definecolor{framecolor}{cmyk}{1,1,1,1}%
\ifthenelse{\equal{#1}{r}}{%
\definecolor{darkred}{rgb}{0.6,0,0}
\definecolor{semidarkred}{rgb}{0.8,0,0}
\floatname{ETbox}{\color{darkred}\large Box}%
\renewcommand{\titlecolor}{darkred}%
\renewcommand{\captioncolor}{darkred}%
\renewcommand{\contentcolor}{semidarkred}%
\renewcommand{\ccolor}{\definecolor{citecolor}{rgb}{0,0,0.8}}%
\renewcommand{\lcolor}{\definecolor{linkcolor}{rgb}{.0,.8,0}}%
\renewcommand{\ucolor}{\definecolor{urlcolor}{rgb}{0,0.8,0.8}}%
\definecolor{shadecolor}{rgb}{1,0.9,0.7}}{%
\ifthenelse{\equal{#1}{i}}{%
\floatname{ETbox}{\color{blue}\large Box}%
\renewcommand{\titlecolor}{blue}%
\renewcommand{\captioncolor}{blue}%
\renewcommand{\contentcolor}{blue}%
\renewcommand{\ccolor}{\definecolor{citecolor}{rgb}{0,.5,0}}
\renewcommand{\lcolor}{\definecolor{linkcolor}{rgb}{.8,0,0}}
\renewcommand{\ucolor}{\definecolor{urlcolor}{rgb}{0.9,0.,0.9}}
\definecolor{shadecolor}{cmyk}{.07,.01,0,0}}{%
\ifthenelse{\equal{#1}{h}}{%
\floatname{ETbox}{\color{blue}\large Box}%
\renewcommand{\titlecolor}{blue}%
\renewcommand{\captioncolor}{blue}%
\renewcommand{\contentcolor}{black}%
\renewcommand{\ccolor}{\definecolor{citecolor}{rgb}{0,.5,0}}%
\renewcommand{\lcolor}{\definecolor{linkcolor}{rgb}{.8,0,0}}%
\renewcommand{\ucolor}{\definecolor{urlcolor}{rgb}{0,0,.7}}%
\definecolor{shadecolor}{cmyk}{0,0,.3,0}}{}{}}}%
\begingroup
\begin{center}   
\begin{minipage}[t]{\etboxwidth} 
\leftskip=\etboxskip
\rightskip=\etboxskip
\definecolor{framecolor}{cmyk}{1,1,1,1}
\begin{framed} 
\vspace{-9pt} 
\arrayrulecolor{\contentcolor}
\begin{shaded}       
\lcolor
\ccolor
\ucolor
\captionsetup{labelfont={color=\captioncolor,bf}, textfont={color=\captioncolor}, labelsep=colon, margin=\etboxskip}%
\captionof{ETbox}[#3]{\textcolor{\captioncolor}{\large\bf#3}} 
\captionsetup{labelfont={color=\titlecolor,bf}, textfont={color=\contentcolor},labelsep=colon, margin=\etboxskip}
\textcolor{\contentcolor}{#4} 
\label{#2} 
\end{shaded} 
\vspace{-9pt} 
\end{framed} 
\end{minipage} 
\end{center} 
\endgroup
\vspace{9pt} 
\arrayrulecolor{black} 
\color{black} 
}
\usepackage{ifthen}
\usepackage[cfg, intoc]{nomencl}
\renewcommand{\nomgroup}[1]{%
 \ifthenelse{\equal{#1}{S}}{\vspace{30pt}\item[]\hspace*{-\leftmargin}%
 \rule[2pt]{0.45\linewidth}{1pt}%
 \hfill \textbf{Symbols} \hfill
 \rule[2pt]{0.45\linewidth}{1pt}\vspace{-5pt} \item[]\hspace*{-\leftmargin}\vspace{5pt}\hspace{1.1cm}Please note that some symbols might stand for more than one quantity depending on the context}{%
\ifthenelse{\equal{#1}{A}}{\item[]\hspace*{-\leftmargin}
\rule[2pt]{0.45\linewidth}{1pt}%
\hfill \textbf{Abbreviations} \hfill
\rule[2pt]{0.45\linewidth}{1pt}}{%
\ifthenelse{\equal{#1}{G}}{\vspace{30pt}\item[]\hspace*{-\leftmargin}
\rule[2pt]{0.45\linewidth}{1pt}%
\hfill \textbf{Glossary} \hfill
\rule[2pt]{0.45\linewidth}{1pt}}{}{}}}}

\makenomenclature

\definecolor{antiquewhite}{rgb}{0.98, 0.92, 0.84}
\definecolor{aliceblue}{rgb}{0.94, 0.97, 1.0}
\definecolor{blanchedalmond}{rgb}{1.0, 0.92, 0.8}
\definecolor{beaublue}{rgb}{0.74, 0.83, 0.9}
\definecolor{beaubluedark}{rgb}{0.79, 0.85, 0.95}
\definecolor{amber(sae/ece)}{rgb}{1.0, 0.49, 0.0}
\definecolor{airforceblue}{rgb}{0.36, 0.54, 0.66}
\definecolor{aliceblue}{rgb}{0.94, 0.97, 1.0}
\definecolor{alizarin}{rgb}{0.82, 0.1, 0.26}
\definecolor{almond}{rgb}{0.94, 0.87, 0.8}
\definecolor{amaranth}{rgb}{0.9, 0.17, 0.31}
\definecolor{amber}{rgb}{1.0, 0.75, 0.0}
\definecolor{amber(sae/ece)}{rgb}{1.0, 0.49, 0.0}
\definecolor{americanrose}{rgb}{1.0, 0.01, 0.24}
\definecolor{amethyst}{rgb}{0.6, 0.4, 0.8}
\definecolor{anti-flashwhite}{rgb}{0.95, 0.95, 0.96}
\definecolor{antiquebrass}{rgb}{0.8, 0.58, 0.46}
\definecolor{antiquefuchsia}{rgb}{0.57, 0.36, 0.51}
\definecolor{antiquewhite}{rgb}{0.98, 0.92, 0.84}
\definecolor{ao}{rgb}{0.0, 0.0, 1.0}
\definecolor{ao(english)}{rgb}{0.0, 0.5, 0.0}
\definecolor{applegreen}{rgb}{0.55, 0.71, 0.0}
\definecolor{apricot}{rgb}{0.98, 0.81, 0.69}
\definecolor{aqua}{rgb}{0.0, 1.0, 1.0}
\definecolor{aquamarine}{rgb}{0.5, 1.0, 0.83}
\definecolor{beaver}{rgb}{0.62, 0.51, 0.44}
\definecolor{blizzardblue}{rgb}{0.67, 0.9, 0.93}
\definecolor{azure}{rgb}{0.0, 0.5, 1.0}

\begin{document}


\begingroup


\centering
\par\normalfont\fontsize{50}{50}\sffamily\selectfont
\textcolor{violet}{\textbf{E--TEST prototype}}\\
\vskip0.5cm
\par\normalfont\fontsize{25}{25}\sffamily\selectfont
\textsc{\textcolor{violet}{{\textbf{Design Report}}}}\par 
\vskip1.25cm

\textsc{\begin{singlespace} \small\fontsize{10}{10} A Sider$^1$, L Amez-Droz$^1$, A Amorosi$^1$, F Badaracco$^8$, P Baer$^4$, A Bertolini$^6$, G Bruno$^8$, P Cebeci$^4$, C Collette$^1$, J Ebert$^4$, B Erben$^4$, R Esteves$^{11}$, C Di Fronzo$^1$, E Ferreira$^8$, A Gatti$^{11}$, M Giesberts$^4$, T Hebbeker$^9$, J-S Hennig$^7$, M Hennig$^7$, S Hild$^7$, M Hoefer$^4$, H-D Hoffmann$^4$, R Jamshidi$^1$, T Kuhlbusch$^{10}$, L Jacques$^2$, R Joppe$^9$, M H Lakkis$^1$, C Lenaerts$^2$, J-P Locquet$^{12}$, J Loicq$^{2,3}$, B Long Le Van$^2$, P Loosen$^4$, M Nesladek$^5$, M Reiter$^4$, A Stahl$^{10}$, J Steinlechner$^7$, S Steinlechner$^7$,  M Teloi$^1$, J van Heijningen$^8$, J Vilaboa Pérez$^2$, M Zeoli$^{1,8}$\end{singlespace}} 

\vskip 1cm

\textsc{\small $^1$ Precision Mechatronics Laboratory, Université de Liège, 9 allée de la découverte B-4000 Liège, Belgium\\
$^2$ Centre Spatial de Liège, University of Liège, Avenue du Pré Aily, 4031 Angleur, Belgium\\
$^3$ Faculty of Aerospace Engineering Delft University of Technology Kluyverweg 1, 2629 HS Delft\\
$^4$ Fraunhofer Institute for Laser Technology ILT Steinbachstraße 15, 52074, Aachen, Germany\\
$^5$ Universiteit Hasselt, Martelarenlaan 42, 3500 Hasselt, Belgium\\
$^6$ Nikhef, Science Park, 1098 XG Amsterdam, The Netherlands\\
$^7$ Faculty of Science and Engineering, Maastricht University, 6200 MD Maastricht, The Netherlands\\
$^8$ Centre for Cosmology, Particle Physics and Phenomenology (CP3), UCLouvain, 1348 Louvain-la-Neuve, Belgium\\
$^9$ Physikalisches Institut A, RWTH Aachen, Germany\\
$^{10}$ Physikalisches Institut B, RWTH Aachen, Germany\\
$^{11}$ Elektronische Circuits en Systemen, Leuven, Kasteelpark Arenberg 10, 3001 Leuven, Belgium\\
$^{12}$KU Leuven, Semiconductor Physics, Celestijnenlaan 200d, 3001 Leuven\\}

\textsc{\small christophe.collette@uliege.be}



\pagestyle{fancy}
\fancyhf{}
\rfoot{\thepage}
\lhead{\begin{picture}(1,1) {\includegraphics[width=1\textwidth]{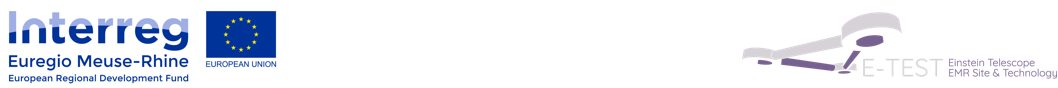}} \end{picture}}
\lfoot{\begin{picture}(0,0)\put(0,-70) {\includegraphics[width=15cm]{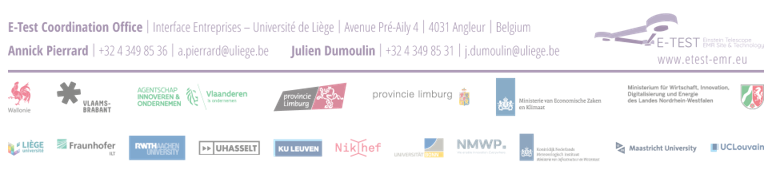}} \end{picture}}

\endgroup


\newpage

\begin{figure}[htp]
   \centering
   \includegraphics[width = 1\linewidth]{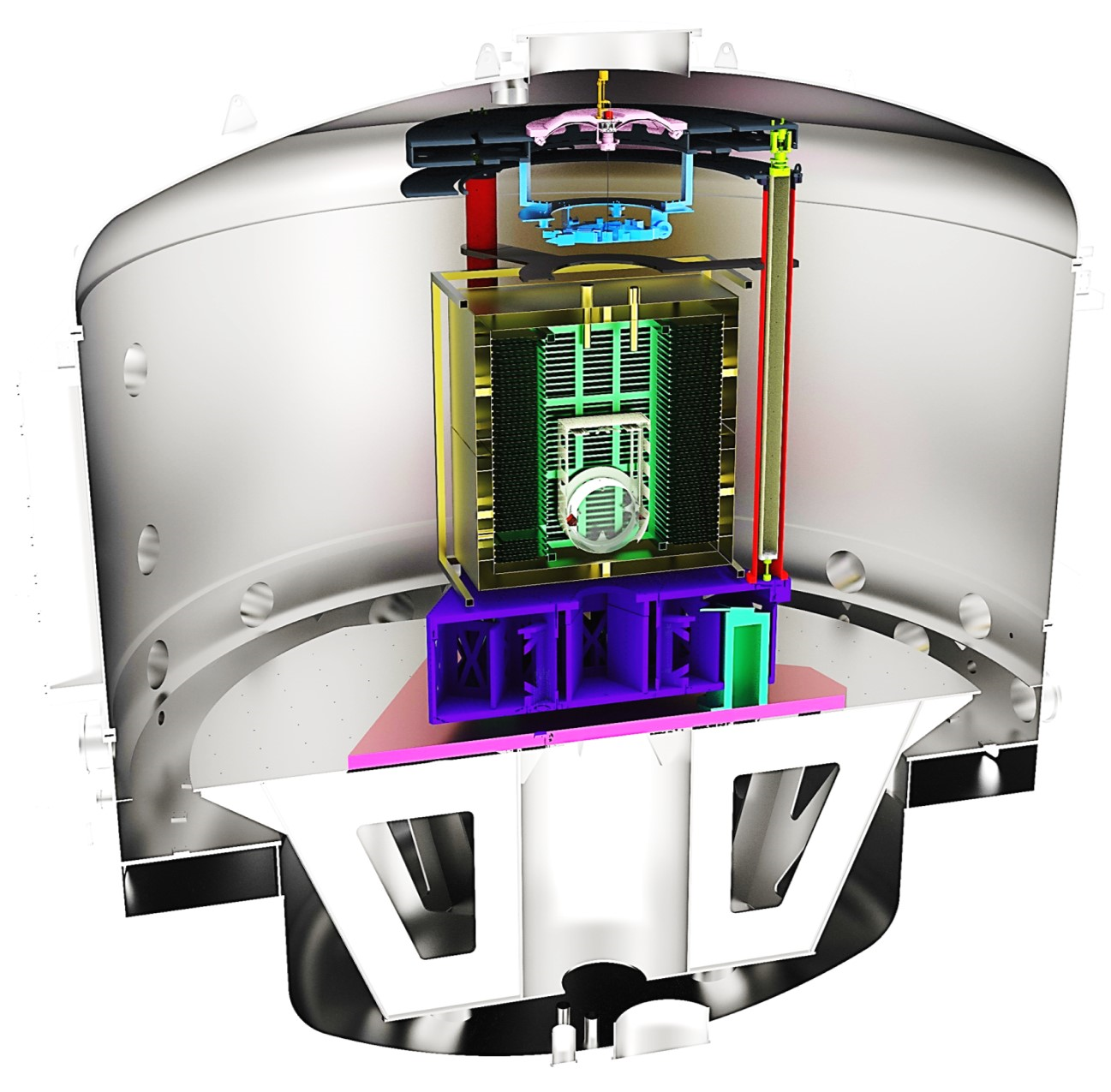}
   \label{fig:ETestPrototype}
\end{figure}

~\vfill
\thispagestyle{empty}
\noindent \textsc{E--TEST Team}\\
\noindent {URL: www.etest-emr.eu}\\ 
\noindent {Contact: christophe.collette@uliege.be}\\ 
\noindent \\ 

\noindent \textit{June 2022} 
\newpage

\chapterimage{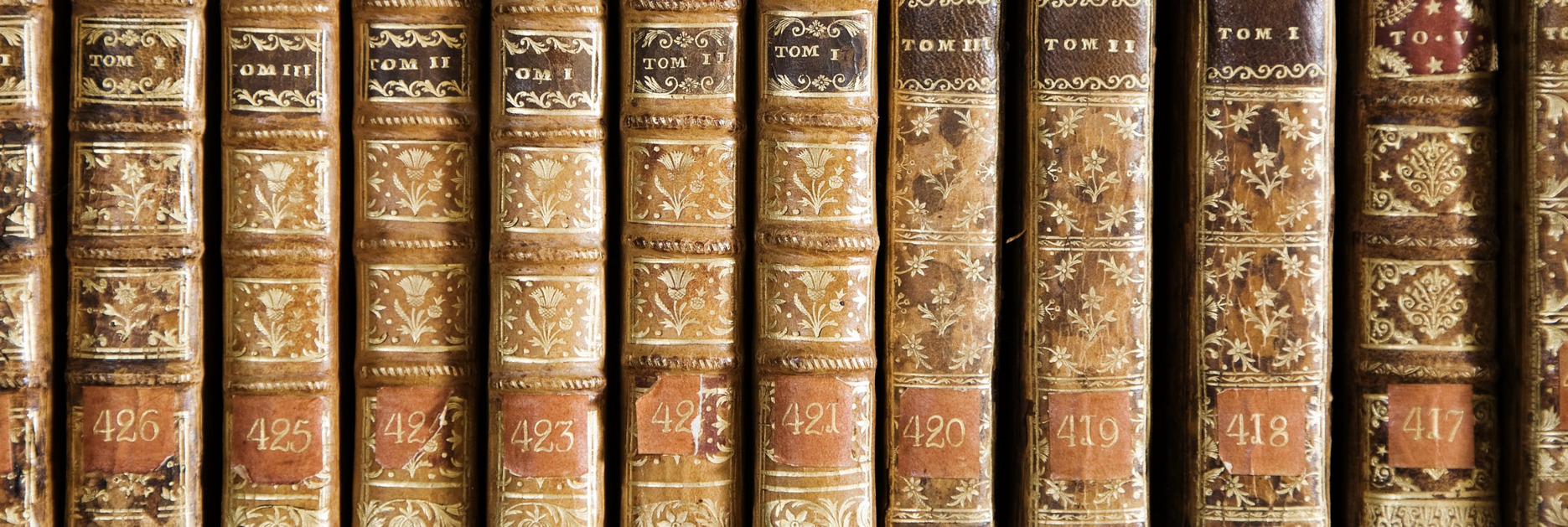} 
\pagestyle{empty} 
\tableofcontents 
\pagestyle{fancy} 

\chapterimage{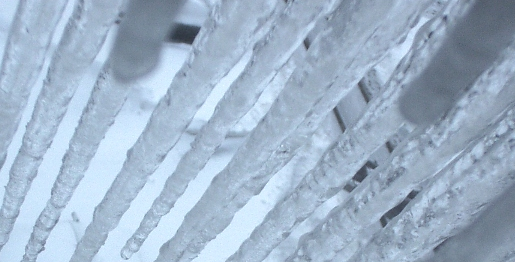} 
\chapter{Aims of the E--TEST Prototype Project}
\label{sec:aims}





\section{Context}
With the recent detections of Gravitational Wave (GW), a new window has been opened on the universe, one century after these waves were predicted by Einstein’s general theory of relativity. Together with these detections, gravitational wave astronomy was born, offering an exciting future for the exploration of the universe through large Michelson interferometers used as gravitational wave detectors, and calling for sensitivity improvement. Although existing instruments do have margins for upgrades, their performance will be ultimately limited by their location and design. A significant improvement of sensitivity can only be obtained using a disruptive technology like it is proposed in 3rd generation gravitational wave detectors e.g. Einstein Telescope (ET) and Cosmic Explorer (CE). 

\begin{figure}[h]
    \centering
    \includegraphics[width = 1\linewidth]{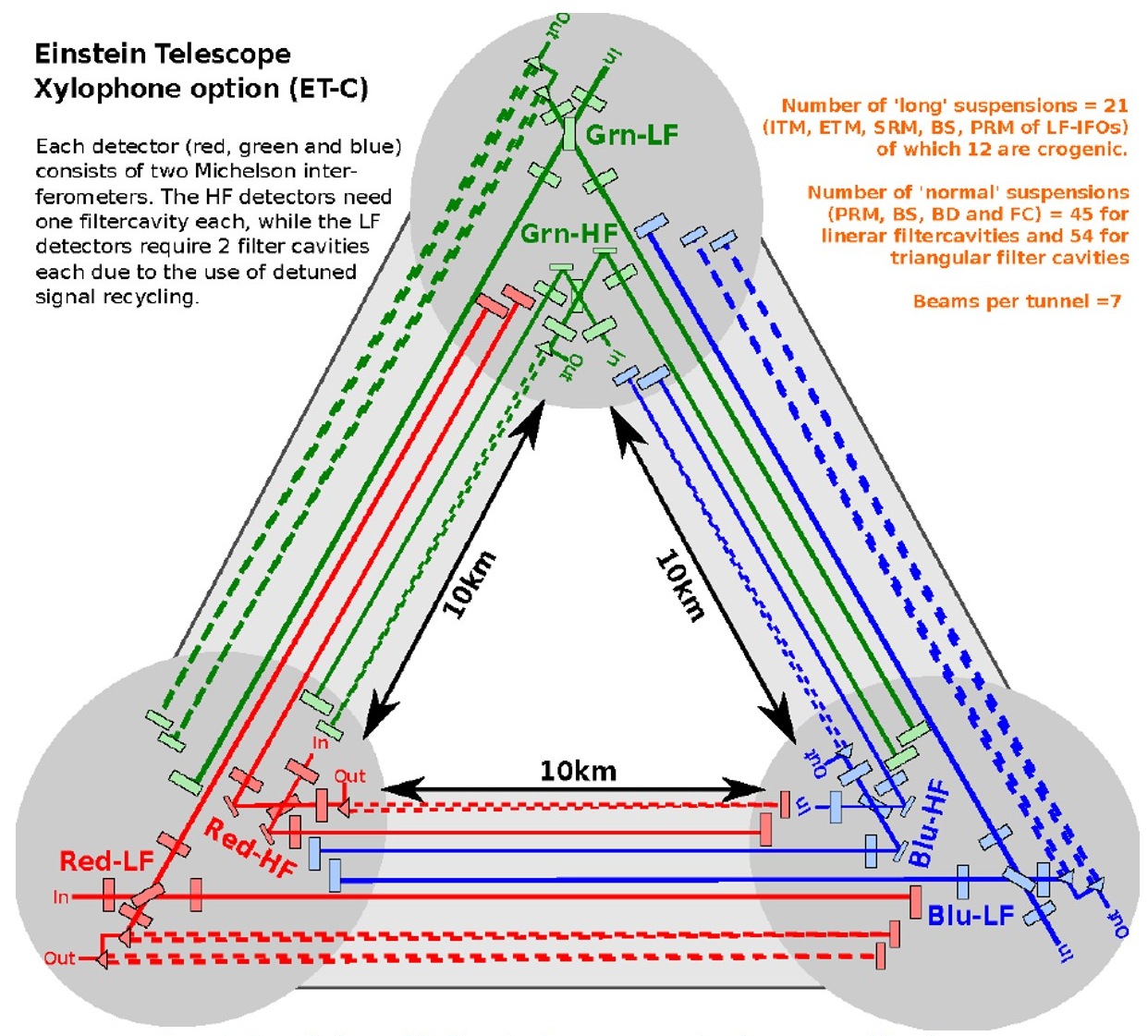}
    \caption{General scheme of the Einstein telescope representing the geometry of the telescope .}
    \label{fig:EinsteinTelescope}
\end{figure}

A simplified sketch of ET is shown in Figure \ref{fig:EinsteinTelescope}. For maximizing the detection capability, it is constituted of 3 detectors (red, blue and green). Each detector consists of two Michelson interferometers whose arms are 10 km long: one designed to maximize the sensitivity at high frequency (HF) and one to maximize the sensitivity at low frequency (LF).  ET-HF will be around room temperature (290K), and use very high power laser. ET-LF will be cryogenic (10K) especially for reducing thermal noise and increase drastically the sensitivity.  

The concept proposed for ET is very different from currently operating GW detectors Advanced LIGO, Advanced Virgo and KAGRA. For example, Advanced Virgo and Advanced LIGO operate with limited size fused silica mirrors at room temperature and use a laser wavelength of 1064\,nm. KAGRA (sometimes referred to as detector of generation 2.5) operates with cryogenic, sapphire mirrors, which allows to stay at 1064\,nm laser wavelength.
In contrast, both ET and CE plan to operate large cryogenic mirrors made from silicon and read out by lasers with wavelength further into the infrared (about 1550\,nm to about 2100\,nm). Another characteristic of 3rd generation detectors is to be much more sensitive at low frequency, which implies to improve the isolation of the mirror from seismic motion and gravity fluctuations.  

In order to prepare and validate the technology necessary for these future instruments, several prototypes and facilities are being constructed all over the world. The prototype presented in this report is one of them. Next section gives an overview of the project which is funding the prototype, namely the \textit{E-TEST} project. 

\section{The E-TEST project}

E-TEST (Einstein Telescope Euregio-Meuse-Rhin Site and Technology) is a project recently funded by the European program Ineterreg Euregio Meuse-Rhine. This program is dedicated to innovative cross boarder activities between Belgium, The Netherlands and Germany. With a total budget of 15M€ and a consortium of 11 partners from the three countries, the objective of the project is twofold. Firstly, develop an eco-friendly and non-invasive imaging of the geological conditions as well as the development of an observatory of the underground in the EMR region.
Secondly, develop technologies necessary for 3rd generation gravitational wave detectors. In particular, it is proposed to develop a prototype of large suspended cryogenic silicon mirror, isolated from seismic vibrations at low frequency. The total budget of the project is equally spread over the two activities. The first activity is not discussed at all in this report.

The E-TEST prototype will have some key unique features: 
\begin{itemize}
    \item A silicon mirror of 100\,kg
    \item A radiative cooling strategy (non contact)
    \item A low frequency hybrid isolation stage
    \item Cryogenic sensors and electronics
    \item A laser and optics at 2 microns
    \item A low thermal noise coating
\end{itemize}

 A sketch of the initial proposal is shown in Fig.\ref{fig:ETestPrototype}.

\begin{figure}[h]
    \centering
    \includegraphics[width = 1\linewidth]{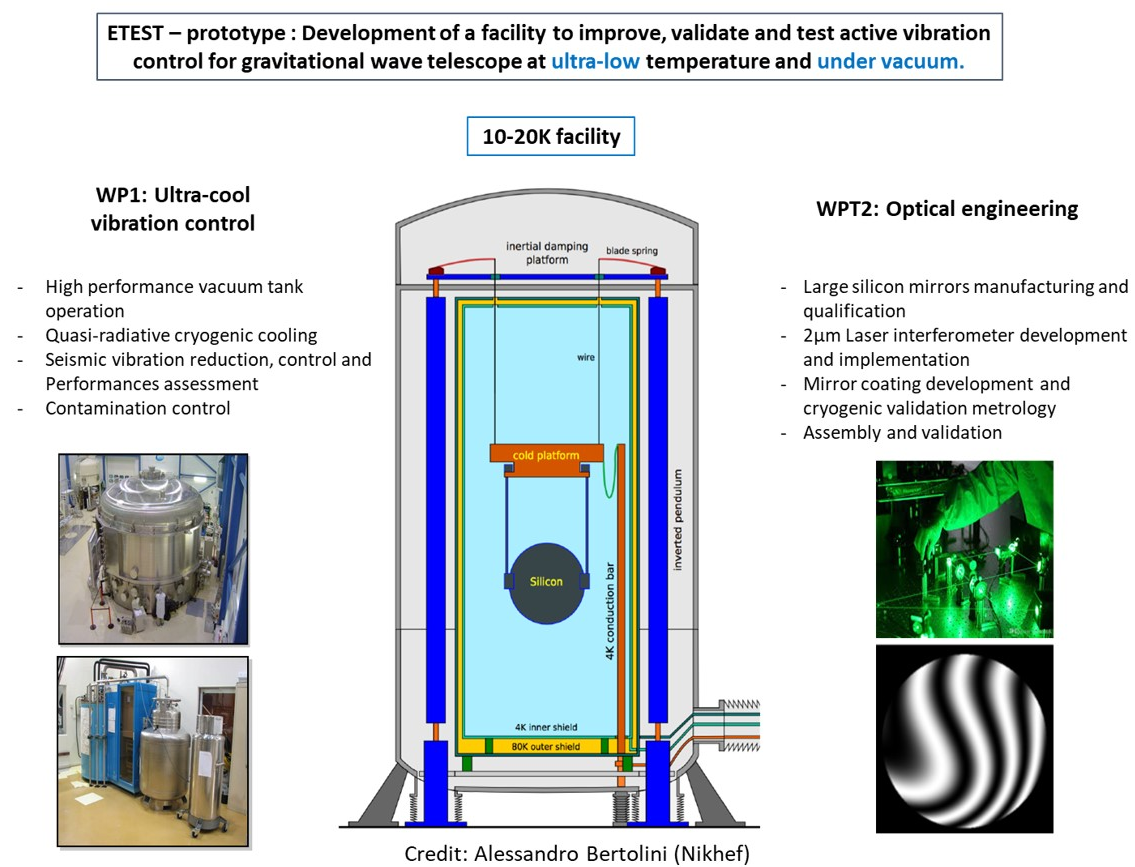}
    \caption{Sketch of E-TEST prototype.}
    \label{fig:ETestPrototype}
\end{figure}
The activities were broken down in two workpackages, WP1 and WP2. WP1 will focus on the design and the realization of a full scale seismic attenuator, combining passive filtering with at least one active pre-isolation stage, with the goal of boosting the performance in the 0.1-10 Hz frequency band. 
The development and characterization of cryogenic sensors and electronics is also part of WP1. The activities in WP2 are focused on optical engineering, and encompass the development of a large silicon mirror, laser and optics at 2 microns, photodetectors and low noise coating technologies. 
The work proposed in WP1 and WP2 are described in details in Chapter 2 and 3. 

In order to take advantages of the existing expertise and infrastructure available in the EMR region, it has been proposed to mount the E-TEST prototype in existing facilities at the Space Centre of Liège (CSL), which has a strong know-how in cryogenic qualification of Space Instruments.

\section{Ecosystem in the Euregio Meuse Rhin}

Einstein telescope is very strategic for the EMR region as it is one site candidate for hosting ET. More than the scientific and technologic return, the operational aspects of such facility will cover a huge spectrum of economical domains.

Two other large scale Interreg projects have been recently funded in the EMR: ETPathfinder and ET2SME.

\begin{figure}[h]
    \centering
    \includegraphics[width = 1\linewidth]{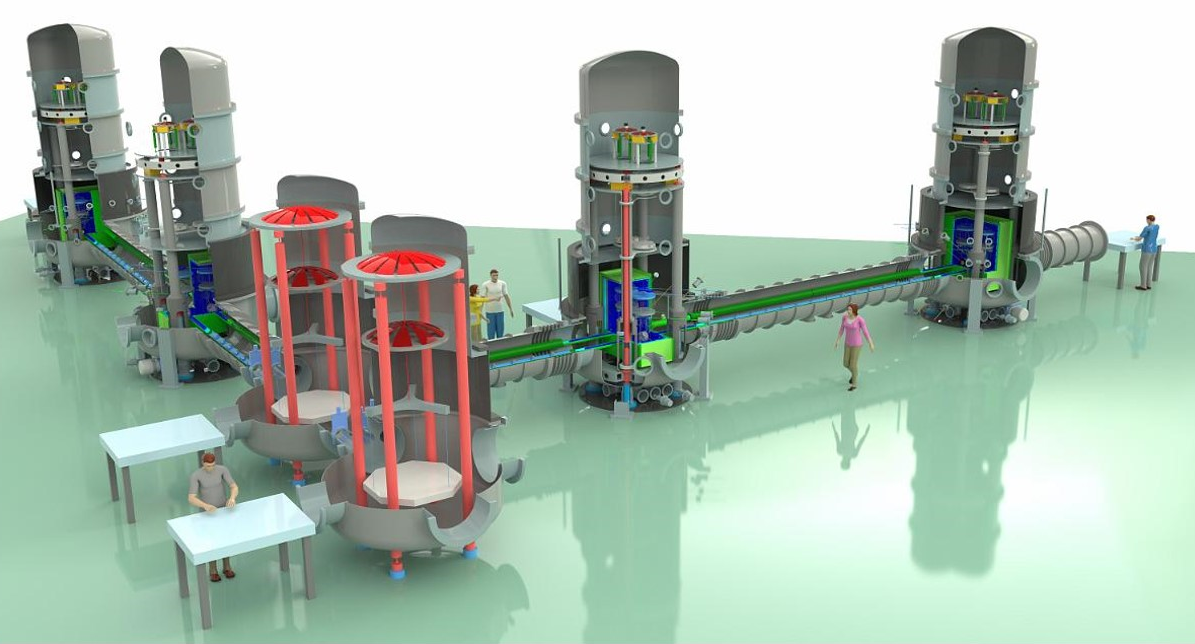}
    \caption{Sketch of ETpathfinder.  More information on ETpathfinder can be found here: \url{https://www.einsteintelescope.nl/etpathfinder/.}}
    \label{fig:ETPathfinder}
\end{figure}

ETpathfinder will provide a new facility to \emph{holistically prototype} the following aspects:
\begin{itemize}
    \item \textbf{New Temperature, i.e. Cyogenics:} Cryogenic temperature of the main test masses (120\,K and 15\,K); low-noise cryo-coolers, decoupling of cryo-coolers from test masses while providing sufficient cooling power, cryo-shields etc;
    \item \textbf{New Mirror Material, i.e. Silicon:} Changing from fused silica (amorphous material, electrically insulating) to silicon (crystalline semi-conductor) will be a major change with many implications for optical aspects, polishing, coatings etc as well as mechanical aspects.
    \item \textbf{New Wavelength, i.e. 1550 - 2100\,nm:} Moving to silicon as test mass material requires to go to longer laser wavelength to reduce optical absorption. Lasers, optical and electro-optical components (such as high-efficiency photo detectors) need to be developed and qualified for GW detector relevant aspects. Laser stabilisation loops (power, frequency, jitter, etc) need to be developed.  
\end{itemize}
All of these strands are interlinked and ultimately will be tested altogether in ETpathfinder, at displacement sensitivity not too dissimilar to a the Einstein Telescope. A sketch of ETpathfinder is shown in Figure \ref{fig:ETPathfinder}.

\subsection{Synergies between ETpathfinder and E-TEST}

Towards the development of the full Einstein Telescope, ET-Pathfinder and E-TEST are two independent but very complementary building blocks. The focus of both projects is clearly different as they are dedicated to different parts of the full telescope but with the final objective to provide the innovative technologies required for ET. 

Both initiatives can be seen as the development of expertise centers to make the links between gravitational wave science, its technological aspects and the huge number of economical actors involved into such development, present in the EMR region. 

The main difference between the two projects is that ETpathfinder results in a full facility laser-interferometer that can be used by knowledge institutes and (SME) companies for developing and testing new technologies.  On the other side E-TEST will result in a prototype in a vacuum chamber and optical instruments working at 20 Kelvin, which corresponds to 20 degrees above the absolute zero. E-TEST also results in a seismic model that can be used to prove that ET can be built in the EMR region. Knowledge institutes and companies can use E-TEST results to develop and test new products or gain new insights/knowledge. 
ET-pathfinder will focus on the high temperature technologies of the Einstein Telescope while the E-TEST will develop innovative and challenging components and practices for ultra-low temperature. 
It is important to note that the technologies developed in E-TEST could also be transposed to other gravitational wave detectors, placing the E-TEST actors in the forehand of this community. 
Due to their complementarity, both projects will work in synergy. Moreover, at the end, the outputs of both will be merged and implemented into the ET. The region will gain a massive competitive edge in this field via ETpathfinder and E-TEST. EMR based companies can use this advantage to create and deliver products and services to ET, even if ET could not be built in the EMR region.



\textbf{Added value of ETpathfinder for E-TEST:}

As ETpathfinder started more than half a year before E-TEST, the last one will benefit from the expertise and lessons learnt from ETpathfinder consortium. From the technical point of view, some of the groups are involved in the two projects (Nikhef, UMaastricht, KULeuven). Their experience will be particularly useful for ensuring the feasibility of E-TEST. E-TEST will also benefit from ETpathfinder in the management of an Interreg project, whose aim is to validate experimentally technology for ET.   

\textbf{Added value of E-TEST for ETpathfinder:}

In E-TEST, we will have only one suspended mirror, which means that we will be able to measure its stability with a precision limited to $10^{-14}m$. After the completion of the E-TEST project, the performance of the suspended mirror and the whole vibration control chain would be tested and even transferred to the ETpathfinder facility in order to assess precisely the residual vibrations of the cryogenic suspension in a full interferometer. And finally the ETEST will bring to the whole community the geophysical model mandatory the build the Einstein Telescope.

\subsection{ET2SME}

ET2SMEs is an Interreg project which supports the development of new innovative products and services by promoting transnational R$\&$D projects established by SMEs collaborating across borders in the EMR. It can be seen as the economic counterpart of the E-TEST project. With the conviction that the technology of 3rd generation gravitational wave detectors need a close cooperation between research institutes and industries, ET2SME provides innovation vouchers for relevant cross boarder projects launched by SME. 
Interested companies will also receive direct support in expanding their business network into neighbouring countries, especially in finding their suitable SME partner. They will also have the opportunity to present their know-how and special competences in a virtual 3D model of the Einstein Telescope (Mapping). Of particular interest here are the Einstein Telescope operating technologies in cryogenics, vacuum, precision mechanics/mechatronics, sensors, optics and optical metrology, mirrors coating, lasers and advanced control algorithms.
Finally, there is the possibility for companies of participating in an ET Industrial Advisory Board, organized jointly by E-TEST, ETpathfinder and ET2SMEs, between Business and Research, for the large-scale facility as a dialogue partner and advisor from now and in the coming years.

\chapterimage{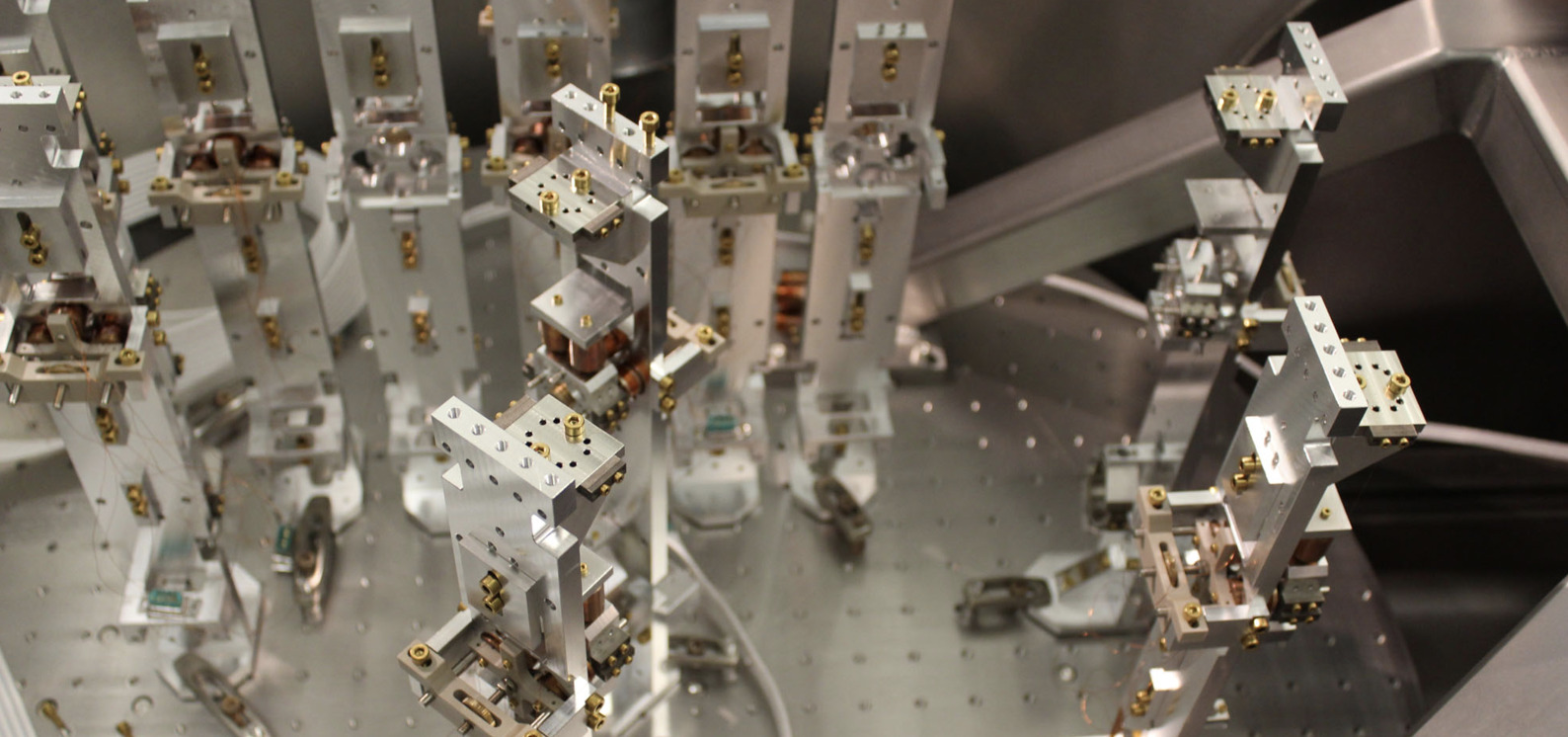} 
\chapter{Ultra--cold vibration control}
\label{sec:cryo}



\section{Review of ET Superattenuator}
\label{sei_concept}

\begin{figure}[ht]
    \centering
    \includegraphics[width = 1\linewidth]{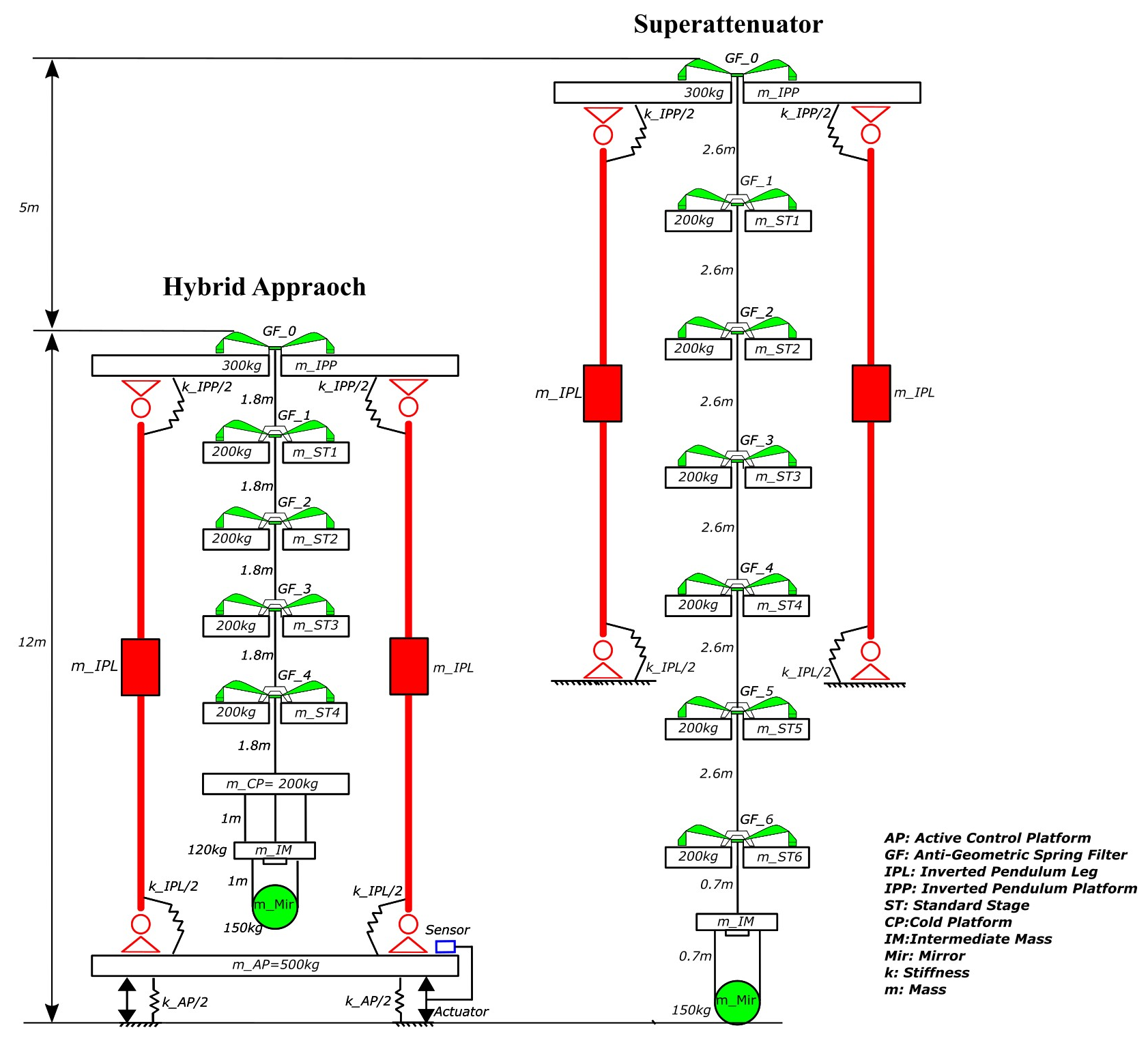}
    \caption{Isolation Systems: ET Superattenuator (Right) and Hybrid Isolation System (Left).}
    \label{fig:HybridSuperattenuator}
\end{figure}

Many investigation campaigns have been conducted since 2010 to define the conceptual design of the ET \cite{sathyaprakash2012scientific,Thoughts2019}. The baseline suspension system is based on the configuration of the Superattenuator isolation system that has been developed by Virgo group but with an upgraded model in order to achieve the desired sensitivity requirement of the ET \cite{accadia2011seismic}. The selection of Virgo Superattenuator as a reference for the ET is because it is more complaint with the ET requirements particularly above 3 Hz and more importantly because its behavior is well confirmed experimentally over several years \cite{abernathy2011einstein}. The desired cross-over frequency (between the antenna sensitivity and horizontal seismic noise) of the ET in the horizontal direction was set to be around 1.8 Hz whereas the one obtained in the Virgo is about 3 Hz \cite{abernathy2011einstein,accadia2011seismic}. Therefore, a maximization algorithm has been applied to attain the performance which ends up with having 17-m long Superattenuator, as shown in Figure \ref{fig:HybridSuperattenuator} (right), that involves six standard multi-cascaded pendulums as well as a payload (marionette and test mirror) \cite{abernathy2011einstein,press1992numerical}. It was noticed that the only way to reduce cross-over frequency so as to extend ET bandwidth below 3 Hz, upon proposed approach, is by increasing the length of the Superattenuator and adds extra pendulum.        
The Virgo Superattenuator however contains one Inverted Pendulum Platform (IPP) stage (7m long with resonance frequency of 40 mHz) \cite{accadia2011seismic}, five cascaded pendulum-masses of standard Geometric Anti-Spring (GAS) filters (each reduces seismic noise by 40 dB in the horizontal and in the vertical Degree Of Freedom (DOF)) and three cascaded pendulum-masses for the payload with an overall height of about 10.5m \cite{acernese2014advanced}. Principally, the Superattenuator was built on the idea of adding pendulums where a typical amount of isolation above a certain cut-off frequency can be achieved just by cascading multiple pendulums. Consequently, the horizontal displacement of the suspension points in an N-stage pendulums is transmitted to the last stage with an attenuation that is proportional to $f^{-2N}$ above its resonance frequencies. Therefore, the ratio between the linear spectral density of the last suspended mass displacement and the linear spectral density of the suspension point displacement decreases as $A/f^{-2N}$ where A = $f_0^{2}$.$f_1^{2}$.$f_2^{2}$.$f_3^{2}$....etc. Hence, cascading multi-pendulums is an efficient way to passively isolate seismic noise above the resonance frequency where the better attenuation is obtained by employing longer pendulums as the resulted resonances are lower \cite{abernathy2011einstein,finn2013gravitational}. 

An IP stage is implemented to limit the amount of seismic noise by offering very low frequency horizontal filtering stage. The IP stage is also used to provide the isolation system with a suspension point positioning system as well as to allow for low-noise control of the mirror by reducing its swinging \cite{losurdo1999inverted,abernathy2011einstein,takamori2002low}. These objectives are achieved by implementing IP which is somehow can be considered (ideal IP) as a massless vertical bar with a certain length that is supporting a mass on its top end and connected to the ground at its bottom end by means of a flexure. The resulting resonance then can be tuned to reach up to 40 mHz \cite{losurdo1999inverted}. The IP however should be carefully designed in order to avoid bistability state if the IP payload is overloaded, and from other side, to avoid flexure’s creep or buckling if materials and total weight are not properly selected \cite{takamori2002low}.   

\section{E-TEST Isolation Approach}
\label{sei_concept}
\subsection{System architecture}

There is a high interest among ET community to reduce the overall height of the proposed Superattenuator of the ET while maintaining the aforementioned performance requirements. This is mainly to reduce the cost and overall complexity of the infrastructure \cite{ET20202} and also the possibility to obtain better isolation performance if other approaches are engaged in the suspension system as highlighted in \cite{Thoughts2019}. In this reference, it has been proposed to involve an active platform in the suspension system or to design a suspension system with two IP platforms instead of only using one IP platform \cite{Thoughts2019}. 
These early researches inspired the suspension concept proposed in E-TEST. It consists of inserting active inertial platform below the Superattenuator suspension system. In doing so, the new isolation system of the E-TEST prototype, shown in Figure \ref{fig:HybridSuperattenuator} (left), combines the LIGO and Virgo approaches, by having active control platform as well as having IP alongside the standard multi-cascaded pendulums, in one unique system.  
As shown in Figure \ref{fig:HybridSuperattenuator}, the new proposed hybrid suspension system contains an Active Platform (AP) that moves freely in 6 DOFs. This active platform is designed on the success of HAM-ISI active platform that was developed for LIGO \cite{kissel2010calibrating}. Then three IPs are mounted on the AP using flexures that move in rotational directions only. Similarly, the upper top of the IPs are attached with the Inverted Pendulum Platform (IPP) by flexures that also move in rotational directions only.The horizontal motion of the IPP is resulted from the mechanism behavior of the IP and therefore IPP is not really restricted from moving in horizontal direction. The Geometric Anti-Spring (GAS) filter afterwards is mounted on the IPP, forming the first standard GAS filter (GF-0) which then followed by other four standard stages of GAS filter. The Cold Platform (CP) is suspended from last standard GAS filter (GF-4). On the other side, the CP is attached to the next stage that is called Intermediate Mass (IM) by three connected wires. This platform is mainly proposed to provide a means for compensating the drift as well as steering the mirror in the desired position. Lastly, the Mirror (Mir) is hanging from IM via 4 wires as shown in Figure \ref{fig:HybridSuperattenuator}. An analytical model using lumped masses connected by springs can be derived to study the dynamics of the system.

The individual stiffnesses that are used in the model for the vertical direction are further defined as follows (the variables are defined in Figure \ref{fig:HybridSuperattenuator} and Table \ref{tab:E-TEST_abbreviation}): 

\begin{equation}\label{APzStiffness}
K_{AP_z} = m_{AP}(2\times\pi\times1.7)^{2}
\end{equation}

\begin{equation}\label{IPLzStiffness}
K_{IPL_z} = m_{IPL}(2\times\pi\times100)^{2}
\end{equation}

\begin{equation}\label{IPPzStiffness}
K_{IPP_z} = m_{IPP}(2\times\pi\times100)^{2}
\end{equation}

\begin{equation}\label{GF0zStiffness}
K_{GF0_z} = (m_{ST1}+m_{ST2}+m_{ST3}+m_{ST4}+m_{CP}+m_{IM}+m_{Mir})(2\times\pi\times0.25)^{2}
\end{equation}

\begin{equation}\label{GF1zStiffness}
K_{GF1_z} = (m_{ST2}+m_{ST3}+m_{ST4}+m_{CP}+m_{IM}+m_{Mir})(2\times\pi\times0.25)^{2}
\end{equation}

\begin{equation}\label{GF2zStiffness}
K_{GF2_z} = (m_{ST3}+m_{ST4}+m_{CP}+m_{IM}+m_{Mir})(2\times\pi\times0.25)^{2}
\end{equation}

\begin{equation}\label{GF3zStiffness}
K_{GF3_z} = (m_{ST4}+m_{CP}+m_{IM}+m_{Mir})(2\times\pi\times0.25)^{2}
\end{equation}

\begin{equation}\label{GF4zStiffness}
K_{GF4_z} = (m_{CP}+m_{IM}+m_{Mir})(2\times\pi\times0.25)^{2}
\end{equation}

\begin{equation}\label{IMzStiffness}
K_{IM_z} = \frac{n\times E\times\pi\times d^{2}}{4\times l}
\end{equation}

\begin{equation}\label{MirzStiffness}
K_{Mir_z} = \frac{1}{\frac{1}{K_{Mir_{Wire}}}+\frac{1}{K_{Mir_{Cantilever}}}}
\end{equation}

\begin{equation}\label{MirWirezStiffness}
K_{Mir_{Wire}} = \frac{n\times E\times\pi\times d^{2}}{4\times l}
\end{equation}

\begin{equation}\label{MirCantileverStiffness}
 K_{Mir_{Cantilever}}=m_{Mir}(2\times\pi\times13)^{2}
\end{equation}

It is noted from equation \ref{APzStiffness} that the resonance frequency of the active platform is selected to be 1.7 Hz. This is achieved by applying proper designed flexure similar to those implemented by LIGO in HAM-ISI active platform \cite{HAMISIBrian}. However, the resonances of the IP (equation \ref{IPLzStiffness} and equation \ref{IPPzStiffness}) are set to 100 Hz. This is because the IP is stiff in the vertical direction. On the other side, equations \ref{GF0zStiffness}-\ref{GF4zStiffness} show the values of the stiffnesses for the GAS filters which depend on the suspended masses from the filters and also depend on their individual resonance frequency which is here set to $0.25$ Hz. Such resonance frequency of the GAS filter is experimentally achieved in \cite{BERTOLINI1999475}. Furthermore, the stiffness value of the intermediate stage depends mainly on the wire property as seen in equation \ref{IMzStiffness}. So far, the wire selected for that is based on fused silica with a Young modulus E of $67 \times 10^9$  N/m$^{2}$ and wire diameter d of $1$ X $10^{-3}$ m. Lastly, equation \ref{MirzStiffness} shows that the stiffness value relies also on the wire property as well as triangular cantilever spring. This cantilever spring is attached in parallel to the suspended wire which adds extra isolation in the vertical direction. The experimental result shows that this cantilever spring can be designed to achieve a resonance frequency up to $13$ Hz. This cantilever spring is a property of the payload and not of the seismic platform (pre-isolator). The mirror is suspended by means of machined silicon rods that are very rigid vertically. Therefore, in order to make the mirror suspension assembly possible, metal cantilever springs have been introduced at the top end of each rod to provide some vertical compliance, resulting in a 13 Hz bounce mode of the last suspension stage. 

Similarly, the individual stiffnesses for the horizontal direction can be further presented as follows: 
\begin{equation}\label{APxStiffness}
K_{AP_x} = m_{AP}(2\times\pi\times1.7)^{2}
\end{equation}

\begin{equation}\label{IPLxStiffness}
K_{IPL_x} = m_{IPL}(2\times\pi\times30)^{2}
\end{equation}

\begin{equation}\label{IPPxStiffness}
K_{IPP_x} = (m_{IPL}+m_{IPP}+m_{ST1}+m_{ST2}+m_{ST3}+m_{ST4}+m_{CP}+m_{IM}+m_{Mir})(2\times\pi\times0.07)^{2}
\end{equation}

\begin{equation}\label{GF0xStiffness}
K_{GF0_x} = (m_{ST1}+m_{ST2}+m_{ST3}+m_{ST4}+m_{CP}+m_{IM}+m_{Mir})(\frac{g}{l_{(IPP-SF1)}})
\end{equation}

\begin{equation}\label{GF1xStiffness}
K_{GF1_x} = (m_{ST2}+m_{ST3}+m_{ST4}+m_{CP}+m_{IM}+m_{Mir})(\frac{g}{l_{(SF1-SF2)}})
\end{equation}

\begin{equation}\label{GF2xStiffness}
K_{GF2_x} = (m_{ST3}+m_{ST4}+m_{CP}+m_{IM}+m_{Mir})(\frac{g}{l_{(SF2-SF3)}})
\end{equation}

\begin{equation}\label{GF3xStiffness}
K_{GF3_x} = (m_{ST4}+m_{CP}+m_{IM}+m_{Mir})(\frac{g}{l_{(SF3-SF4)}})
\end{equation}

\begin{equation}\label{GF4xStiffness}
K_{GF4_x} = (m_{CP}+m_{IM}+m_{Mir})(\frac{g}{l_{(SF4-CP)}})
\end{equation}

\begin{equation}\label{IMxStiffness}
K_{IM_x} = (m_{IM}+m_{Mir})(\frac{g}{l_{(CP-IM)}})
\end{equation}

\begin{equation}\label{MirxStiffness}
K_{Mir_x} = (m_{Mir})(\frac{g}{l_{(IM-Mir)}})
\end{equation}

Similarly, the stiffness value presented in equation \ref{APxStiffness} shows that the resonance frequency of the AP in the horizontal direction is set to 1.7 Hz. This can be achieved by implementing cantilever spring similar to those employed by LIGO in HAM-ISI suspension system \cite{HAMISIBrian}. On the other side, the stiffness value between the AP and IP leg is selected to be 30 Hz as depicted in equation \ref{IPLxStiffness}. Again, this resonance frequency is selected to be large enough since the motion is restricted in the horizontal direction. However, the resonance frequency of the IPP, shown in equation \ref{IPPxStiffness}, is chosen to be 0.04 Hz as this value is experimentally obtained in Virgo Superattenuator as presented in \cite{accadia2011seismic}. Lastly, equations \ref{GF0xStiffness}-\ref{MirxStiffness} show that the stiffness values are depending on the cascaded pendulum masses and wire length under gravity force effects. 
\subsection{Closed-Loop Performance}
In order to examine the closed loop performance of the E-TEST suspension system, two controllers are designed; one for horizontal, shown in equation \ref{ControllerH}, and one for vertical, shown in equation \ref{ControllerV}. Although quite aggressive, it is believed that these controllers remain realistic with a careful mechanical design and low noise instruments.  
\begin{equation}\label{ControllerH}
C_{h} = 8.9691\times10^{7}\times\frac{(s+69.59)(s+0.446)}{(s+3809)(s+0.009174)}
\end{equation}
\begin{equation}\label{ControllerV}
C_{v} = 8.9675\times10^{7}\times\frac{(s+54.99)(s+0.1318)}{(s+1563)(s+0.005429)}
\end{equation}

\begin{figure}[ht]
    \centering
    \includegraphics[width = 0.50\linewidth]{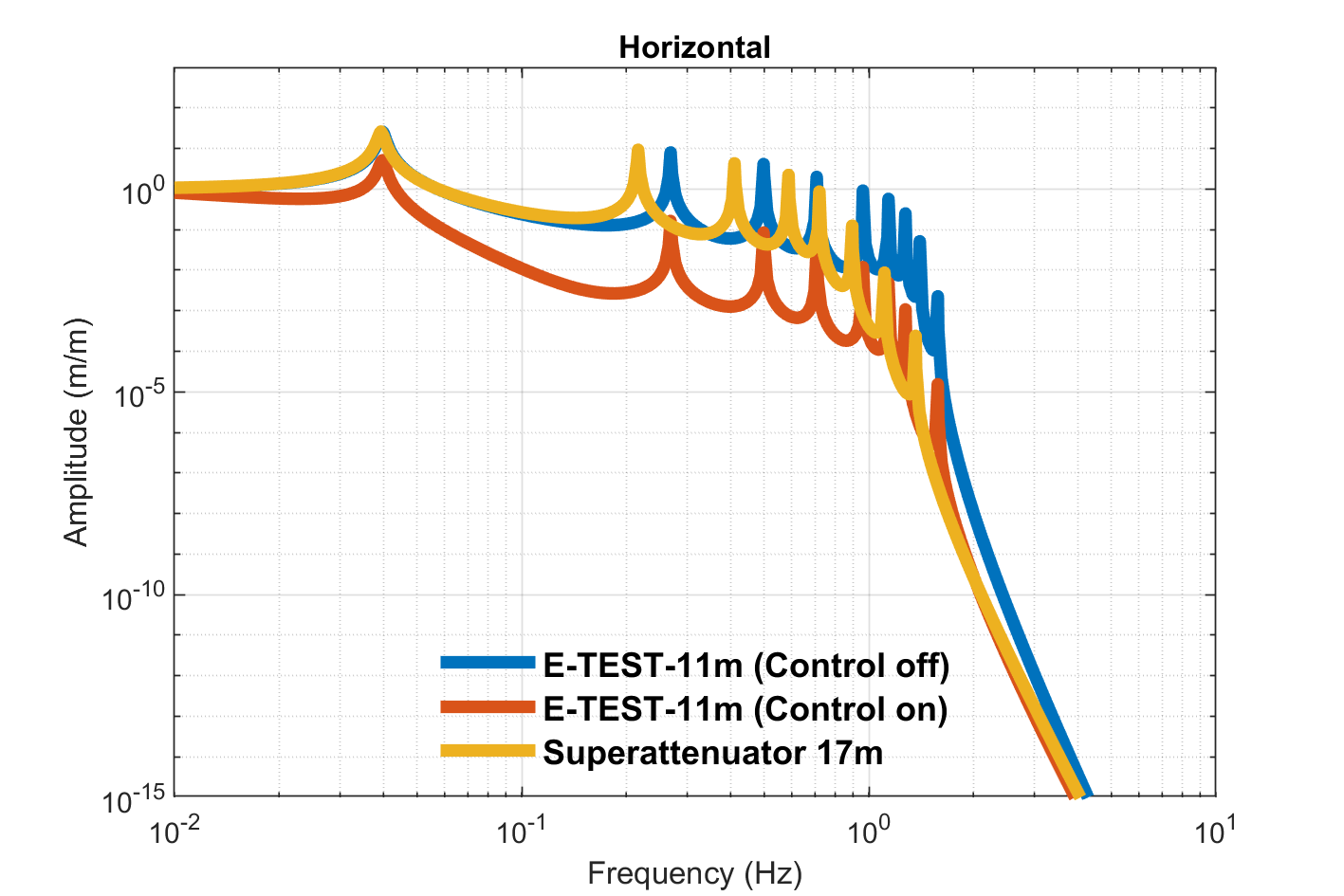}
    \includegraphics[width = 0.49\linewidth]{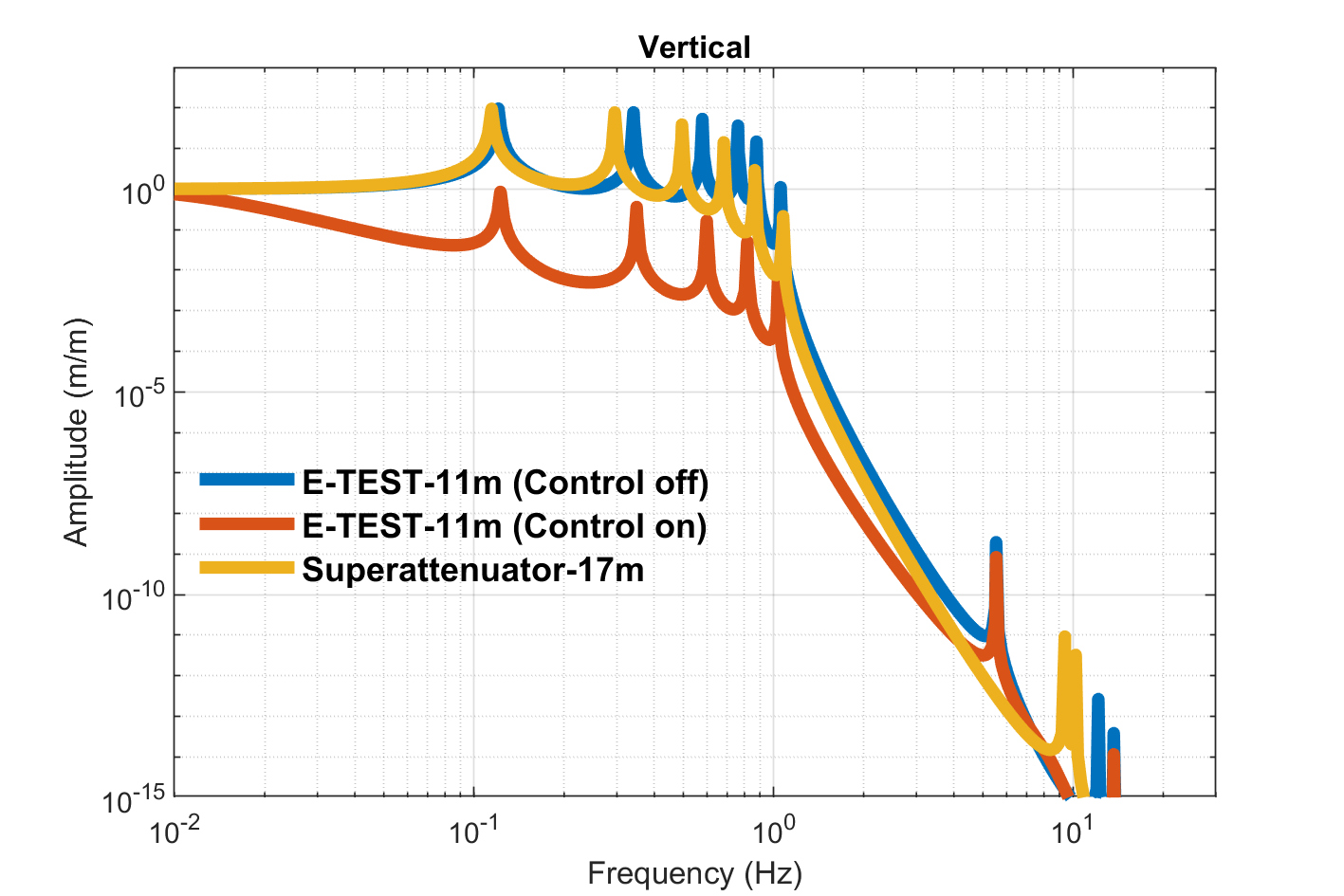}
    \caption{Transfer Function (Ground/Mirror) of Superattenuator and Hybrid Isolation System.}
    \label{fig:Transmissibility}
\end{figure}

\begin{figure}[H]
    \centering
    \includegraphics[width = 0.50\linewidth]{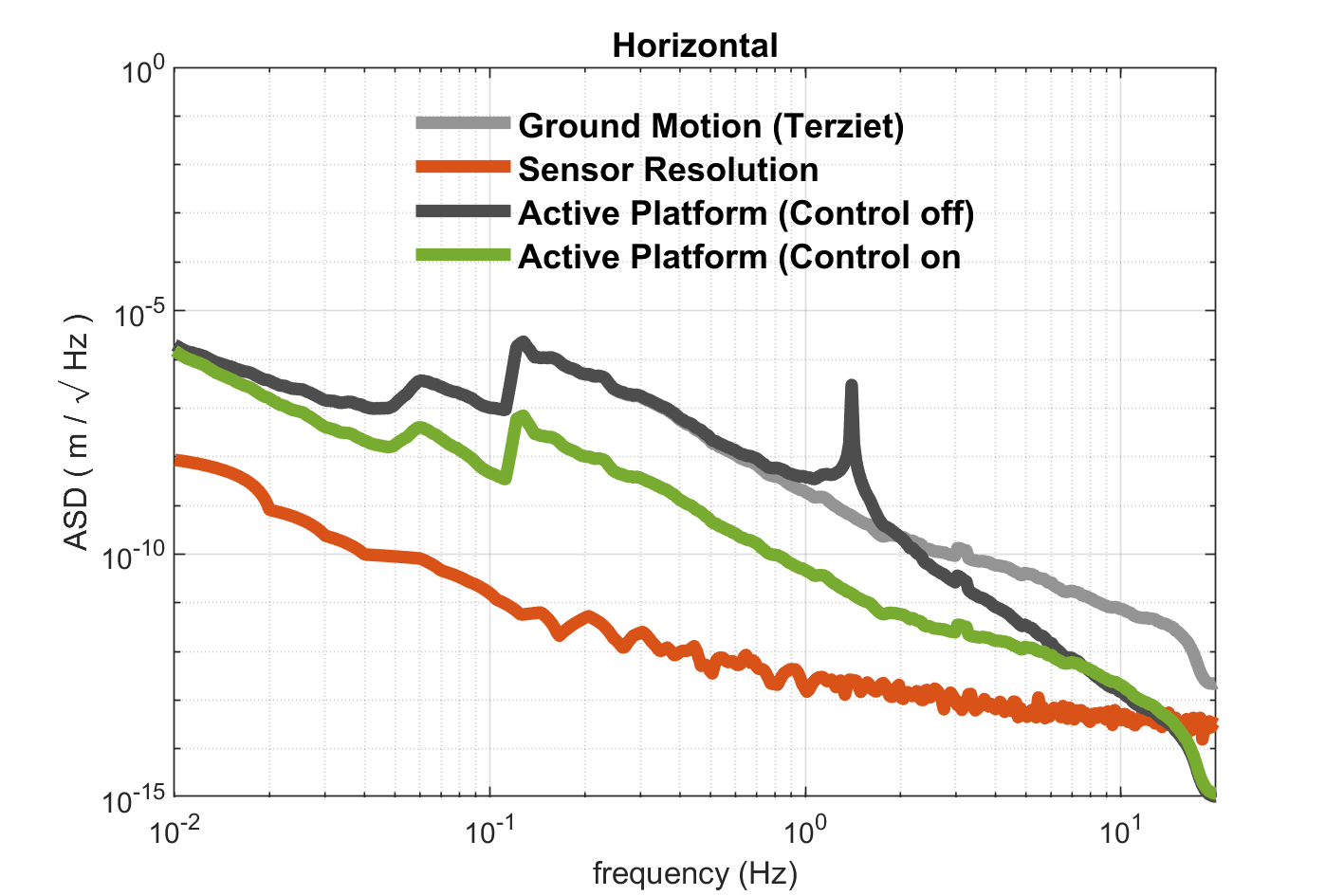}
    \includegraphics[width = 0.49\linewidth]{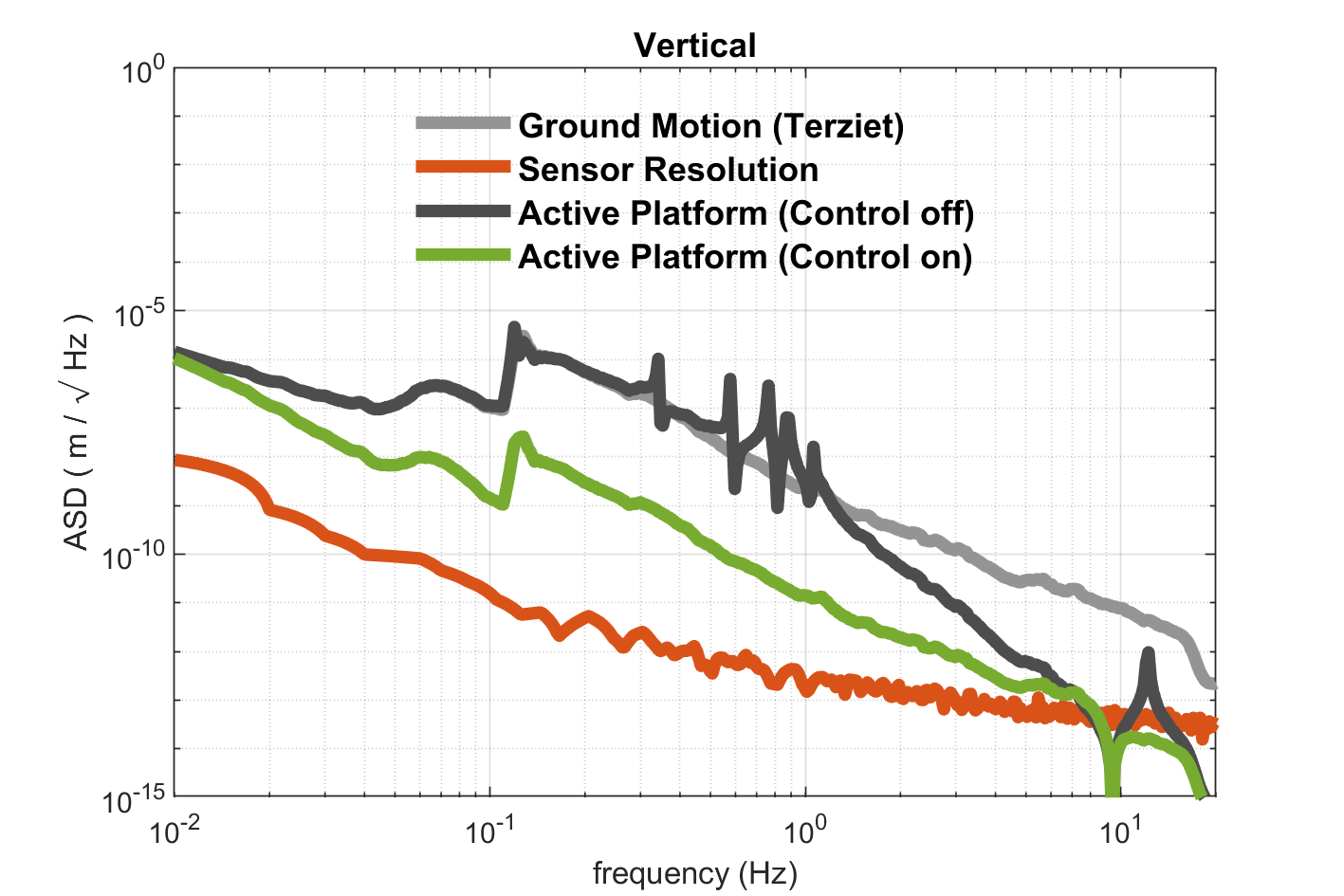}
    \caption{Active Platform Motion of E-TEST Isolation System.}
    \label{fig:ASDAP}
\end{figure}

Figure \ref{fig:Transmissibility} presents the output performance of the transmissibility (from ground to the last stage (mirror)) for the open loop (blue curve) as well as for the closed loop (red curve) of the E-TEST suspension system. In addition, the transmissibility behavior of the ET Superattenuator isolation system (yellow curve) is presented in the same figure. It is clearly seen from Figure \ref{fig:Transmissibility} that the ET Superattenuator obtains the desired cross over frequency (1.8 Hz @ 10$^{-9}$) with on overall suspension chain of about 17m. However, the E-TEST suspension system can achieve the same cross over frequency with only 11m suspension chain.
Furthermore, the E-TEST approach introduces extra low frequency isolation of about two orders of magnitude around 1 Hz for the horizontal and also for the vertical directions. This isolation is only attainable by implementing active inertial control such as in the E-TEST approach. This is because ET Superattenuator approach depends only on the passive isolation strategy which can add a certain amount of isolation but only above the resonance of the structure.    
The Amplitude Spectral Densities (ASD) of the active inertial platform motion is shown in Figure \ref{fig:ASDAP}. For comparison, the figure also shows the sensor resolution as well as the ground motion. The sensor resolution which is considered in this study is based on the Horizontal Inertial Sensor (HINS) and Vertical Inertial Sensor (VINS) that were developed at Precision Mechatronics Laboratory (PML) – Belgium \cite{ding2021development}. Moreover, the ground motion used in this study  was measured in Terziet, 250m below the surface of the ground. In addition to seismic isolation, Figure \ref{fig:ASDAP} clearly shows that the active control successfully damp the resonance of the platform. 
 
\begin{figure}[H]
    \centering
    \includegraphics[width = 0.50\linewidth]{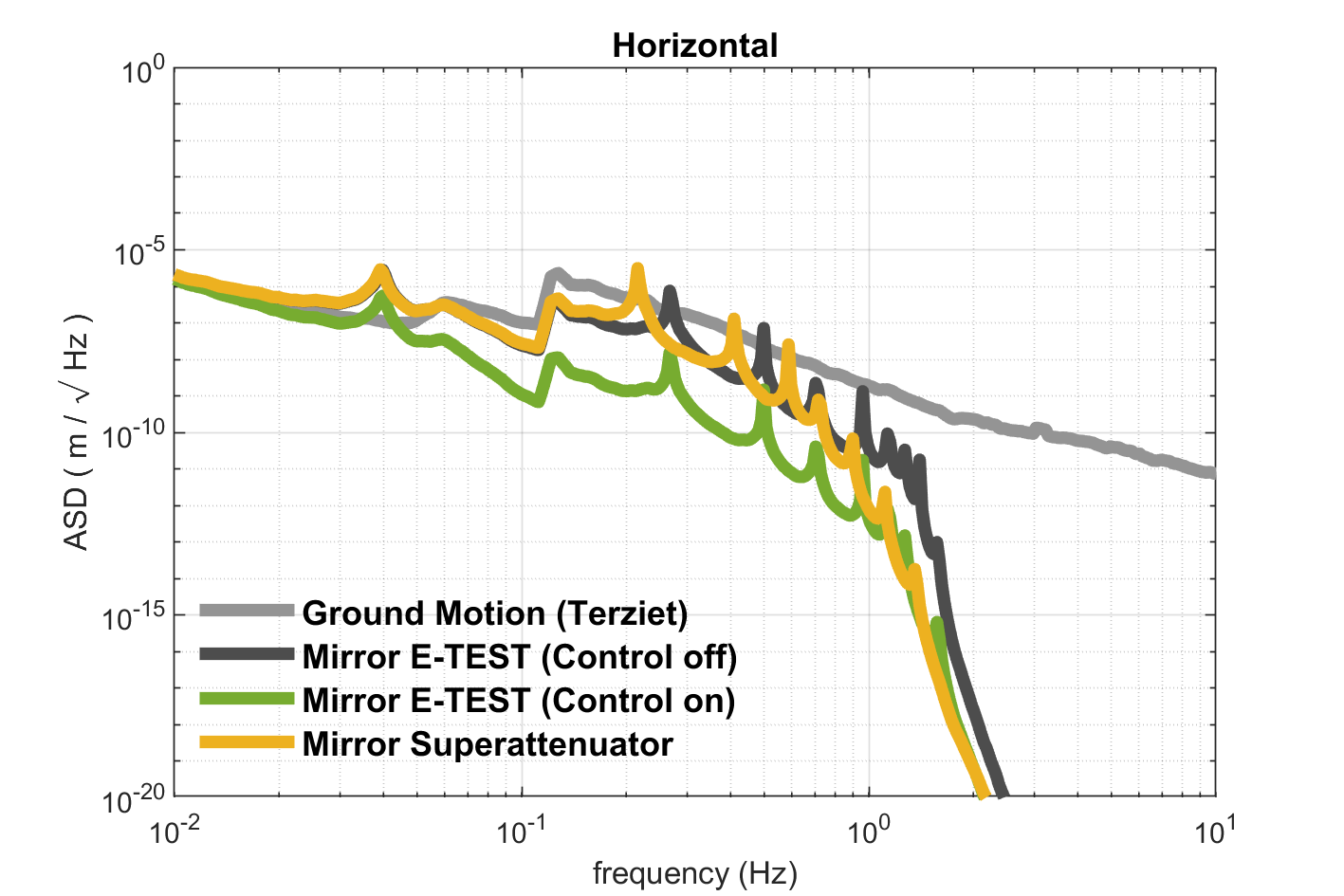}
    \includegraphics[width = 0.49\linewidth]{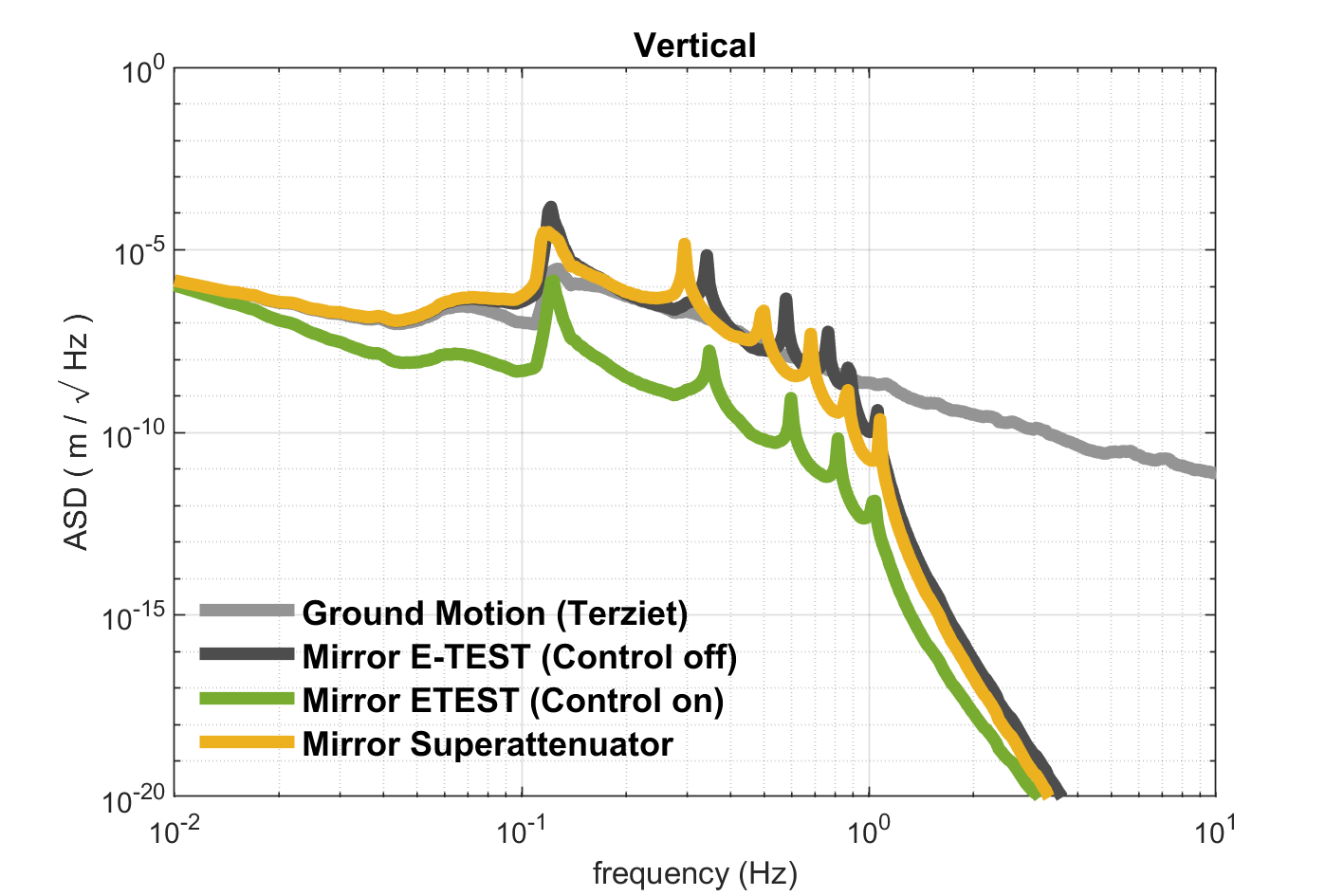}
    \caption{Test Mirror Motion of Superattenuator and Hybrid Isolation System.}
    \label{fig:ASDMirror}
\end{figure}

Furthermore, the ASD for the last stage (test mirror) when considering the entire suspension chain of the ET Superattenuator as well as the entire E-TEST suspension system are further shown in Figure \ref{fig:ASDMirror}. It is shown that the isolation performance of the E-TEST system in the closed loop (green curve) is superior compared to the performance of the open loop (black curve) at low frequency. Consequently, using the E-TEST approach,
the mirror is also more stable in that frequency range.
     
\section{E-TEST Prototype Modeling}
\label{sei_concept}
\subsection{Isolation System}

\begin{figure}[H]
    \centering
     \includegraphics[width = 1\linewidth]{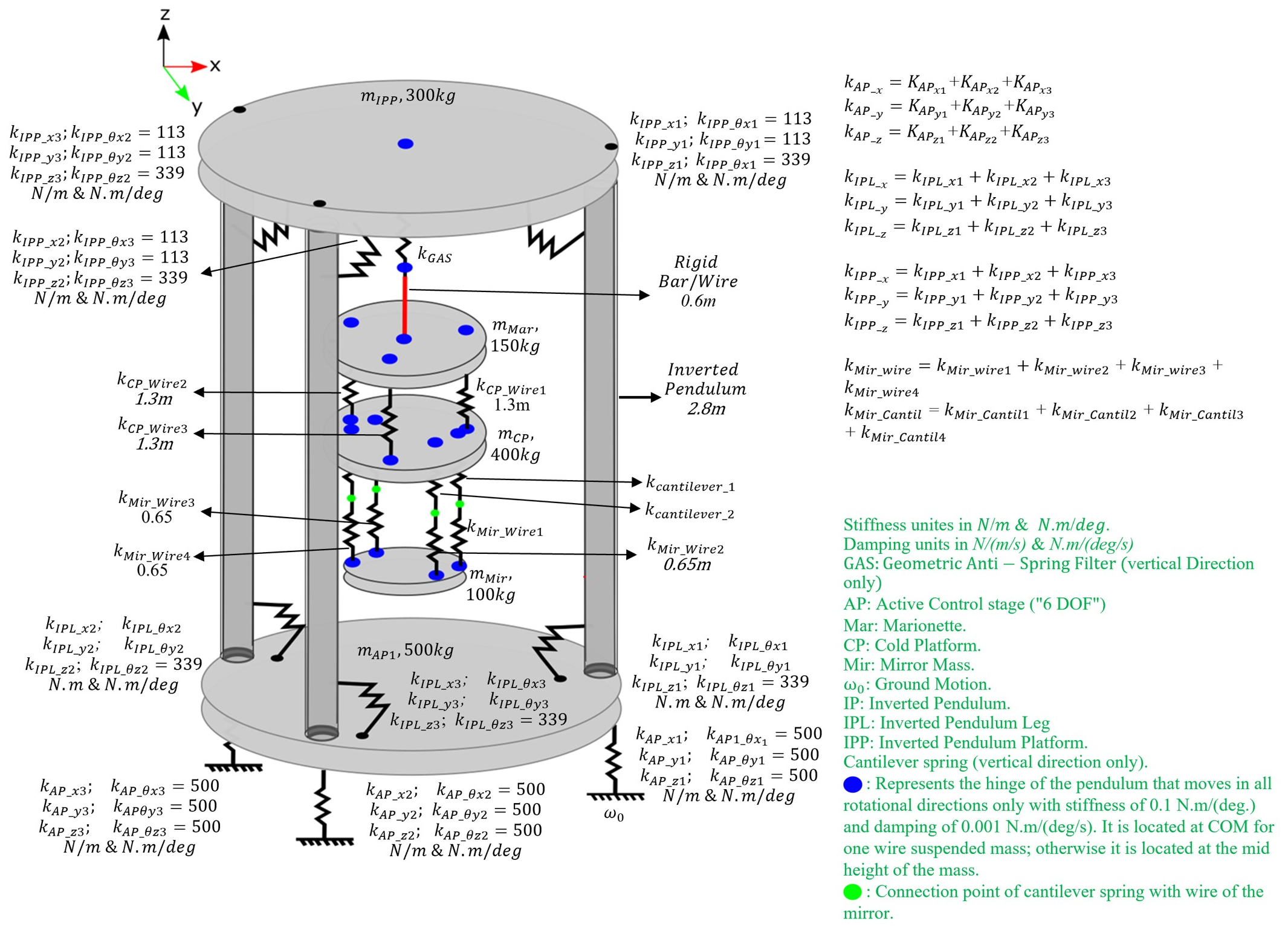}
    \caption{Analytical Schematic Model of E-Test Isolation Prototype.}
    \label{fig:ETestSuspensionSystem6DOF}
\end{figure}

In order to validate experimentally the isolation concept presented in the previous section, it is proposed to build a prototype of hybrid isolation system. The prototype will be full scale with a 100kg cryogenic silicon mirror. However, it will have less stages due to space and budget limitations. A sketch of the prototype is shown in Figure \ref{fig:ETestSuspensionSystem6DOF}. 

The notations of the schematic model are provided in Table \ref{tab:E-TEST_abbreviation}. Overall, the E-TEST suspension system contains an Active Platform (AP) that moves freely in 6 DOFs. It contains three springs aligned horizontally (x-axis or y-axis) and three springs aligned vertically (z-axis). Then three IPs are mounted on the AP through flexural hinges. Similarly, the upper top of the IPs are attached to the IPP by flexures which also move in rotational directions only. The GAS filter is then mounted on the IPP. Thereafter, a Marionette (Mar) platform is suspended from the GAS filter by one wire. This platform is mainly used for the alignment and for drift compensation of the cryogenic part. The cryogenic part involves a Cold Platform (CP) and a large Mirror (Mir) which will be operated at cryogenic temperature (20 Kelvin). The CP is designed to mount the cryogenic sensors. The cryogenic part however is suspended from Mar platform via 3 wires. Lastly, the test mirror is hanging from CP via 4 wires as shown in the Figure  \ref{fig:ETestSuspensionSystem6DOF}. Overall, the entire suspension system will be less than 5m high. 

\begin{table}[H]
    \centering
    \begin{tabular}{|c|c|c|}
        \hline
        AP & Active control platform \\
        \hline
        SF & Standard filter \\
        \hline
        IPL & Inverted pendulum leg \\
        \hline
        IPP & Inverted pendulum platform \\
        \hline
        CP & Cold platform \\
        \hline
        IM & Intermediate Mass\\
        \hline
        Mir & Mirror \\
        \hline
        GAS & Geometric-Anti Spring  \\
        \hline
        $m_{AP}$ & Mass of first active platform  \\
        \hline
        $m_{IPL}$ & Mass of inverted pendulum leg \\
        \hline
        $m_{IPP}$ & Mass of inverted pendulum Platform \\
        \hline
        $m_{ST}$ & Mass of standard filter \\
        \hline
        $m_{CP}$ & Mass of cold mass \\
        \hline
        $m_{IM}$ & Mass of intermediate mass \\
        \hline
        $m_{Mir}$ & Mass of mirror \\
        \hline
        $z_{i}$ & vertical displacement of stage $i$ \\
        \hline
        $x_{i}$ & Horizontal displacement of stage $i$ \\
        \hline
        $K_{AP}$ & Stiffness of AP  \\
        \hline
        $K_{IPL}$ & Stiffness of IPL \\
        \hline
        $K_{IPP}$ & Stiffness of IPP \\
        \hline
        $K_{ST}$ & Stiffness of ST \\
        \hline
        $K_{IM}$ & Stiffness of IM \\
        \hline
        $K_{Mir}$ & Stiffness of Mir \\
        \hline
        $f_{a}$ & External force   \\
        \hline
        E & Young modulus of the wire    \\
        \hline
        l & Pendulum wire length     \\
        \hline
        $w_{0}$ & Ground motion \\
        \hline
        g & Gravity acceleration  \\
        \hline
        \end{tabular}
    \caption{Notations Used in the Sketch of the E-TEST Isolation System.}
    \label{tab:E-TEST_abbreviation}
\end{table}

\label{sei_concept}
\subsection{Dynamics and Modal Analysis}

\begin{figure}[H]
    \centering
     \includegraphics[width = 1\linewidth]{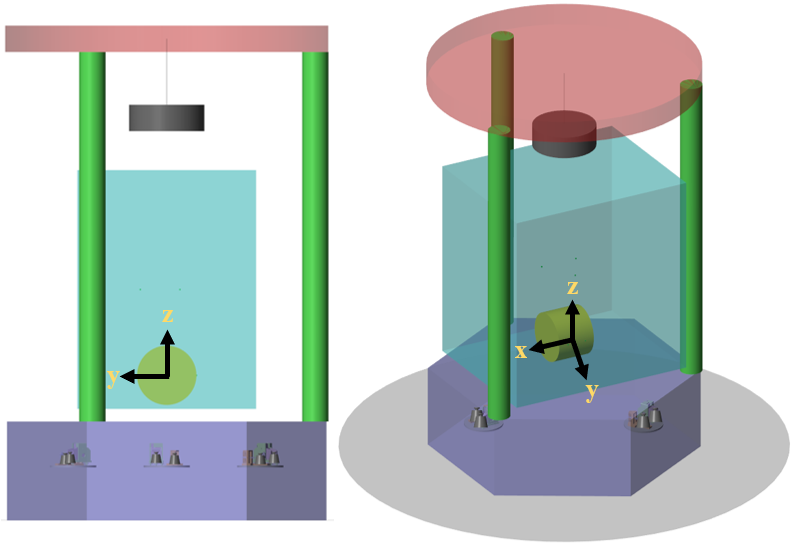}
    \caption{3D View of the E-TEST Simscape Model (Left: Front View) and (Right: Isometric View).}
    \label{fig:Simscape}
\end{figure}

In order to study the dynamics of the prototype in 3D, a multi-body model has been developed using Simscape. It is a Matlab toolbox allowing to study lumped mass systems under gravity. Since it works under the Simulink environment, it is also a convenient tool for implementing feedback control strategies.   
A 3D view of the Simscape model is shown in Figure \ref{fig:Simscape}.

A deeper understanding of the model dynamics is provided by computing the mode shapes. This is achieved by building Finite Element (FE) model on the Structural Dynamic Toolbox (SDT) \cite{SDT}. From the Simscape model, the mode shapes have been obtained in three steps: 1) extraction of a state space model from Simscape, 2) calculation of eigenvalues and eignvectors in Matlab and 3) projection of these modes on a finite element representation of the system.   

\begin{figure}[H]
    \centering
     \includegraphics[width = 1\linewidth]{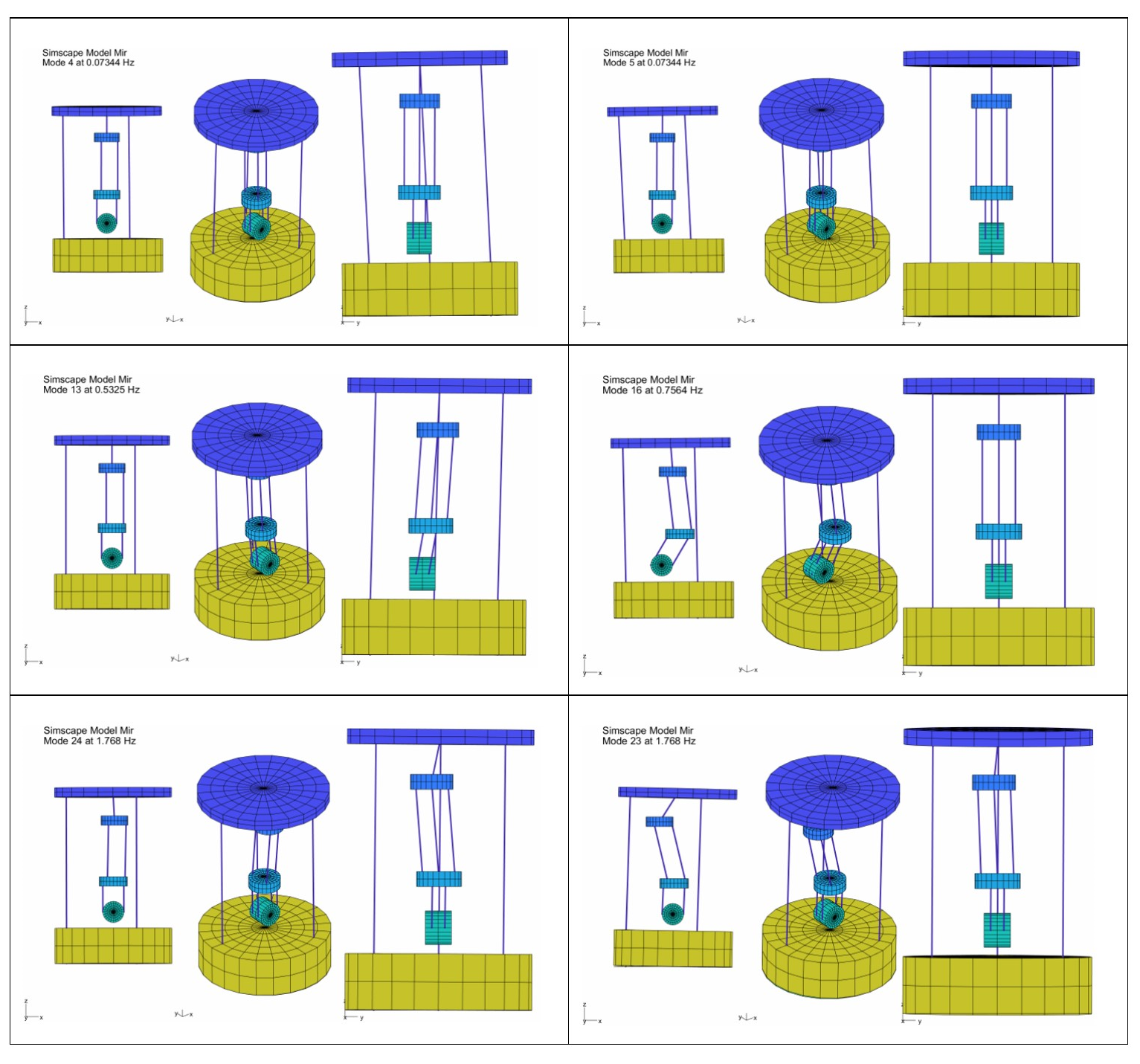}
    \caption{Horizontal Mode Shapes (x-axis).}
    \label{fig:ModeShapeHorizontal}
\end{figure}

\begin{figure}[H]
    \centering
     \includegraphics[width = 1\linewidth]{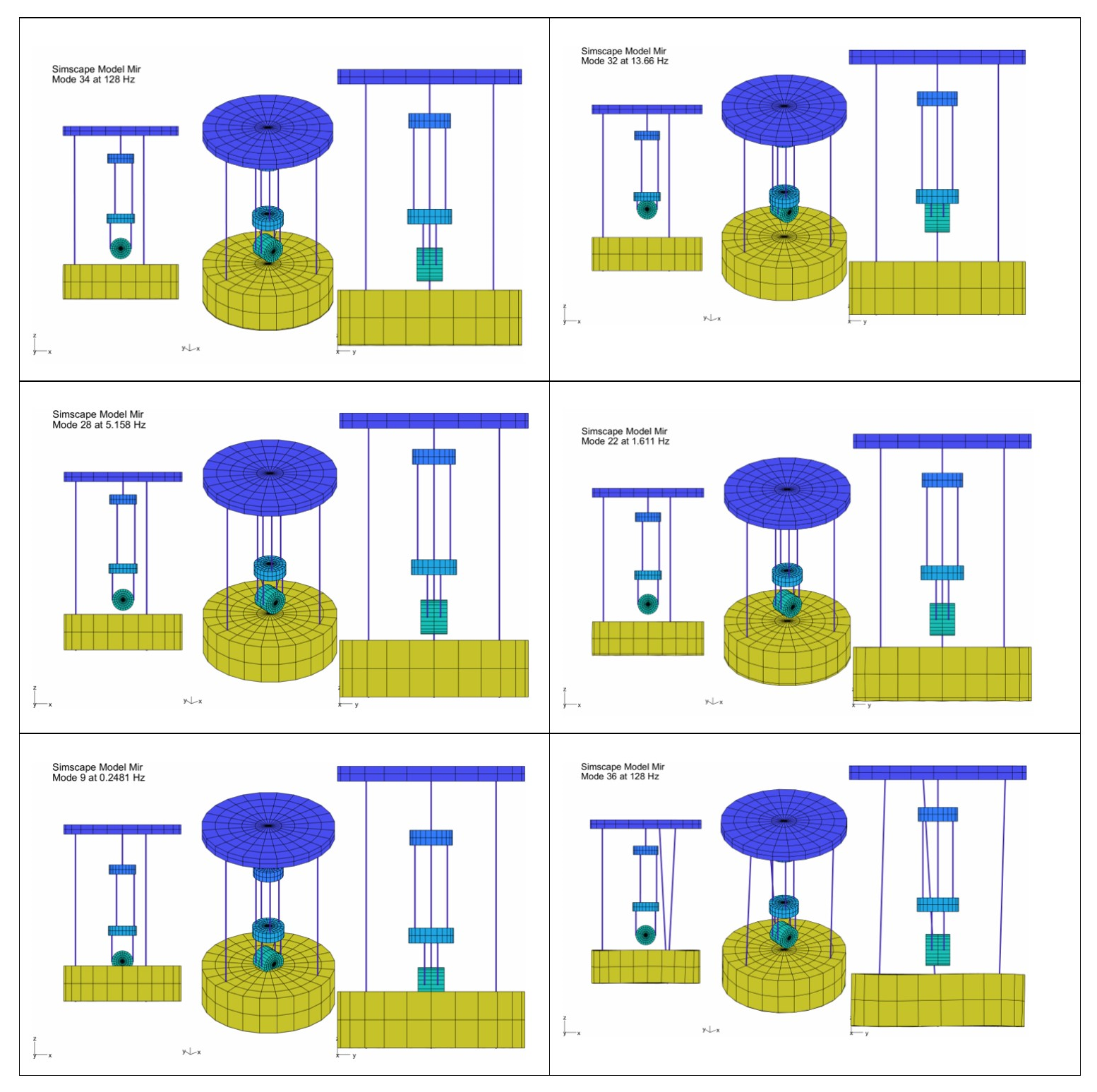}
    \caption{Vertical Mode Shapes (z-axis).}
    \label{fig:ModeShapeVertical}
\end{figure}

As an illustration, horizontal mode shapes (x-axis) are shown in Figure \ref{fig:ModeShapeHorizontal} and vertical (z-axis) mode shapes are shown in Figure \ref{fig:ModeShapeVertical}.   

\subsection{Closed-Loop Performance}
\begin{figure}[H]
    \centering
    \includegraphics[width = 0.8\linewidth]{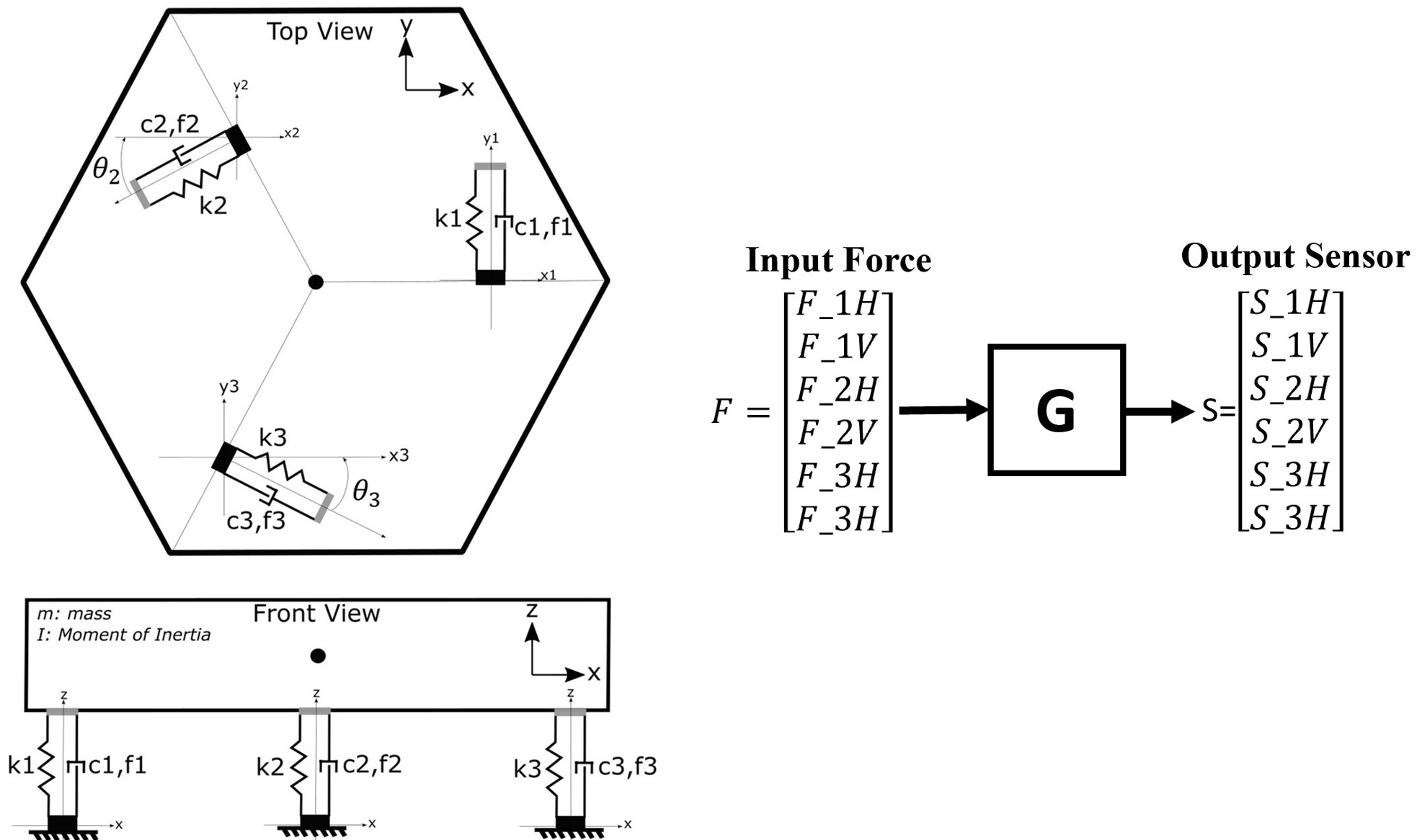}
    \caption{Active Platform (left) and Input/Output Block Representation (right) of the Active Platform.}
    \label{fig:ActivePlatform}
\end{figure}

\begin{figure}[H]
    \centering
    \includegraphics[width = 0.49\linewidth]{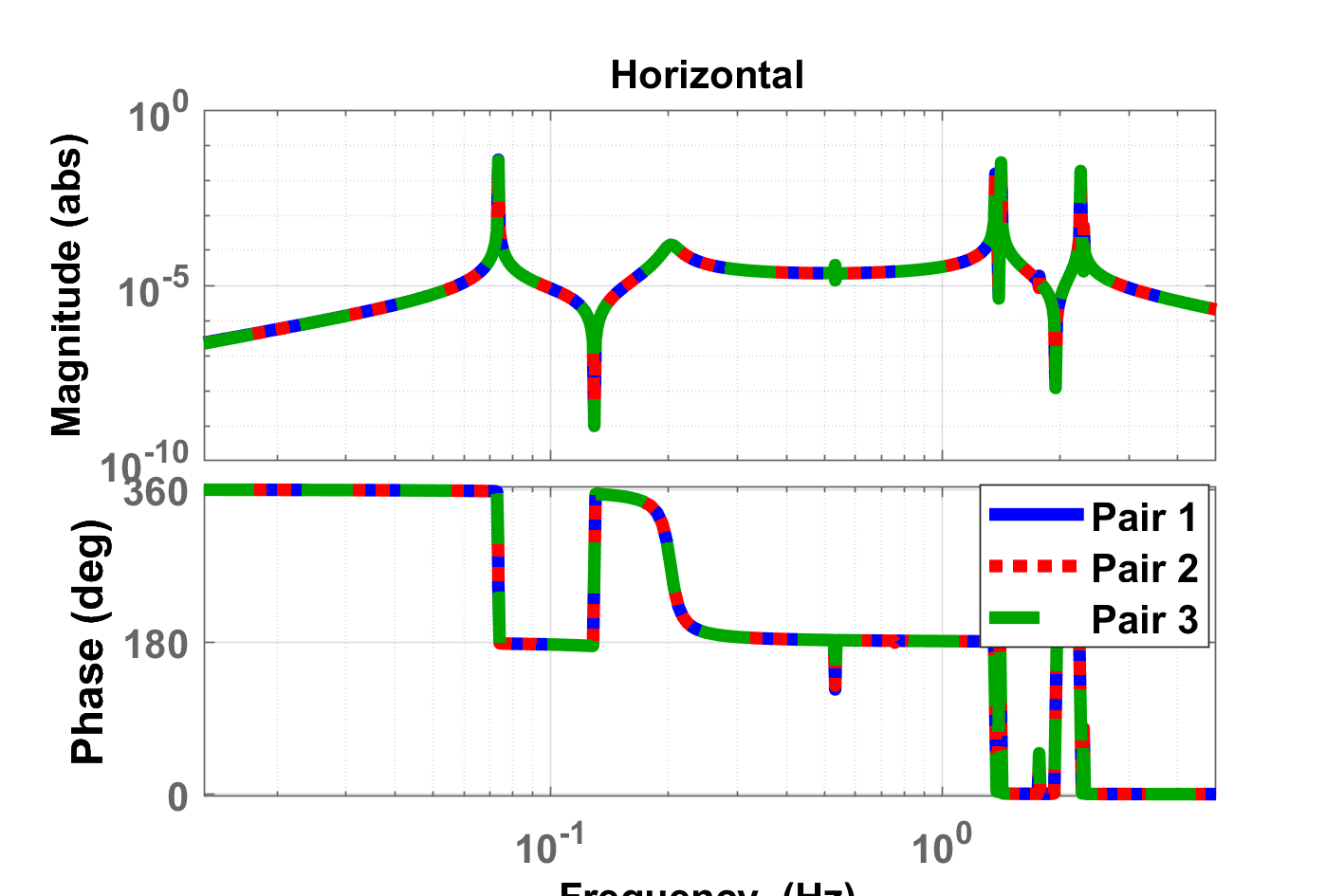}
    \includegraphics[width = 0.49\linewidth]{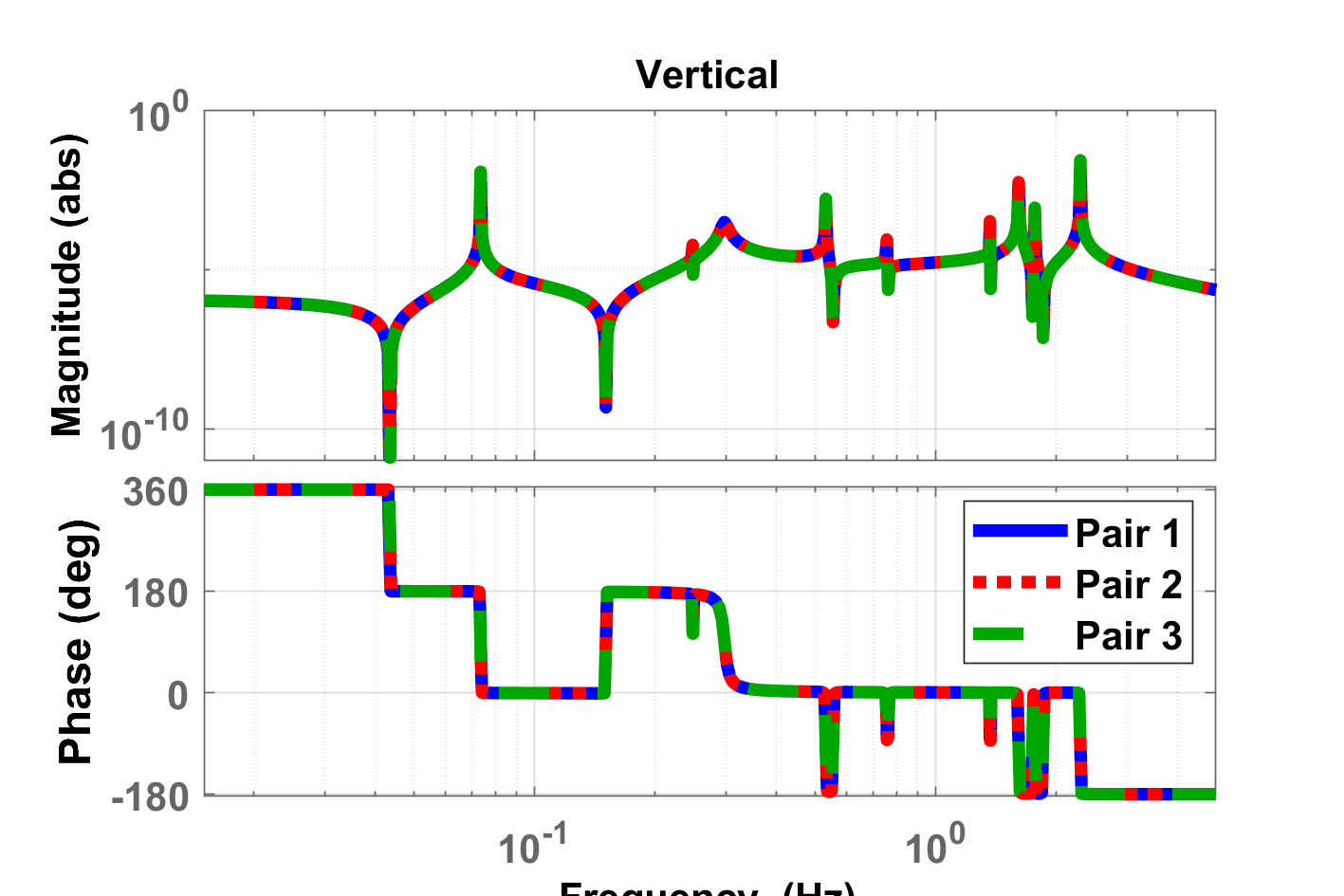}
    \caption{Transfer Function of Each Sensor/Actuator Pair of the Active Platform.}
    \label{fig:TransferFunctionSensorActuator}
\end{figure}

A schematic diagram for the active platform of the E-TEST is further shown in Figure \ref{fig:ActivePlatform} (left). It contains six actuators (3 for horizontal directions and 3 for vertical directions) and six inertial sensors (3 for horizontal directions (HINS) and 3 for vertical directions (VINS)).  The interferometer used in these inertial sensors are re-calibrated very regularly. In addition, High pass filters (with a very low corner frequency, around 10mHz) are used in order to avoid drifts in the control signals. A block diagram of the input/output is depicted in Figure \ref{fig:ActivePlatform} (right).      
The transfer function of each pair of sensor/actuator are shown in Figure \ref{fig:TransferFunctionSensorActuator}. These curves have been extracted from the full model. 
Due to the symmetry, the three vertical ones are identical and the three horizontal ones are identical as well.   
  
\begin{figure}[H]
    \centering
    \includegraphics[width = 0.8\linewidth]{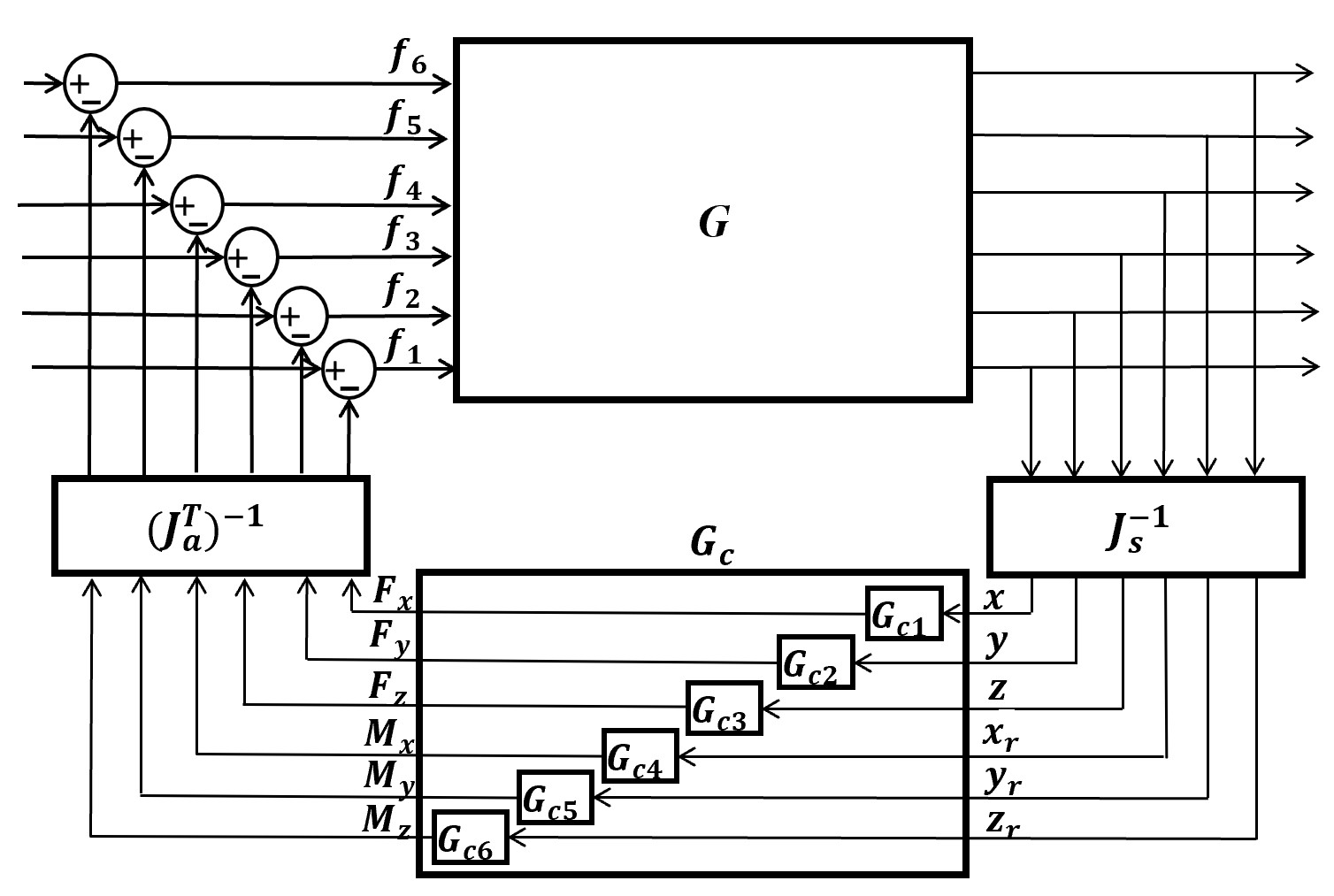}
       \caption{Schematic Representation of the Centralized Control of the AP.}
    \label{fig:ControllerSchematic}
\end{figure}

 The off-diagonal elements of the transfer matrix (not shown in this report) are of the same order of magnitude as the diagonal ones. It means that, in order to control the platform, a decoupling strategy has to be applied.  
 As an example, a decoupling strategy based on a Jacobian approach has been used to minimize the coupling and move from the local frame to the Cartesian frame of the AP. This is schematically represented in Figure \ref{fig:ControllerSchematic}, where $J_a$ is the actuator Jacobian and $J_s$ is sensor Jacobian. These Jacobian matrices are mainly depending on the geometry of the active platform as well as on the position of the sensors and actuators on the active platform.     

Once the plant is decoupled, each degree of freedom can be controlled individually as shown in the same figure. The poles and zeros of the controllers are designed based on a manual tuning (loop shaping) in the SISO tool - Matlab. This method is efficient in terms of understanding the behavior of the system and also the positions of the poles/zeros of the controller. Such a method is considered time consuming and the controller needs to be re-designed in case of any modification in the parameters of the system. Therefore, a more advanced method based on H-infinity loop shaping is going to be implemented later which can define the optimal controller automatically based on the given specifications and restrictions.  

\begin{figure}[H]
    \centering
    \includegraphics[width = 0.49\linewidth]{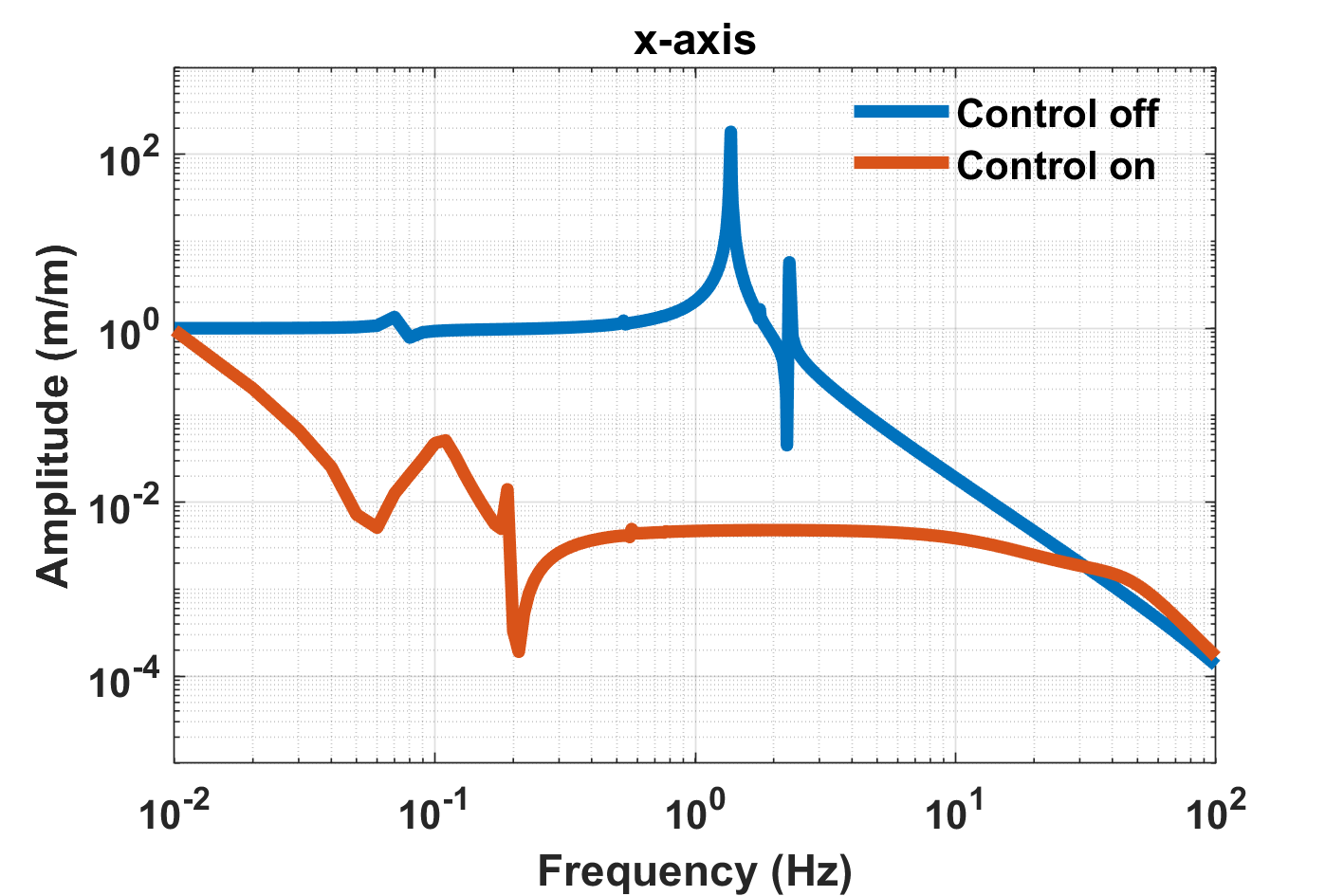}
    \includegraphics[width = 0.49\linewidth]{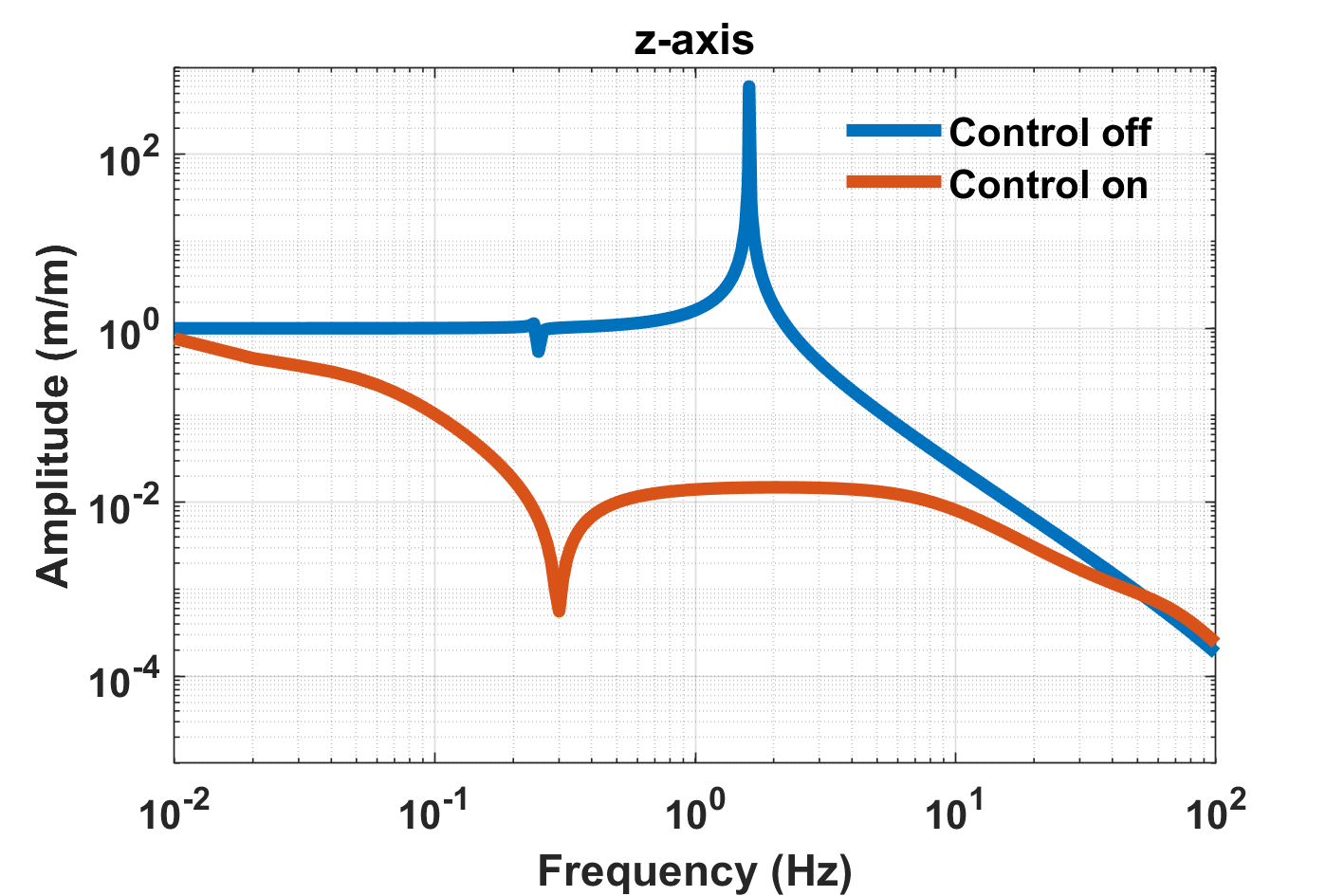}
    \caption{Transmissibilities from Ground to AP.}
    \label{fig:TransmissibilityAPETEST}
\end{figure}

\begin{figure}[H]
    \centering
    \includegraphics[width = 0.49\linewidth]{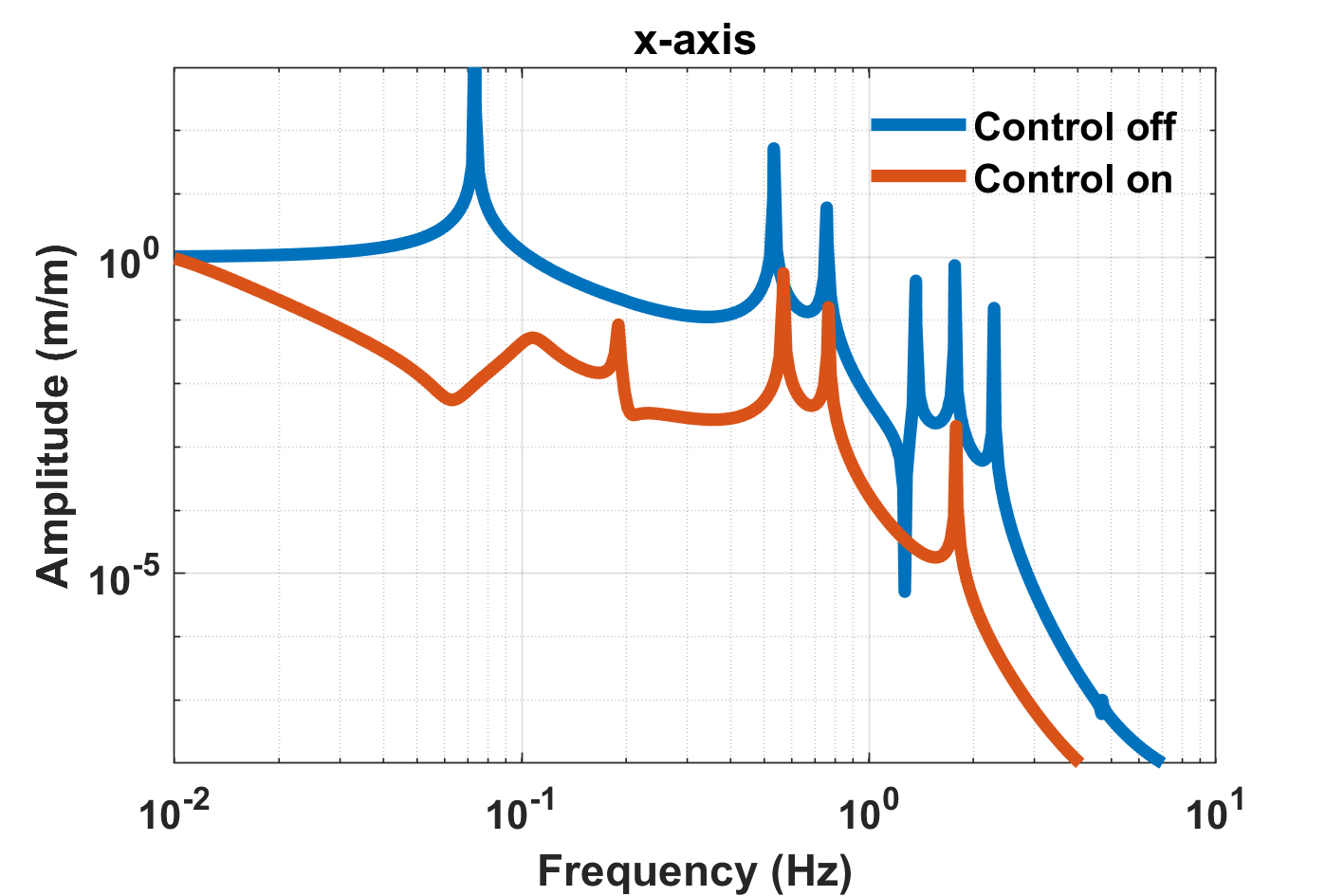}
    \includegraphics[width = 0.49\linewidth]{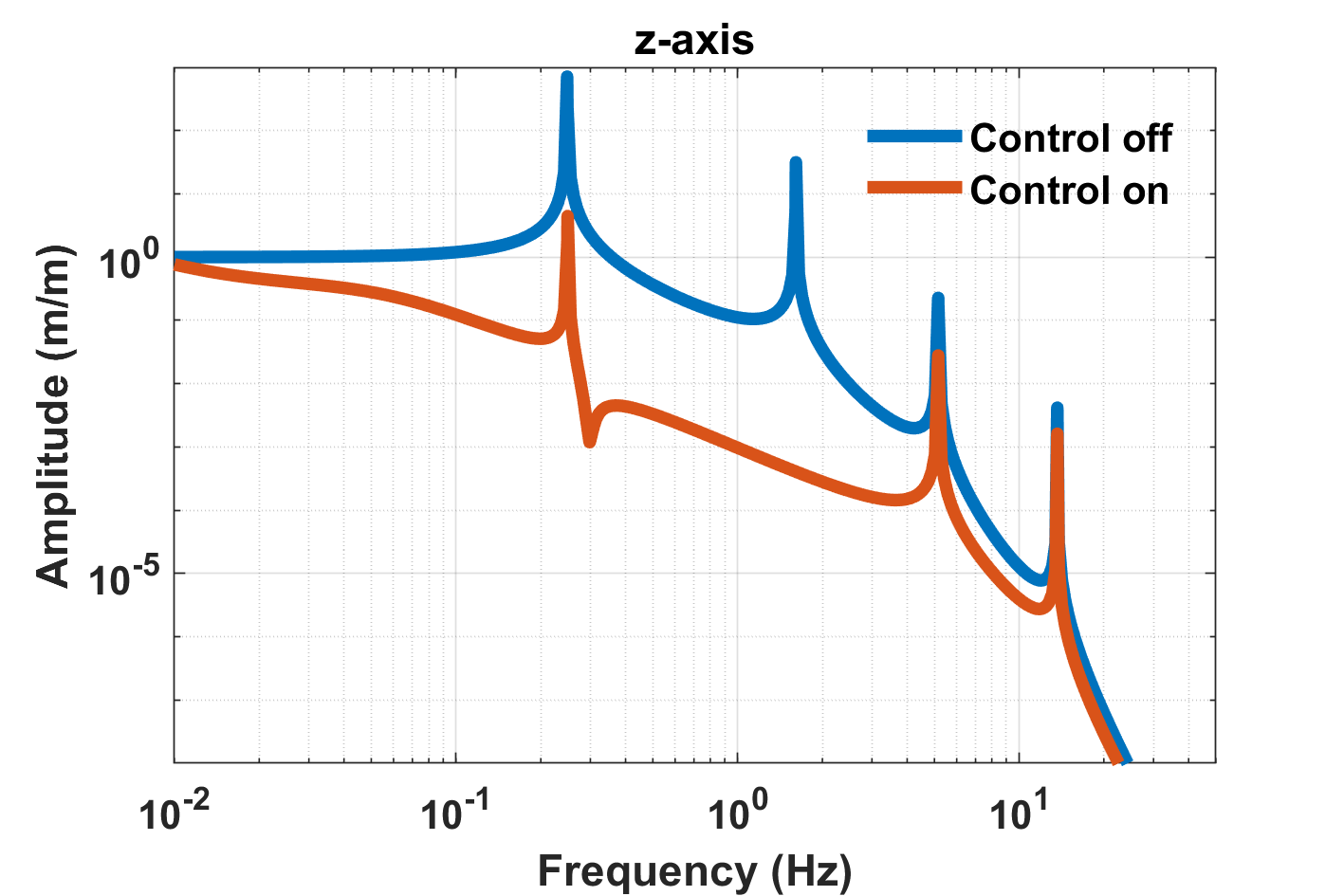}
    \caption{Transmissibilities from Ground to Mirror.}
    \label{fig:TransmissibilityMirrorETEST}
\end{figure}

Figure \ref{fig:TransmissibilityAPETEST} compares the open loop and the closed loop transmissibilities from the ground to the active platform in vertical (z-axis) and horizontal (x-axis) directions. The figure shows that the transmitted motion is reduced by about two orders of magnitude when the inertial control is switched on. Not surprisingly, the same reduction is also visible on the transmissibilities from the ground to the mirror as shown in Figure  \ref{fig:TransmissibilityMirrorETEST}.

\begin{figure}[H]
    \centering
    \includegraphics[width = 0.49\linewidth]{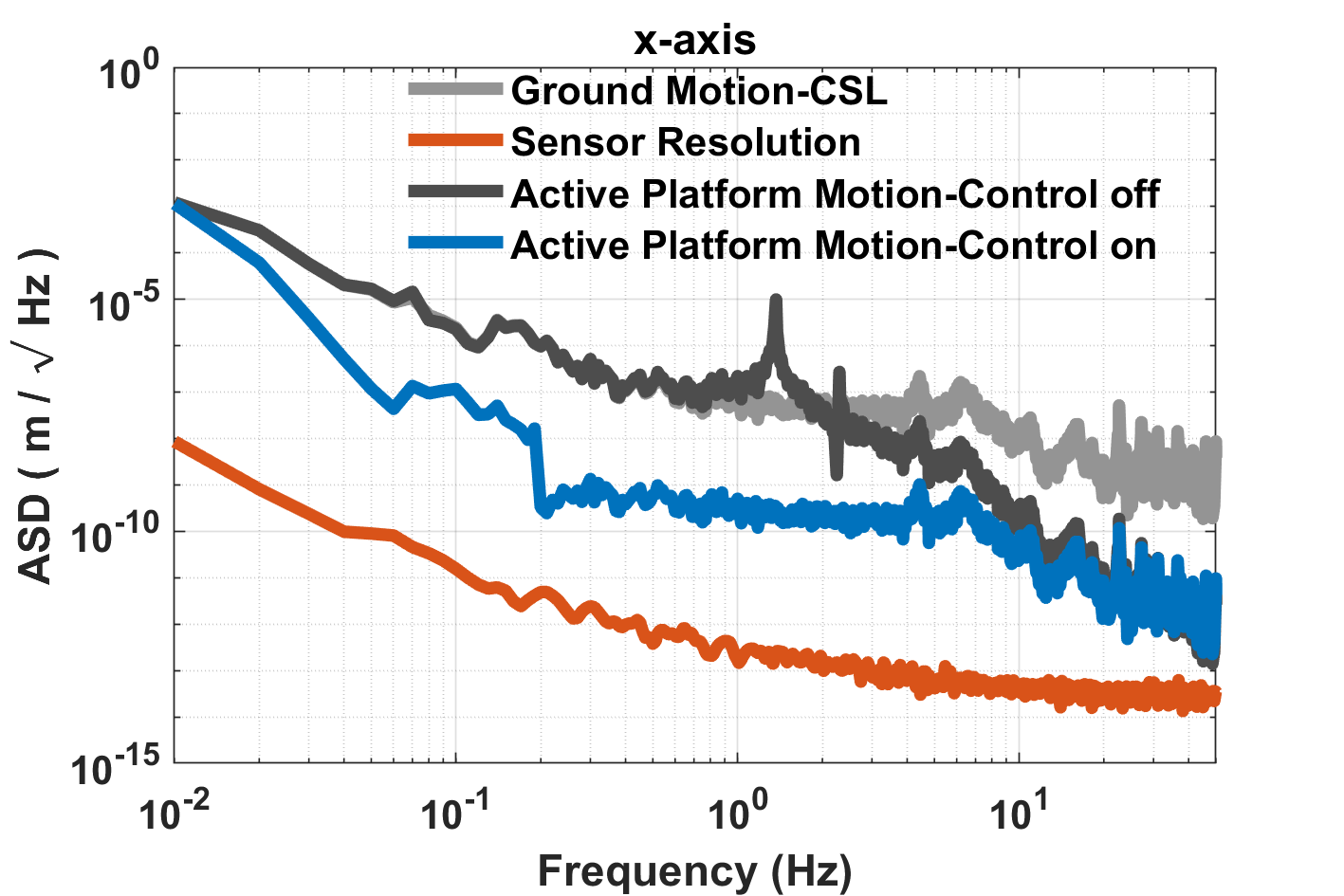}
    \includegraphics[width = 0.49\linewidth]{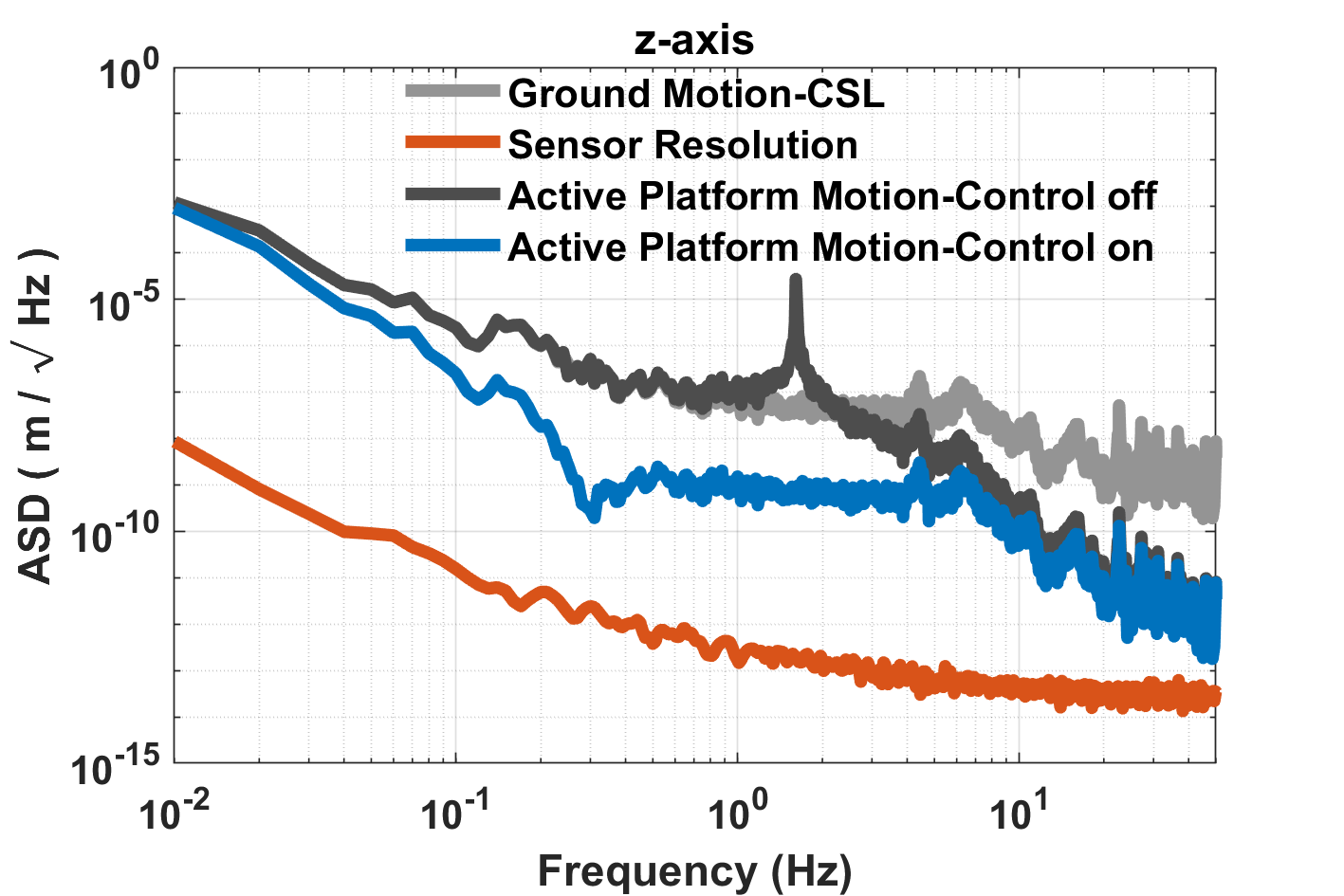}
    \caption{ASD of the Active Platform Motion, the Sensor Noise and the Ground Motion at CSL.}
    \label{fig:ASDAPETEST}
\end{figure}

\begin{figure}[H]
    \centering
    \includegraphics[width = 0.49\linewidth]{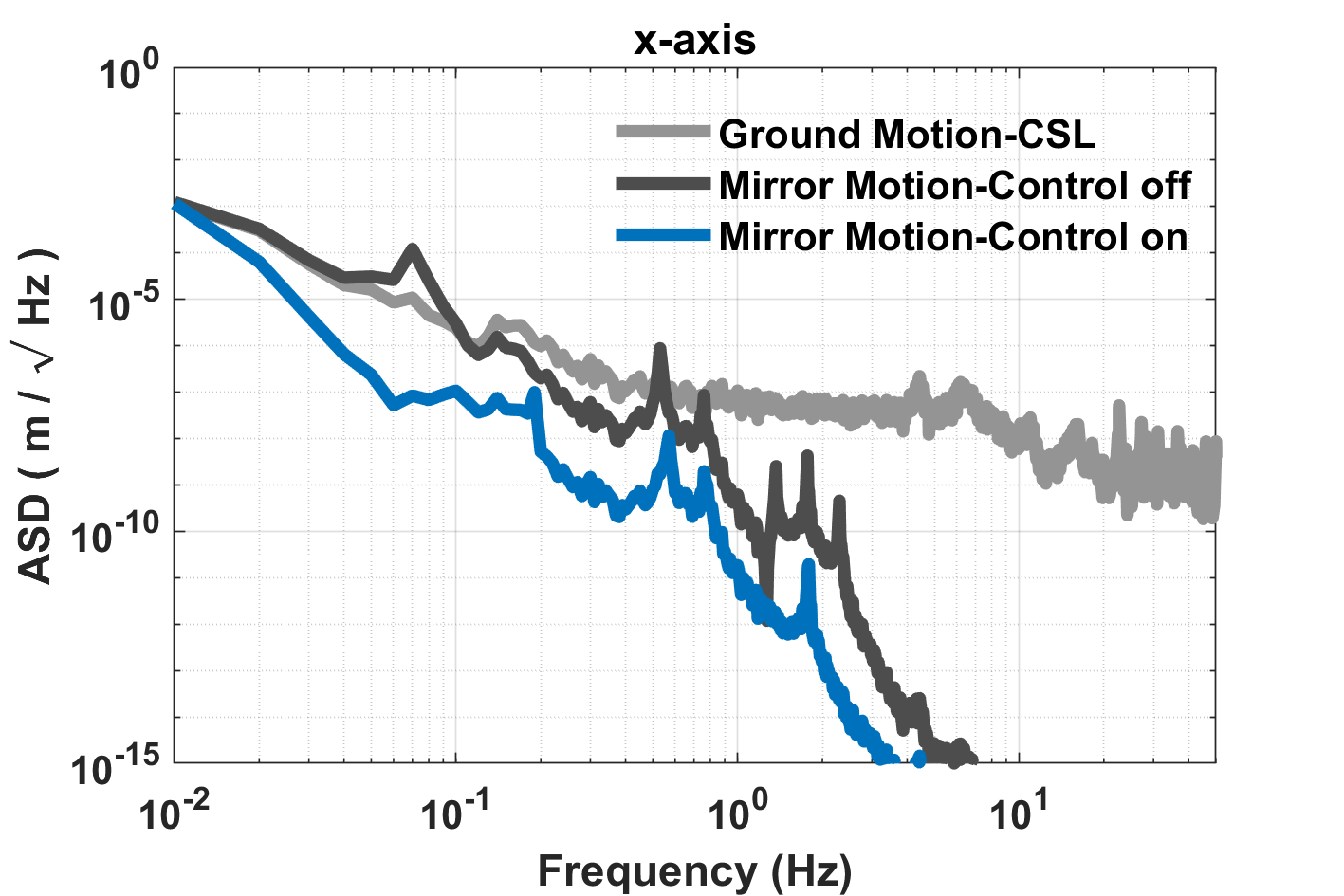}
    \includegraphics[width = 0.49\linewidth]{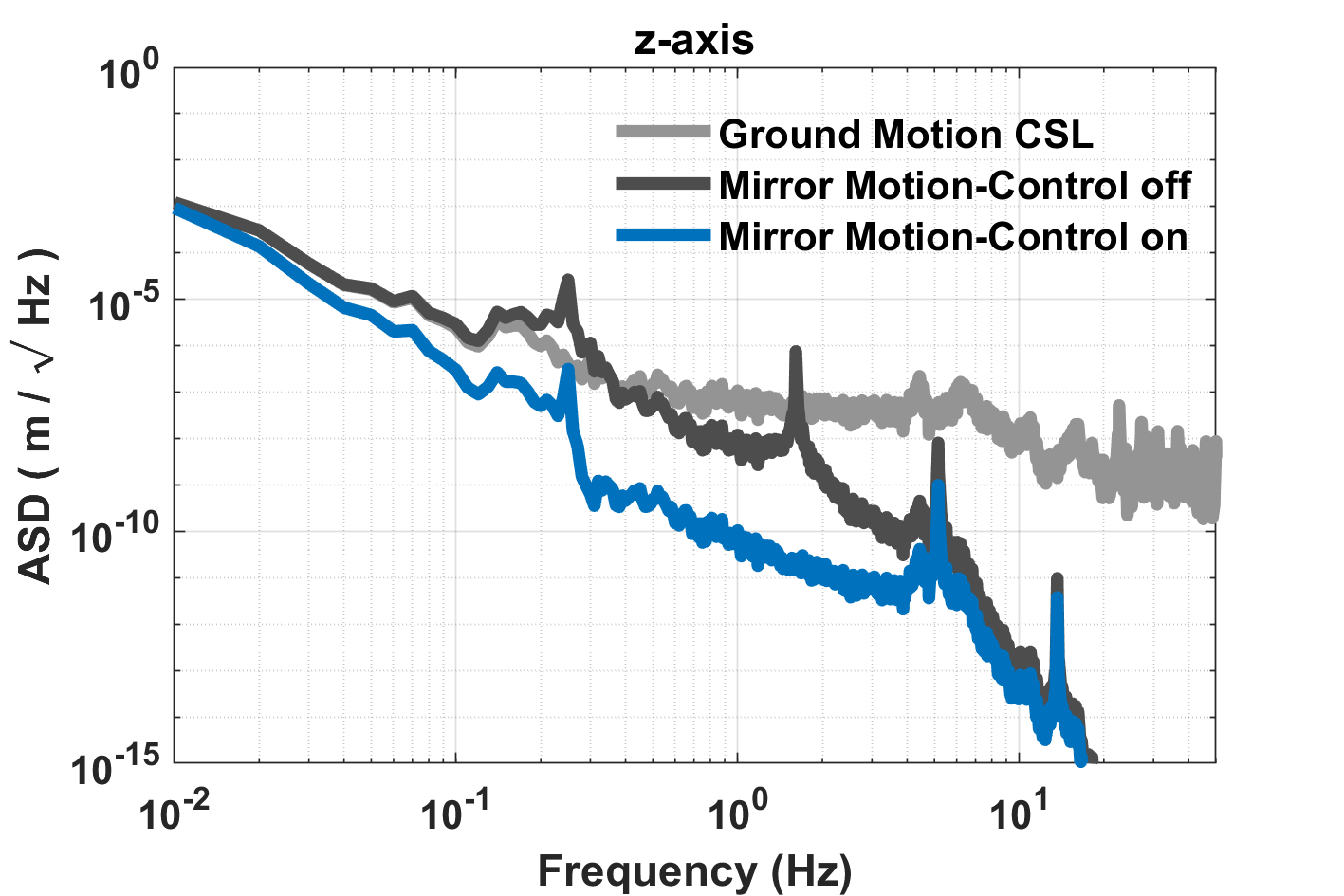}
    \caption{ASD of the Mirror and the Ground Motions at CSL.}
    \label{fig:ASDMirrorETEST}
\end{figure}

Furthermore, the ASD of the active platform as well as of the mirror are illustrated respectively in Figures \ref{fig:ASDAPETEST} and \ref{fig:ASDMirrorETEST}. For reference, the noise floor of the in-loop sensors is also shown in the same figure.  


\subsection{Effect of the center of mass altitude }

In the previous section, it was assumed that the Center of Mass (COM) of the AP was close from the Center of Stiffness (COK), which facilitates the natural decoupling between the different directions. However, in practice, the center of mass of the AP will be much higher. This is mainly due to the safety frames and tubes which are part of the AP body. 

In this section, we illustrate the effect of a modification of the altitude of the COM on the controller design. To this purpose, additional simulations have been conducted where the center of mass of the AP has been shifted 65cm higher than the top of the AP. 

\begin{figure}[H]
    \centering
    \includegraphics[width = 1\linewidth]{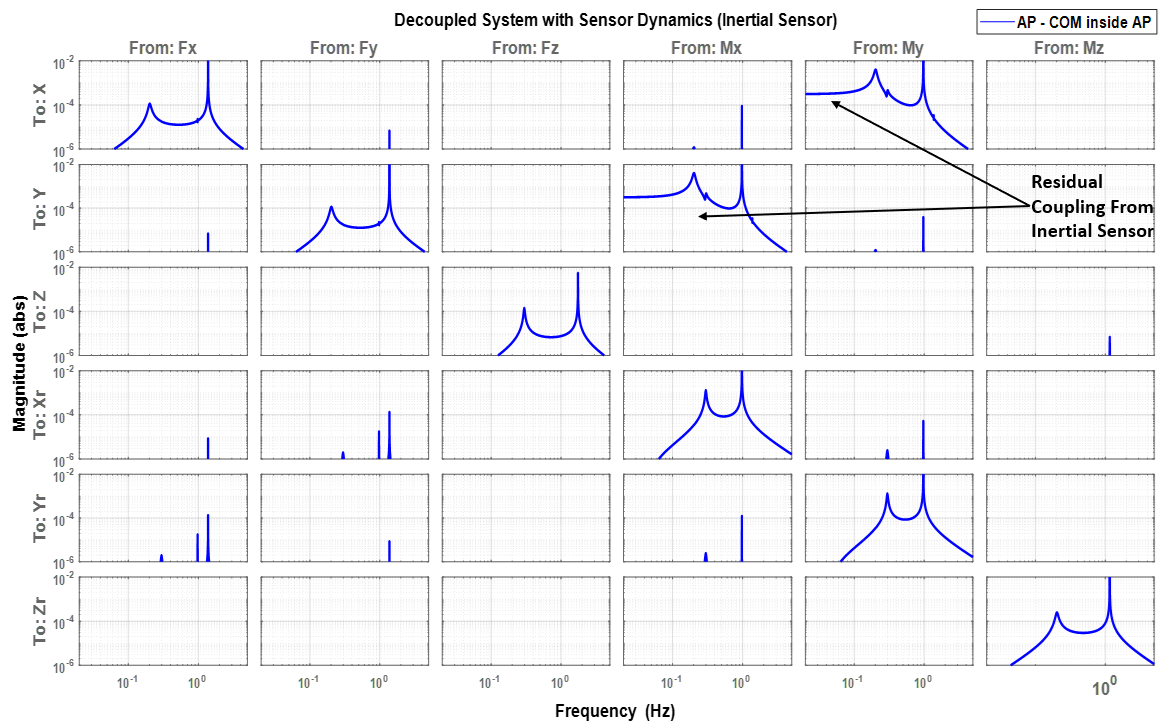}
       \caption{Transfer Matrix of Decoupled Active Platform System only with Implementing Sensor Dynamics (Inertial Sensor).}
    \label{fig:APWithoutIP}
\end{figure}

In order to better understand the effect, Figure \ref{fig:APWithoutIP} shows the control plant of the AP obtained after decoupling with respect to the COM, when the COM coincide with the COK. One sees an excellent decoupling, except for two terms, which is inherent to the nature of inertial sensors. Horizontal sensors are very sensitive to rotation at low frequency. 

\begin{figure}[H]
    \centering
    \includegraphics[width = 1\linewidth]{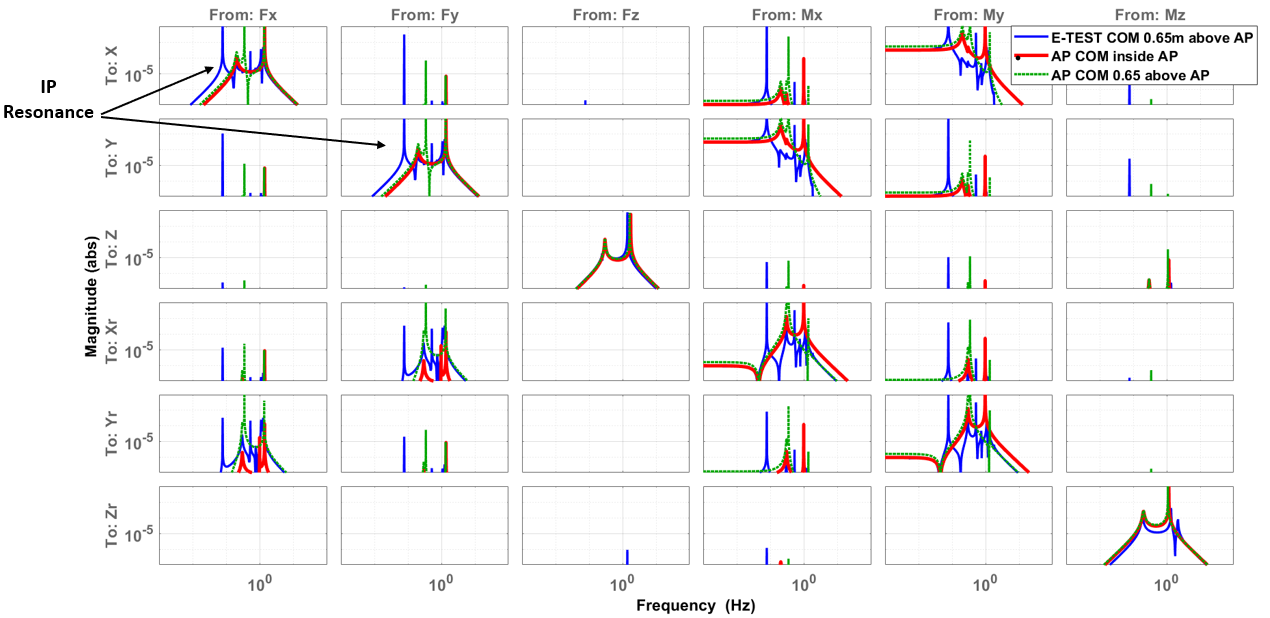}
       \caption{Transfer Matrix of the Decoupled Systems (AP alone and Entire E-TEST) with Implementing Sensor Dynamics (Inertial Sensor).}
    \label{fig:APWithIP}
\end{figure}

\begin{figure}[H]
    \centering
    \includegraphics[width = 1\linewidth]{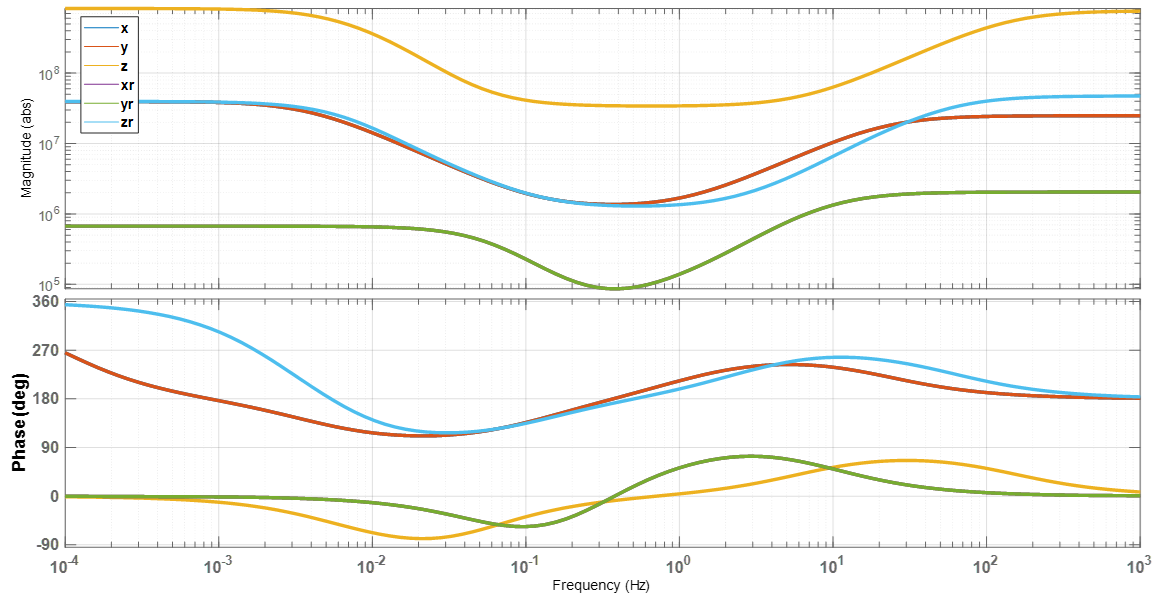}
       \caption{Controller Design for the Closed Loop System of the E-TEST.}
    \label{fig:ControllerDesignETEST}
\end{figure}

\begin{figure}[H]
    \centering
    \includegraphics[width = 1\linewidth]{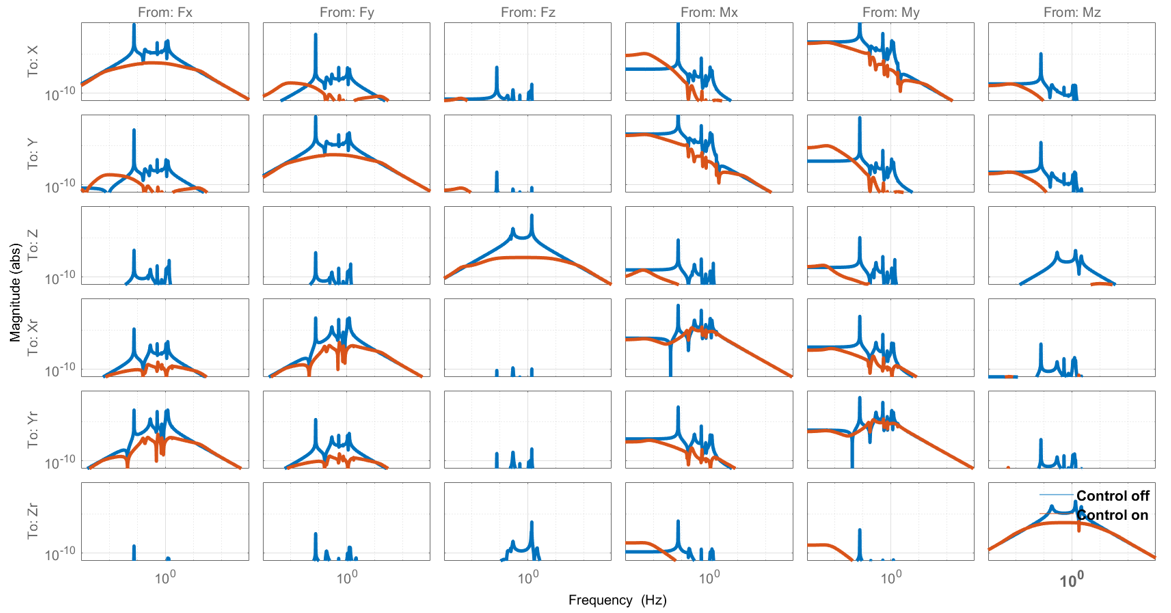}
       \caption{Transfer Matrix Of the Open Loop and Closed Loop of the E-TEST.}
    \label{fig:TransferMatrixETEST}
\end{figure}

\begin{figure}[H]
    \centering
    \includegraphics[width = 1\linewidth]{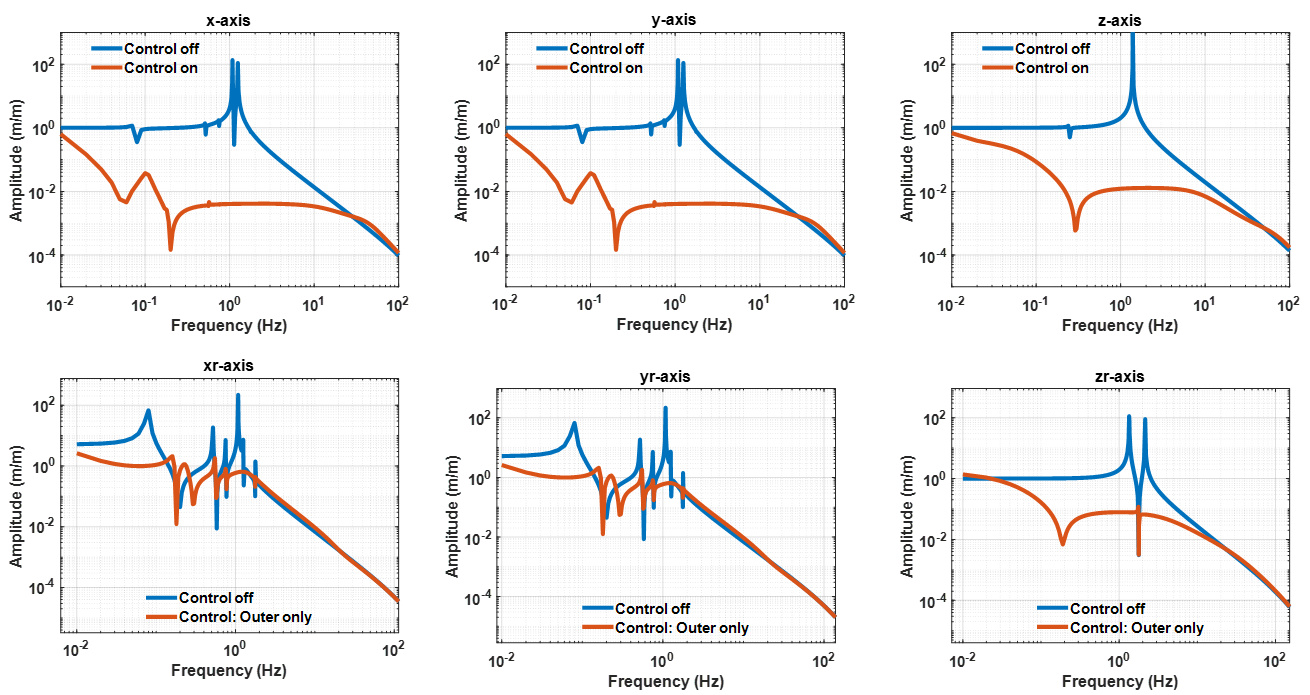}
       \caption{Transmissibility (AP/Ground) where COM is 0.65m above AP.}
    \label{fig:TransmissibilityALL}
\end{figure}

Figure \ref{fig:APWithIP} compares the decoupling obtained after shifting the COM, while still using the same decoupling technique (with respect to the center of stiffness). When the COM is higher, one sees that larger couplings exist between horizontal forces and cross rotations. However, similar performances can still be obtained using PID-like controller on each element of the diagonal control matrix. As an example, the control filters shown in Figure \ref{fig:ControllerDesignETEST} have been used to obtain the closed loop transfer matrix shown in Figure \ref{fig:TransferMatrixETEST} and the transmissibilities shown in Figure \ref{fig:TransmissibilityALL}. 

Through this example, one concludes that, although the altitude of the COM is inducing some additional couplings, it is not inducing any additional difficulty in the design of the AP controllers.

\section{Prototype conceptual design}
\begin{figure}[H]
  \centering
\includegraphics[width=17cm]{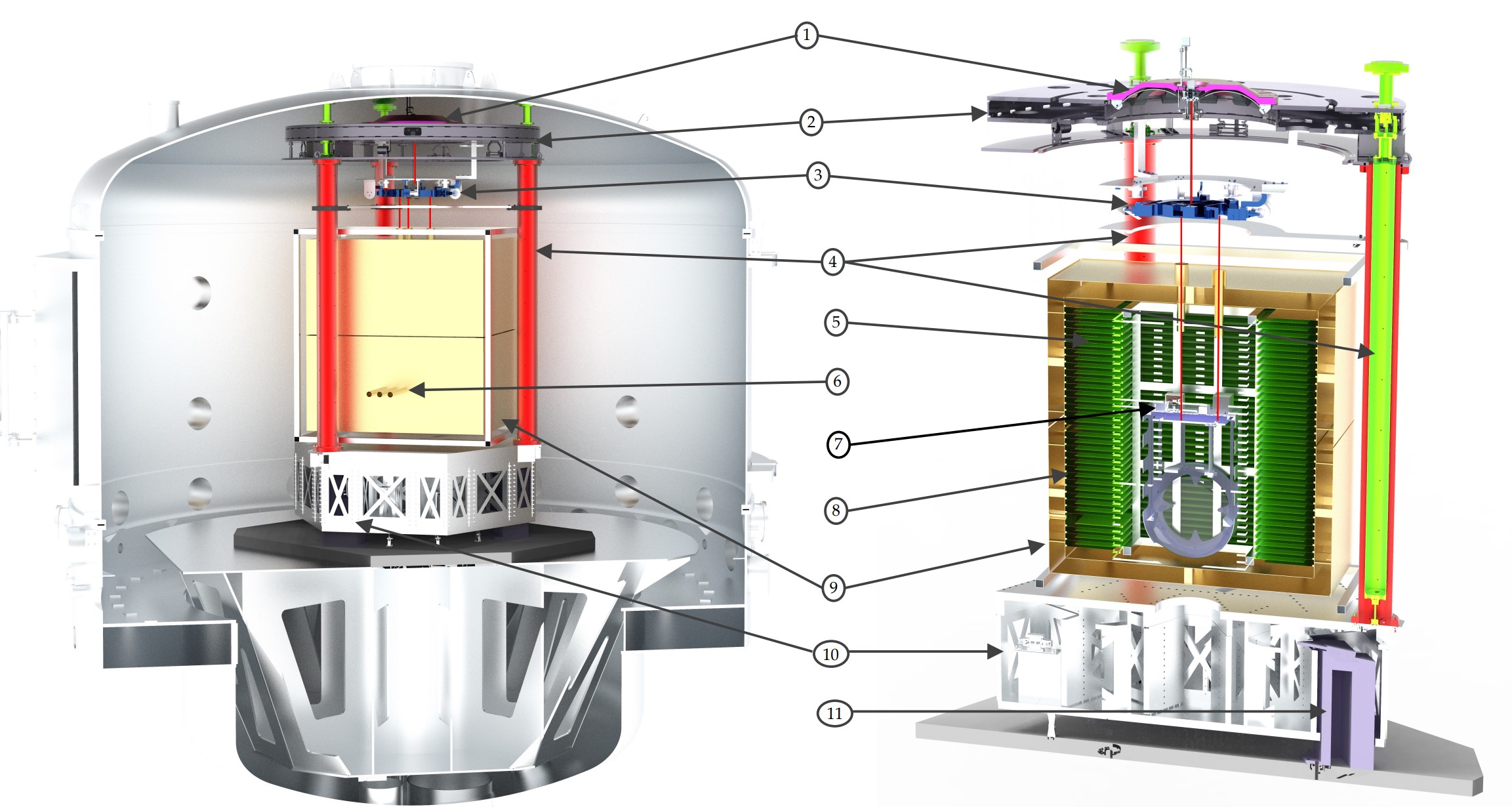}
\caption{Overview of the E-TEST prototype design. A large vacuum tank (left) hosts the cryogenic mirror suspension (right). From top to bottom we can see 1) the top GAS filter, 2) the top stage, 3) the marionette and 4) the inverted pendulum legs within pipes that support a reference ring below the top stage. The cryogenic part features 5) the inner cryostat which has the interlacing fin type heat exchanger. The whole cryostat features (6) three access points for outside experiments to interact with the cryogenic mirror. The inner cryostat is attached to 7) the cold platform. The inner cryostat fins interlace into the fins of the 8) outer cryostat which provides a cold environment and houses the (9) 100\;kg silicon mirror. All of this is supported by 10) an active platform, which provides a stable and quiet environment. In turn, the active platform hangs from three large blades with have a (11) support pillar on the ground.}\label{fig:E-TESTprtOvvw}
\end{figure}

A 3D view of the E-TEST suspension system is shown in Figure \ref{fig:E-TESTprtOvvw}. It consists of one active platform that provides inertial control in the 6 degrees of freedom (detailed in section \ref{sec:ActivePlatformCAD}). The three legs of the IPs are mounted on the active platform and support the top stage (IPP). The IPP provides a large amount of isolation in the horizontal direction as the resonance frequency can be tuned to extremely low values (about 70 mHz in KAGRA \cite{okutomi2019development} and 30 mHz in Virgo \cite{accadia2011seismic}). Additionally, the IPP stage provides means for positioning the parts suspended from the top stage, which can compensate for the tidal drift\;\cite{takamori2002low}.  

\begin{figure}[ht]
    \centering
    \includegraphics[width = 1\linewidth]{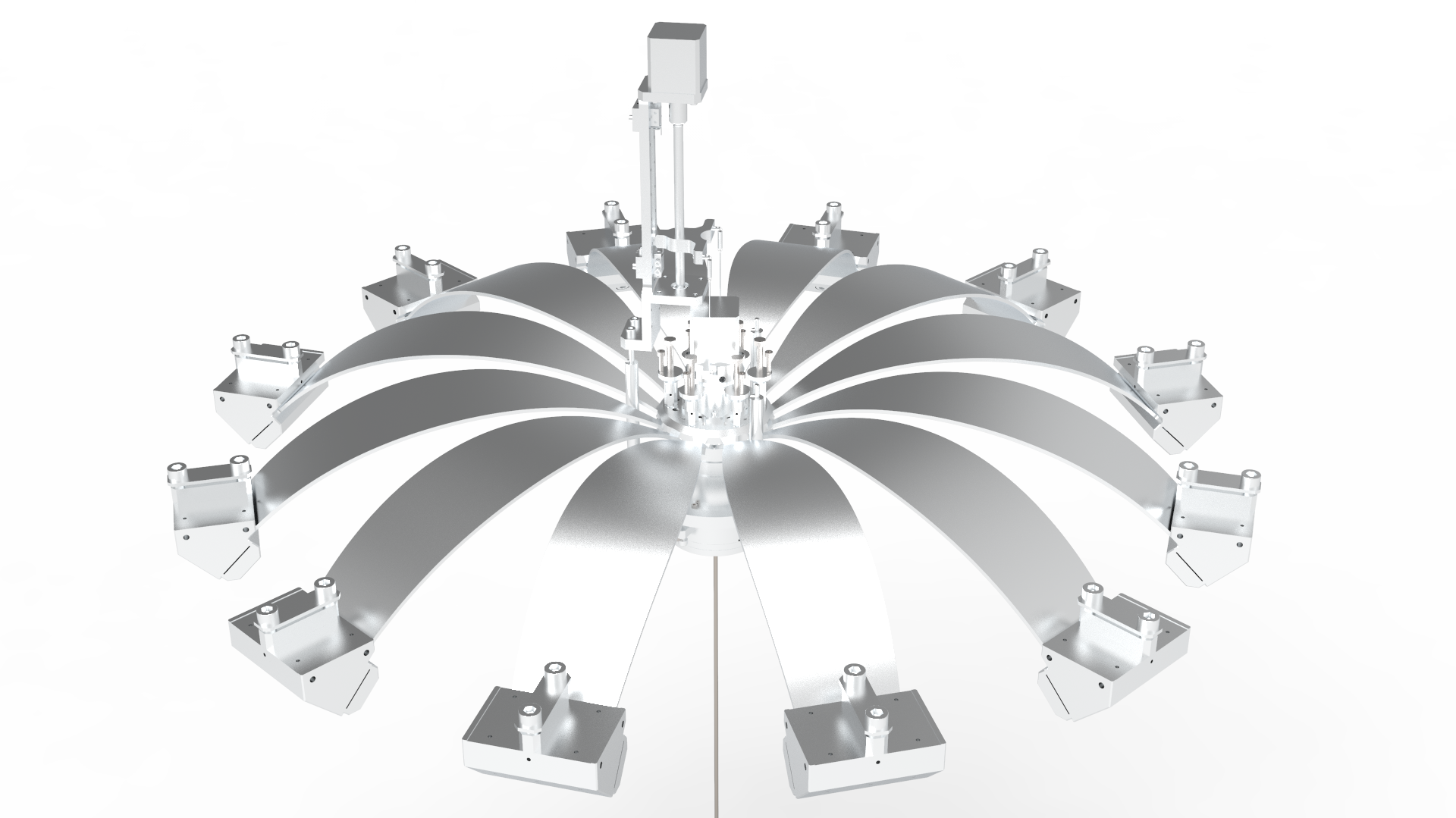}
    \caption{CAD View of Geometric-Anti Spring (GAS) Filter}
    \label{fig:GASFilterCAD}
\end{figure}

\begin{figure}[H]
    \centering
    \includegraphics[width = 0.49\linewidth]{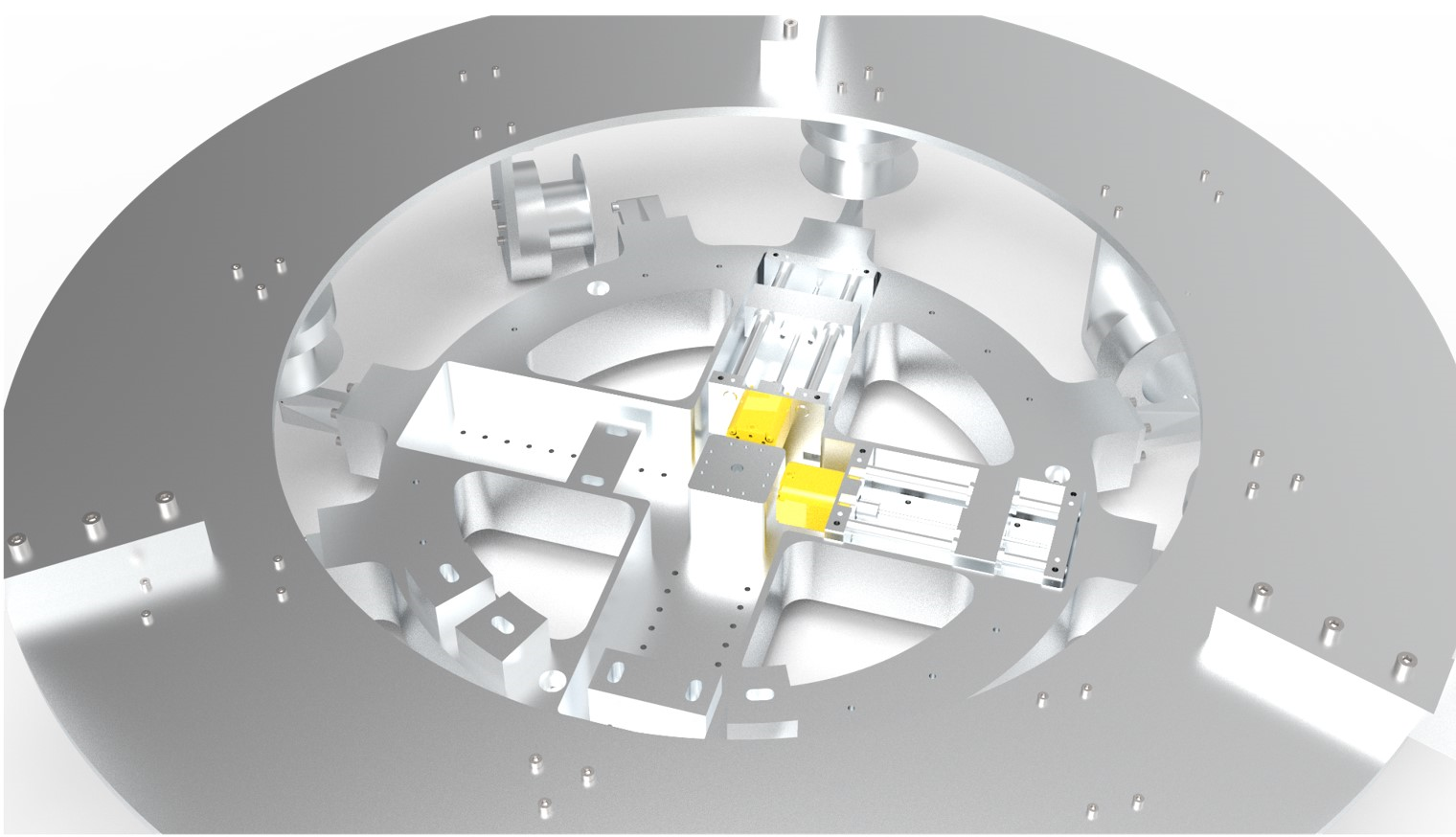}
    \includegraphics[width = 0.49\linewidth]{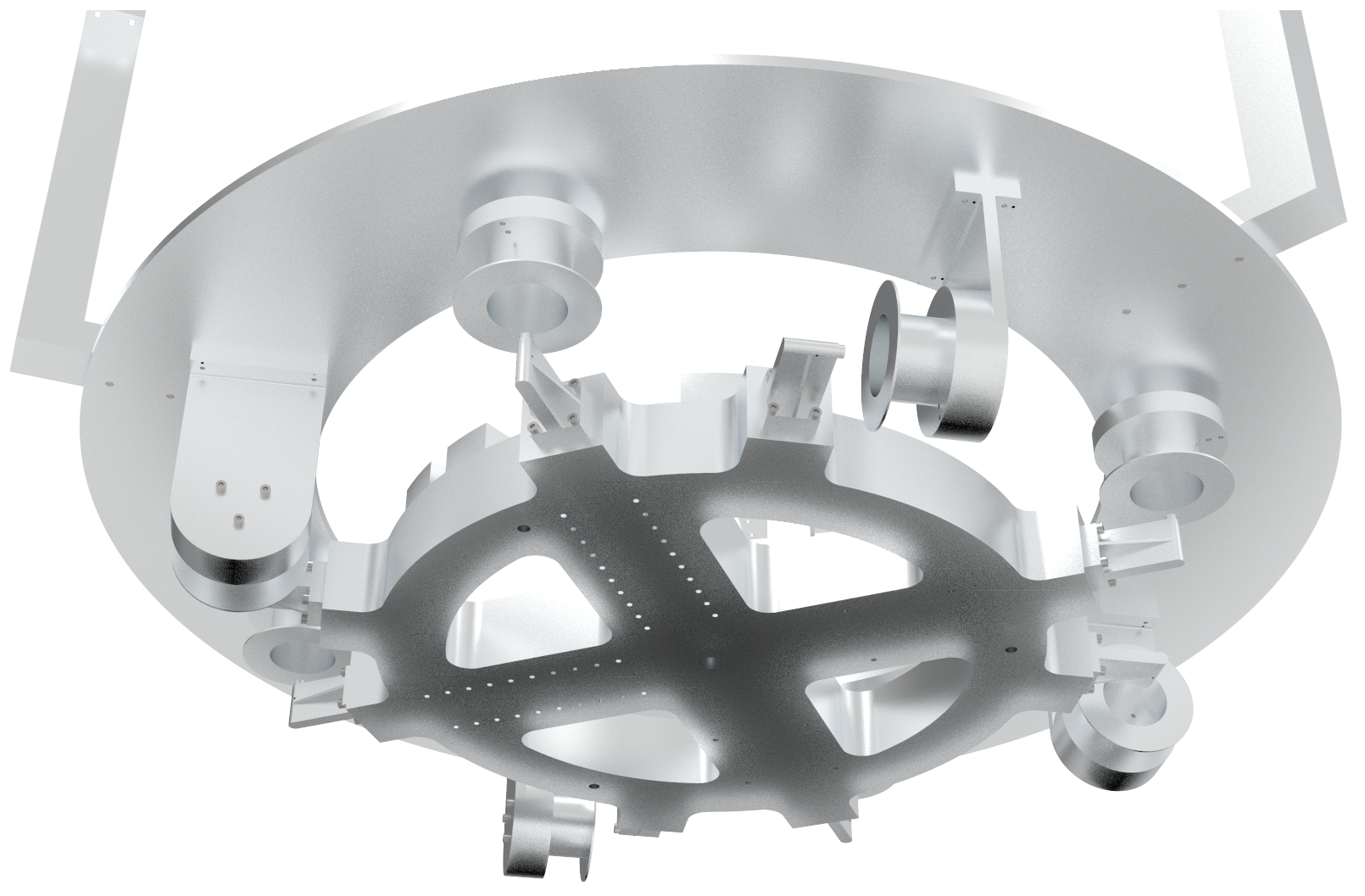}
    \caption{CAD View of Marionette: Top view (left) and Bottom View (right) }
    \label{fig:MarionetteCAD}
\end{figure}

The top stage (IPP) houses a large GAS filter that adds vertical isolation (shown in Figure \ref{fig:GASFilterCAD}). The GAS filter is composed of twelve springs. From the GAS filter, a marionette, shown in Figure \ref{fig:MarionetteCAD}, is suspended, which is mainly used to position a cryogenic payload (presented in section \ref{cryo_rad}). This cryogenic part contains the cold platform and the large mirror which will be operated at cryogenic temperature (20\;K). 
The cold platform is a cryogenic test-bed for cryogenic electronics and sensors presented separately in sections \ref{cryo_cmos} to \ref{cryo_sens}. Before installing sensors or other equipment in the suspended cryostat, a smaller cryostat has been developed with a cylindrical cold volume -- down to 6\;K -- of 15\;cm diameter and 15\;cm height  for testing these instruments (presented in section \ref{cryo_test}).


\begin{figure}[htb]
    \centering
    \includegraphics[width = 1\linewidth]{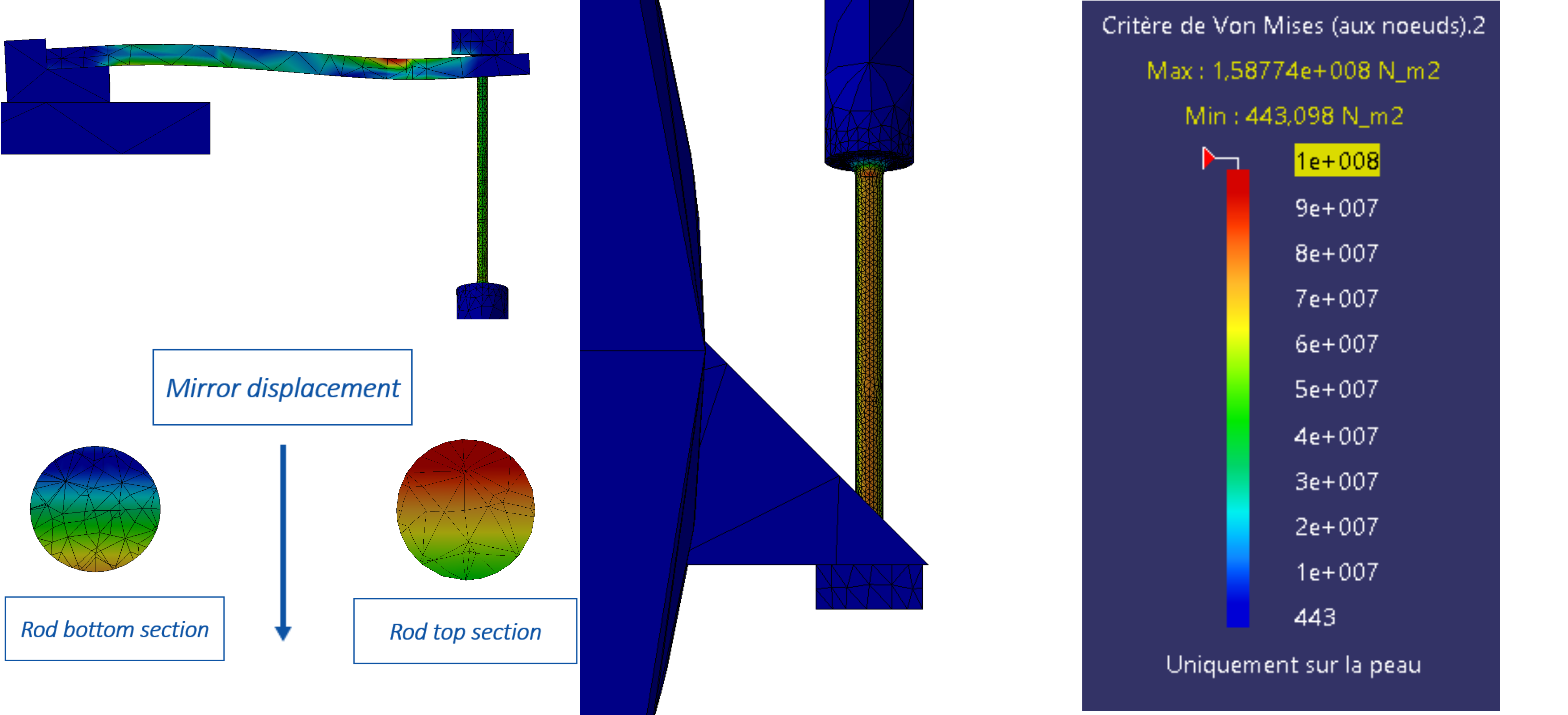}
    \caption{Stress Map Under Gravity in One blade and one Rod Used To Suspend the Mirror.}
    \label{fig:FEMStress}
\end{figure}

\begin{figure}[H]
    \centering
    \includegraphics[width = 1\linewidth]{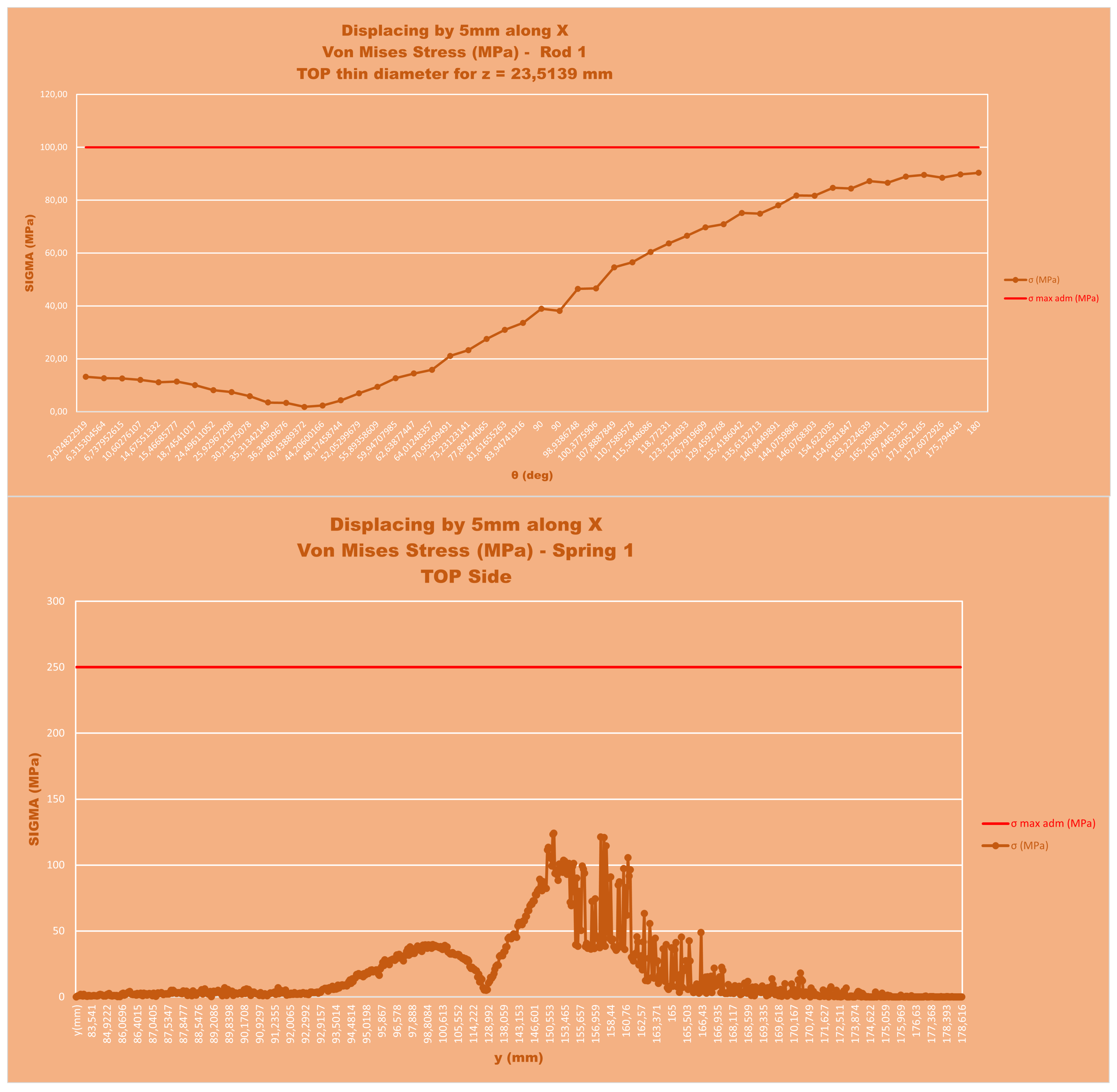}
    \caption{graph FEM}
    \label{fig:FEMGraphe}
\end{figure}

\section{Active Platform}\label{sec:ActivePlatformCAD}
\subsection{Mechanical Design}
\begin{figure}[h]
    \centering
    \includegraphics[width = 1\linewidth]{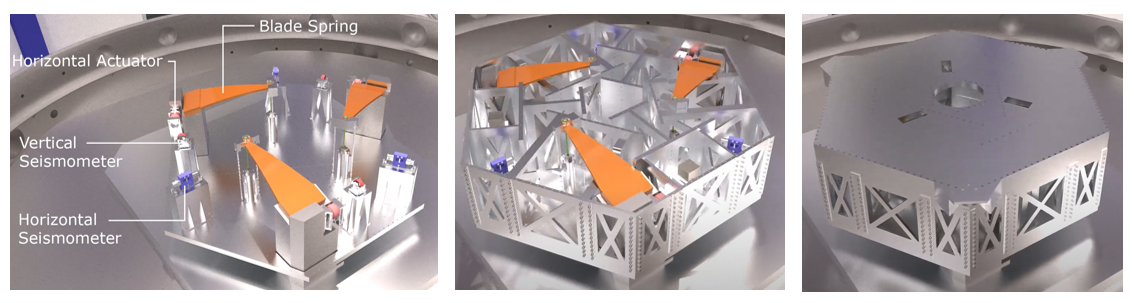}
    \caption{CAD View of Active Platform.}
    \label{fig:ActivePlatformCAD}
\end{figure}

The active platform developed for the E-TEST prototype is shown in Figure \ref{fig:ActivePlatformCAD}.

\begin{figure}[h]
    \centering
    \includegraphics[width = 1\linewidth]{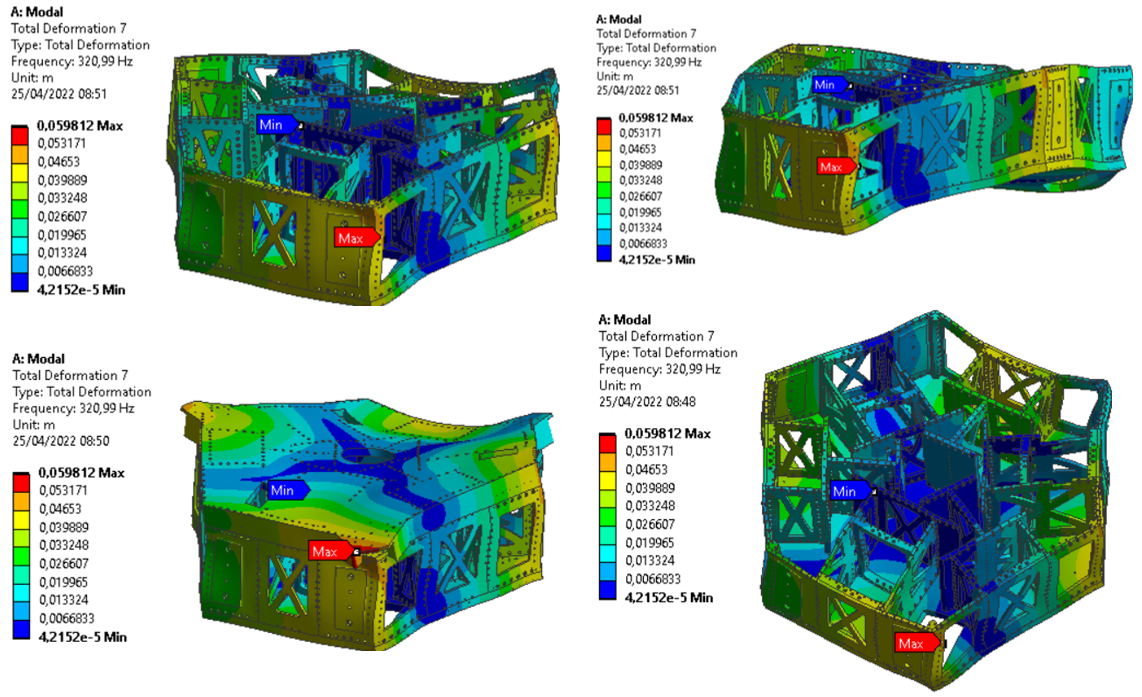}
    \caption{Different Views of Finite Element Analysis of the Active Platform: First Flexible Mode Appears Above 300 Hz }
    \label{fig:FirstModeAP}
\end{figure}

It is a completely redesigned and up-scaled version of the LIGO HAM-ISI platform. In order to host the large cryostat, the platform diameter has been increased to 2.5m.
The sandwich structure of the payload using vertical panels has been reinforced for  maintaining the first flexible mode above 300 Hz. The shape of this mode is shown in Figure \ref{fig:FirstModeAP}.

\begin{figure}[h]
    \centering
    \includegraphics[width = 1\linewidth]{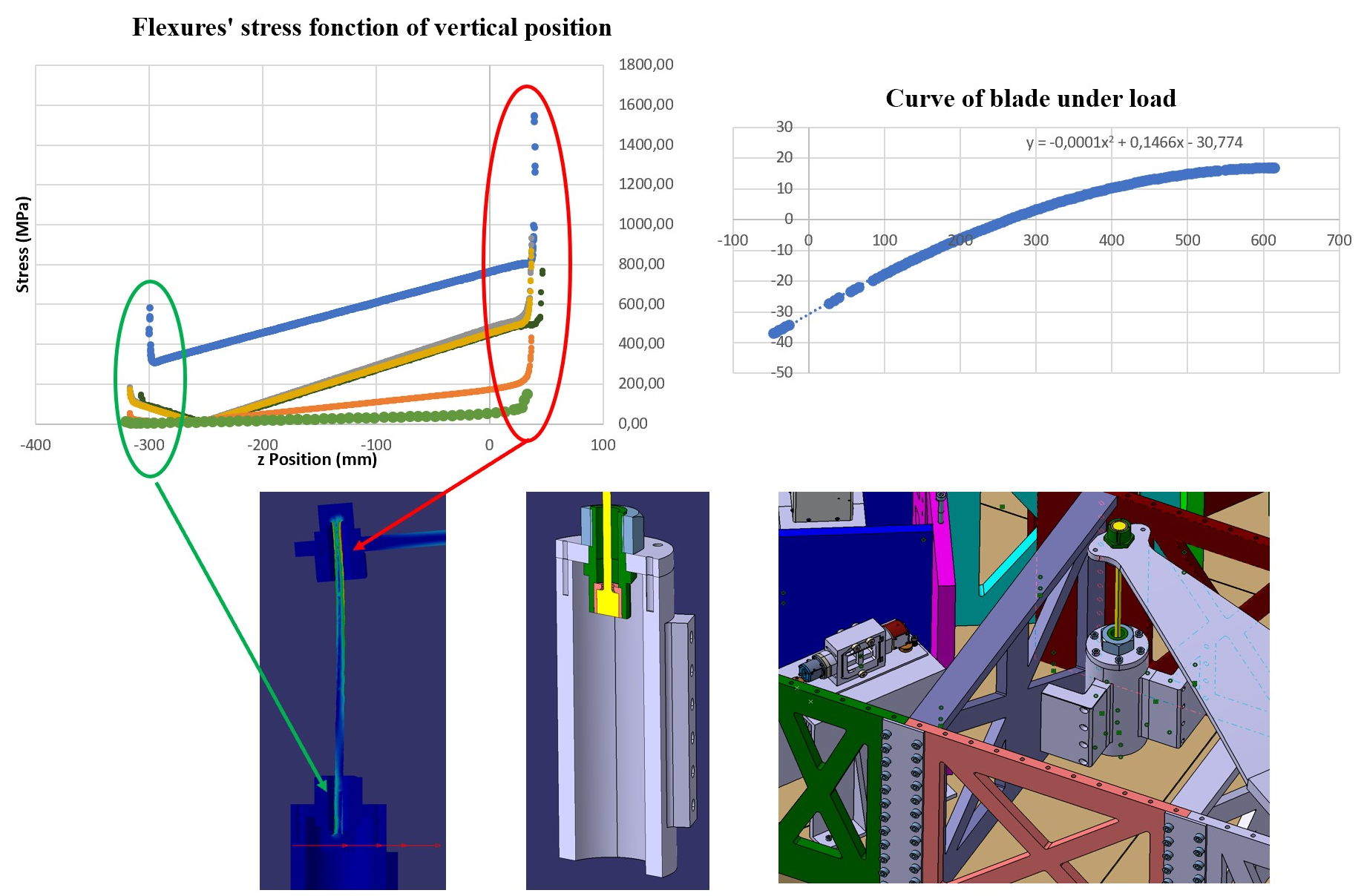}
    \caption{Finite Element Analysis of the Flexure.}
    \label{fig:FluxureAP}
\end{figure}

A particular attention has also been paid to the mass and stiffness repartition in order to facilitate the decoupling of the plant in view of its control. Thanks to this careful design, the rotation modes around the horizontal axis correspond to pure rotations around the center of mass (see Figure \ref{fig:FirstModeAP} (right)). The blade springs used for vertical isolation have been stiffened for supporting a total payload of 1700kg, while maintaining the suspension mode below 2Hz. The material chosen for the blades is steel PH13-8 due to its high yield strength.  The tip of each blade is connected to a flexure in steel C250 ensuring an adequate compliance in the horizontal direction. The blades have been designed in such a way that they are inclined at equilibrium, with the aim of changing the load in pure traction to overcome the high stress that occurs in the flexures due to bending. This is shown in Figure \ref{fig:FluxureAP}.

The positions of the horizontal sensors as well as the horizontal actuators are carefully chosen in order to facilitate the plant decoupling.
The next section presents some recent development in inertial sensors that will be adapted for controlling the active platform.

\subsection{Inertial Sensors}\label{sec:ActivePlatformSensor}
The active stage uses inertial control in order to provide an additional layer of ground isolation, on top of the isolation achieved with the passive stages. The inertial control requires the absolute motion of the platform to be very precisely monitored. As the amount of motion that can be cancelled actively is ultimately limited by the minimum motion that is sensed by the sensors, i.e. their so-called resolution, the active stage will be equipped with ultra-high resolution inertial sensors. The inertial sensors that will be integrated to the AP are similar to those developed at PML \cite{Ding2021,Zhao2020}. These sensors use the well-known STS-1 commercial inertial sensors from the Streckeisen company as base-line for the design. Although the STS-1-V/H sensors have been considered as first choice seismometer for many years, the resolution of these “off-the-shelf” sensors have been further improved.\\

\begin{figure}[H]
    \centering
    \includegraphics[width = 0.60\linewidth]{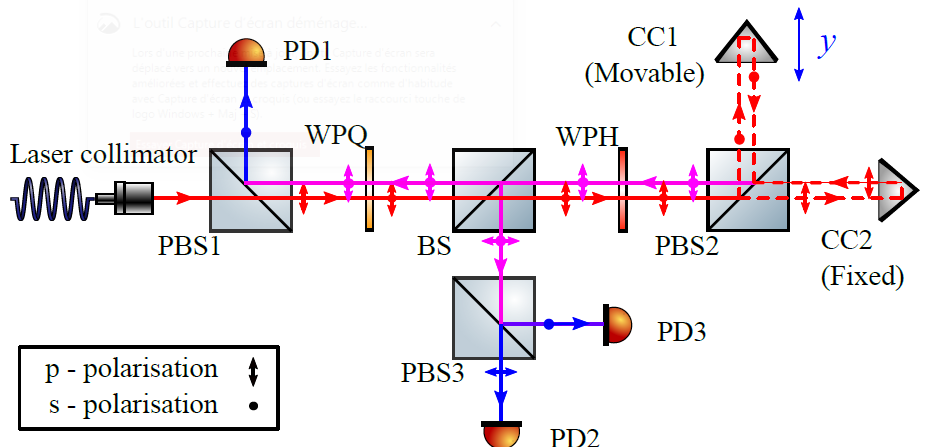}
    \includegraphics[width = 0.39\linewidth]{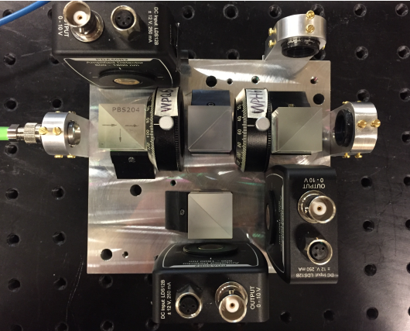}
    \caption{Homodyne, quadrature, Michelson interferometer. Optical scheme (left) and mechanical realization (right). The mechanical design is approximately 14 cm long for 10 cm wide \cite{Ding2021}.}
    \label{fig:InertialSensor1}
\end{figure}

A major improvement was to replace the conventional electrical readout with a high resolution Michelson interferometer. The optical design of the interferometer includes several polarizing elements, such as polarizing beam-splitters and wave-plates, in order to by-pass the fundamental dynamic limitation of classical Michelson interferometer, therefore providing a long range readout \cite{Watchi2018}. The interferometer output also consists of three photodiodes, so that the common-mode noise of the interferometer, especially laser intensity noise, can be safely discarded \cite{Watchi2018}. In the end, this makes up for a long-range, homodyne, quadrature, Michelson interferometer capable of resolving sub-pm resolution at frequencies below 0.1 Hz. Figure \ref{fig:InertialSensor1}. shows the optical scheme of the interferometer and its mechanical realization. Figure \ref{fig:InertialSensor1}.  shows the readout resolution, mostly dominated by photodetector noise ( i.e. shot noise and dark current noise) \cite{Ding2021,Ding2018}. Signal above 1 Hz is suspected to be residual ground motion or internal vibrations of the opto-mechanical components.

\begin{figure}[H]
    \centering
    \includegraphics[width = 0.9\linewidth]{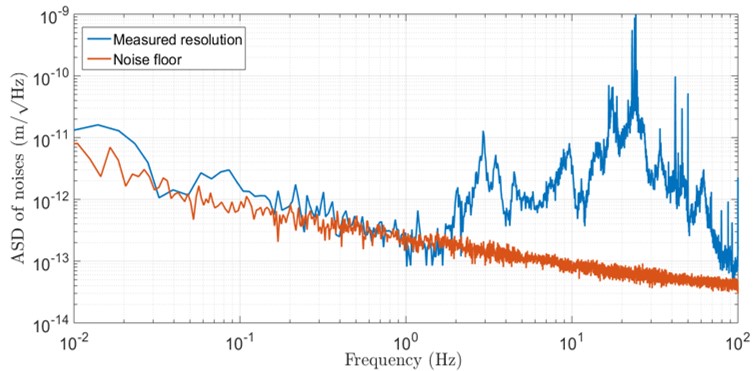}
    \caption{Resolution of the Michelson interferometer, measured during a “blocked-mass test”. The readout reaches a resolution of  1 pm @ 0.1 Hz, 0.2 pm @ 1 Hz and 0.1 pm @ 10 Hz \cite{Ding2021,Ding2018}. The signal below 1 Hz is dominated by photodetector noise while the signal above 1 Hz is dominated by residual ground motion / internal vibrations.}
    \label{fig:InertialSensor2}
\end{figure}

\begin{figure}[H]
    \centering
    \includegraphics[width = 0.30\linewidth]{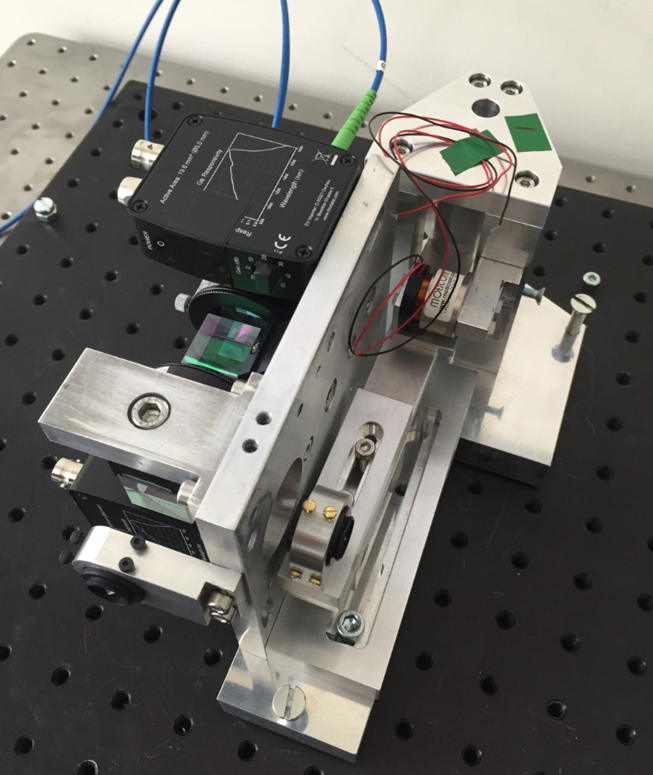}
    \includegraphics[width = 0.69\linewidth]{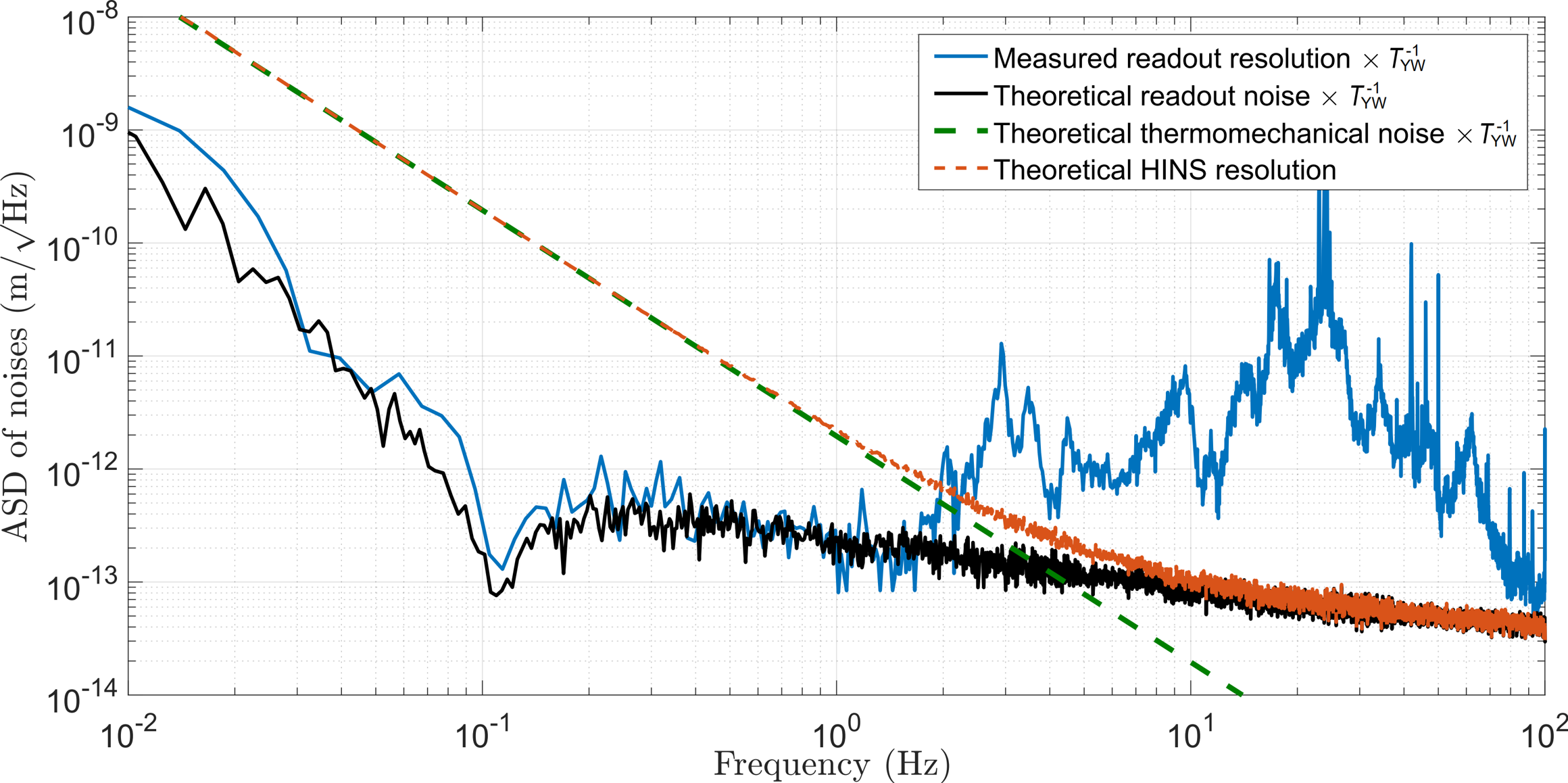}
    \caption{Horizontal Interferometric Inertial Sensor (HINS) \cite{Ding2021}. Sensor (left) and performance (right). It is characterised by a resonance frequency of 0.11 Hz and a Q-factor of 20. It is achieving sub-pm resolution above 0.01 Hz.}
    \label{fig:InertialSensor3}
\end{figure}

The mechanics of the inertial sensors consist of a revised version of the STS-1 mechanics. The HINS senses horizontal motion using a garden-gate pendulum oscillating in an horizontal plane. The joint of the pendulum is a cross-spring hinge formed by three steel clamped plates. Through a ring-down test, the dynamic characteristics of the mechanics have been extracted. It shows a main resonance frequency at 0.11 Hz and a quality factor of 20. The dimensions of the HINS are 200 x 140 x 150 mm$^3$. Coupled to the interferometric readout, the sensor can theoretically provide 2 x 10$^{-13}$ m resolution @ 1 Hz, and better than 10$^{-13}$ m @ 10 Hz. It is limited by mechanical thermal noise below 4 Hz, and by the readout noise above 4 Hz. The HINS is shown in Figure \ref{fig:InertialSensor3}.

\begin{figure}[H]
    \centering
    \includegraphics[width = 0.30\linewidth]{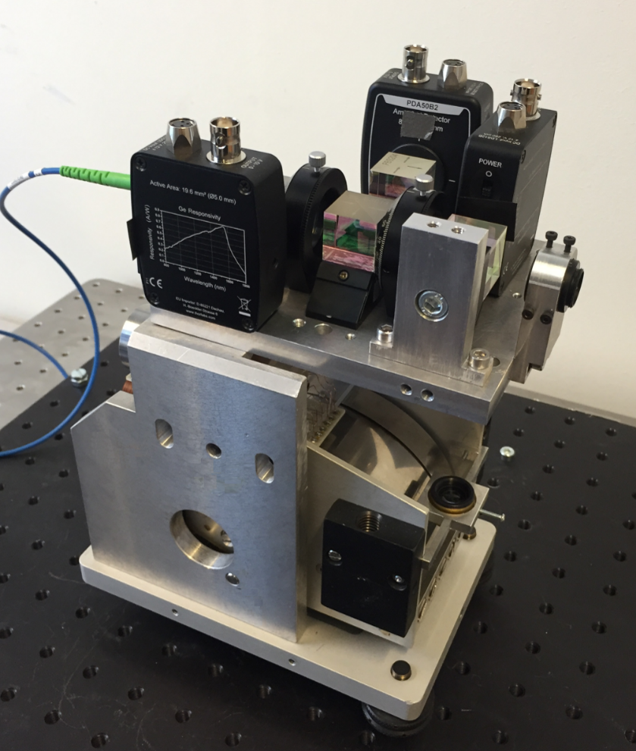}
    \includegraphics[width = 0.69\linewidth]{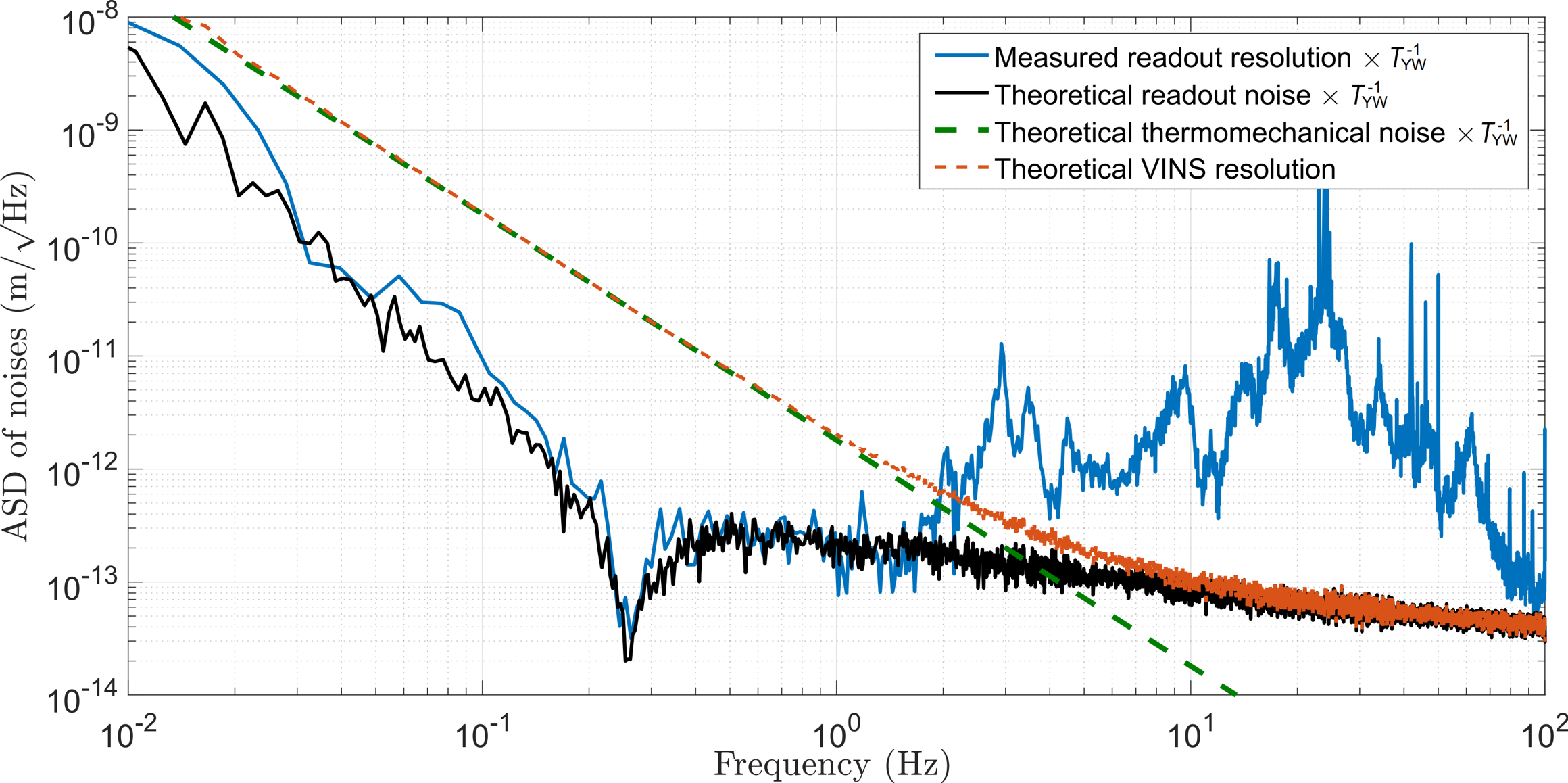}
    \caption{Vertical Interferometric Inertial Sensor (VINS) \cite{Ding2021}. Sensor (left) and performance (right). It is characterised by a resonance frequency of 0.25 Hz and a Q-factor of 30. It is achieving sub-pm resolution above 0.01 Hz.}
    \label{fig:InertialSensor4}
\end{figure}

The VINS is also based on the STS-1-V mechanics. It consists of a pendulum oscillating in a vertical plane (LaCoste pendulum). It is maintained in an horizontal position using a CuBe leaf-spring. The suspension reaches a natural frequency of 0.25 Hz for a size comparable to the HINS. The ring-down tests validate that value of the natural frequency and a Q-factor of 30. Its performances are very similar to the HINS sensor. The VINS is shown in Figure \ref{fig:InertialSensor4}. 

The HINS and VINS sensors have shown satisfying performances in terms of resolution and bandwidth. These sensors have been characterized in open-air and in a  1 mbar (low-vacuum) environment. Noise budgeting shows that the low frequency resolution is dominated by microscopic Brownian motion of surrounding gases. Performances are expected to be magnified when used in the rarefied gas field inside of the vacuum chamber (ultra-high vacuum environment). Even though HINS and VINS are not readily UHV-compatible, they can easily be adapted to fit the vacuum chamber. In addition, research aiming to further improve sensor resolution is currently in progress, in order to bridge the gap between current performances and the targeted measurement noise. The investigated improvement are

\begin{itemize}
    \item[(i)] Changing the metallic joints for low-thermal dissipation material (fused-silica glass flexures). 
    \item[(ii)] Assessing the closed-loop performances for improved linearity.
    \item[(iii)]  Improved photodetector performances and better optical alignment.\\
\end{itemize}

\begin{figure}[H]
    \centering
    \includegraphics[width = 0.34\linewidth]{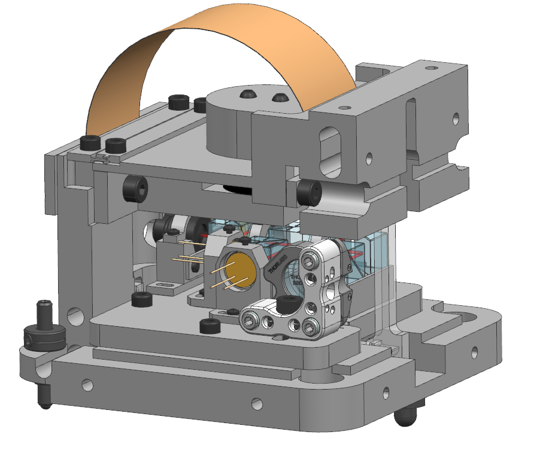}
    \includegraphics[width = 0.65\linewidth]{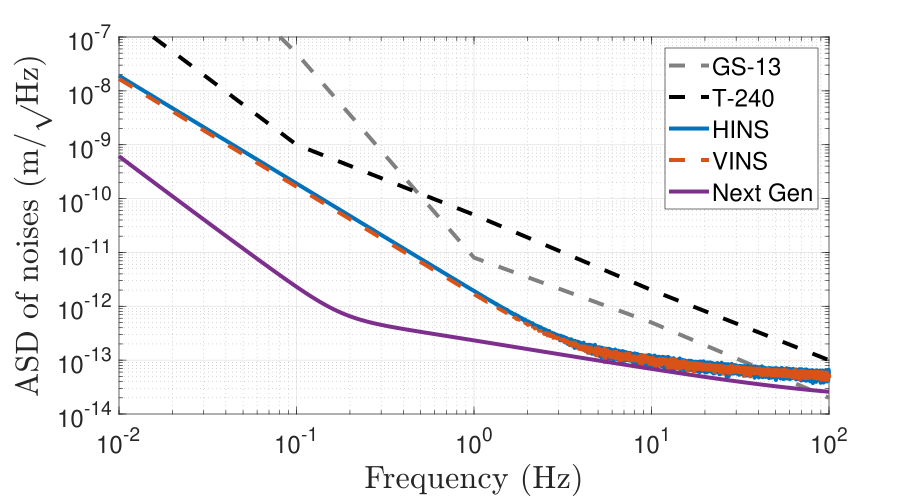}
    \caption{Prototype of the next-generation VINS sensor. Using  new, high-performances, photodiodes and considering reduced thermal-noise in the UHV environment, the performances can be improved by 1 order of magnitude below 1 Hz, and by a factor of 2 above 1 Hz.}
    \label{fig:InertialSensor5}
\end{figure}

CAD view and expected performances of the next-generation HINS and VINS are shown in Figure \ref{fig:InertialSensor5}.

\section{Radiative cooling strategy}
\label{cryo_rad}
\subsection{Existing infrastructure}

Created by the University of Liège, the Centre Spatial de Liège (CSL) is a research center focused on space instrumentation with an environmental test facility serving the European Space Agency (ESA), the space industry and regional companies.

CSL activities inherit more than 50 years of experience focused on scientific space instrumentation.
The common thread of these activities is the optical heart of space missions. From design to qualification testing, CSL is involved in each phase of engineering.

Assembly, integration and verification of space instrumentation are operated on clean rooms fully compatible with the requirements of ESA programs. A majority of scientific optical instruments was tested in our facilities.

CSL provides a customized service to expose instruments, systems and satellites to extreme conditions encountered in space with a unique expertise in optics and thermal regulation, including cryogenics.

The huge size of vacuum chambers (up to 6.5 m diameter) enables the qualification of full satellites. In the recent past, Herschel, Plank, Gaïa, Aeolus and EUCLID were successfully tested.
The expertise is maintained thanks to the research and development activities shared with our university and with numerous partners (academics and industrials) at regional and international levels.

Regarding the E-Test project, the available CSL’s facilities are particularly convenient for the cryogenic test implementation. Indeed, CSL is equipped with cleanrooms, vacuum chambers and He liquefiers. 

\subsection{Cool down strategy}

The E-test experiment will take place at CSL in our biggest vacuum chamber: Focal 6.5. One goal of the experiment is the cooling down to cryogenic temperature of a 100 kg silicon mirror. As explained before in the document, the mirror is suspended in order to avoid any vibration. This vibration isolation implies that no contact can be occurs between the suspended parts and external (to the pendulum) elements. 

This very important constraint forced us to develop an innovative concept allowing the contactless cooling of the mirror. This concept is based on radiative exchanges between two intertwined radiators-like structures called inner and outer cryostats. 

\begin{figure}[H]
    \centering
    \includegraphics[width = 0.75\linewidth]{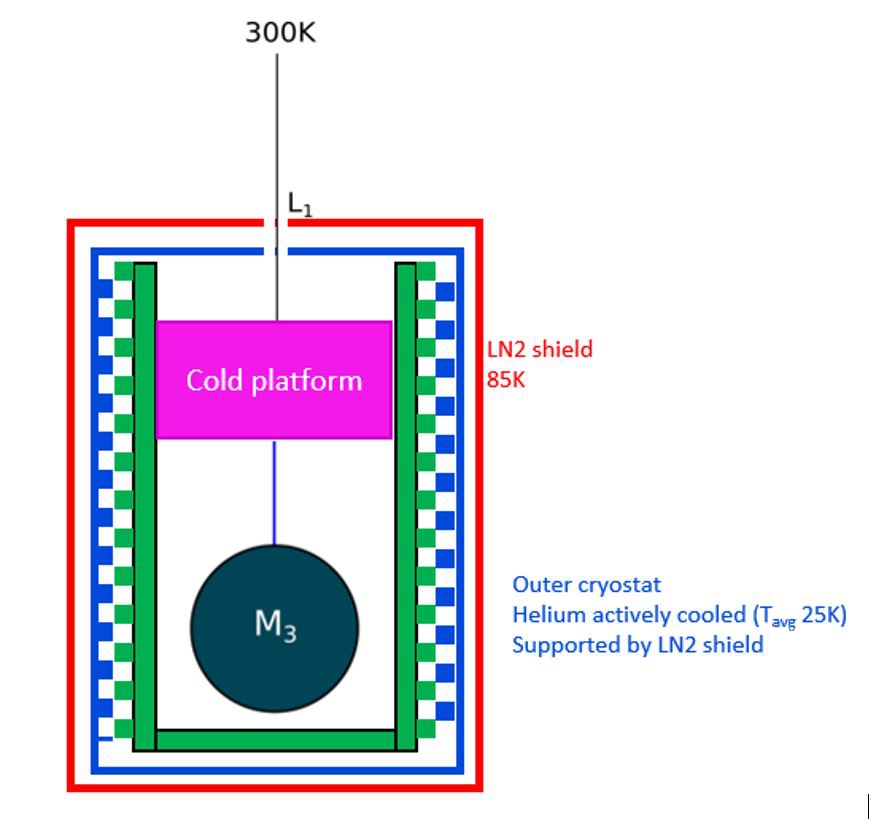}
    \caption{Design of the cooling concept}
    \label{fig:concept}
\end{figure}

Figure~\ref{fig:concept} presents the scheme of the cryogenic concept developed for the E-test project in close collaboration with the partners of the consortium. It is based on the radiative exchanges only between the inner cryostat (in green on the Figure~\ref{fig:concept}) directly attached on the cold platform supporting the suspended mirror (this cryostat is consequently also suspended), and the outer cryostat (in blue on the Figure~\ref{fig:concept}) in direct connection with the He shrouds and pre-cooled by the LN2 shield. It is important to mention that the LN2 shield and the He shrouds as well as the outer cryostat have no contact with any part of the suspended elements in order to avoid vibration transmission to the mirror.

The required size of the radiative heat exchange area depends on two main factors: the target final temperature of the payload and the total heat load to be extracted at that temperature. A third important aspect is the transient behavior of the system: the total mass of the cold suspended. The cooling time should be minimized and so the mass of the cold payload (inner cryostat, mirror, cold platform). Assuming the total heat load to be extracted is about 0.3W (including IR loading through openings, harness heat leaks), the required area then depends on the emissivity at that temperature. As the emissivity decreases drastically with temperature, even for black paint, one can assume a 0.5 emissivity. Under these assumptions and with a sink temperature at 20K (average temperature of the external cryostat), the required flat area of the heat exchanger is about 70m² for a target payload temperature of 25K. Now if the external cryostat is fed with liquid helium at 4K, this reduces to 41m² still for a target of 25K. Because it depends of the fourth power of the temperature, the required area significantly increases if the required target temperature is lower: for 20K, it goes up to 100m² instead of 41m² and to more than 300m² for a 15K target. 

To increase the exchange area while remaining in a compact envelope, the chosen strategy is to work with a stack of horizontal fins for both cryostats: the suspended inner one and the outer one linked to the gaseous Helium fed shrouds. These fins are spaced in order to leave 2cm gap above and bottom each fin when the inner and outer cryostats are in place (interlaced), avoiding any contact between fins and consequently cryostat structure. In order to maximize the emissivity and the cooling efficiency, the fins will be coated with thick layer of black paint on both faces. Finally, the fins are designed in order to minimize their weight while ensuring sufficient rigidity in order to keep the natural bending in the range of 1.5mm. This intertwined radiator geometry improves the radiative exchange area by about a factor 16 compared to the projected flat surface. This means that to reach 25K on the inner payload with an outer cryostat at an average temperature of 20K, the projected area can be reduced from 70m² to 4m². 

\begin{figure}[H]
    \centering
    \includegraphics[width = 0.75\linewidth]{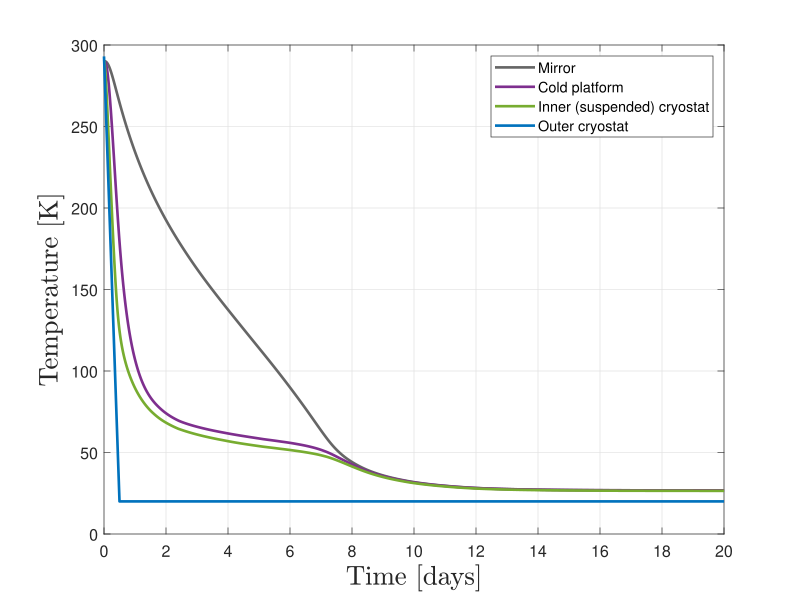}
    \caption{Temperature profiles of silicon mirror (grey), cold platform (purple), inner cryostat (green), outer cryostat (blue), with the temperature of the helium thermal tent cooled to 20K in 12 hours and emissivity of black paint assumed at 0.5.}
    \label{fig:proto_cooling}
\end{figure}

Figure~\ref{fig:proto_cooling} shows the results of the latest thermal simulations which give a stabilized (dT/dt << 1 K/day) temperature of the mirror of 27K after 14 days. The whole cryostat is modelled with ESATAN-TMS. The geometrical mathematical model (GMM) is based on the latest CAO available to compute the view-factors between the surfaces. The thermo-optical properties of the materials are obtained by heritage from previous verified projects. For the black paint of the radiator fins however, a worst-case approach was preferred to be conservative regarding its dimensions.

Then the thermal mathematical model (TMM) is solved by taking into account both thermal conductivities and specific heat in function of the temperature. Some of those properties, including the thermal conductivity of the fibres in sapphire were provided by Nikhef.

\begin{figure}[H]
    \centering
    \includegraphics[width = 0.75\linewidth]{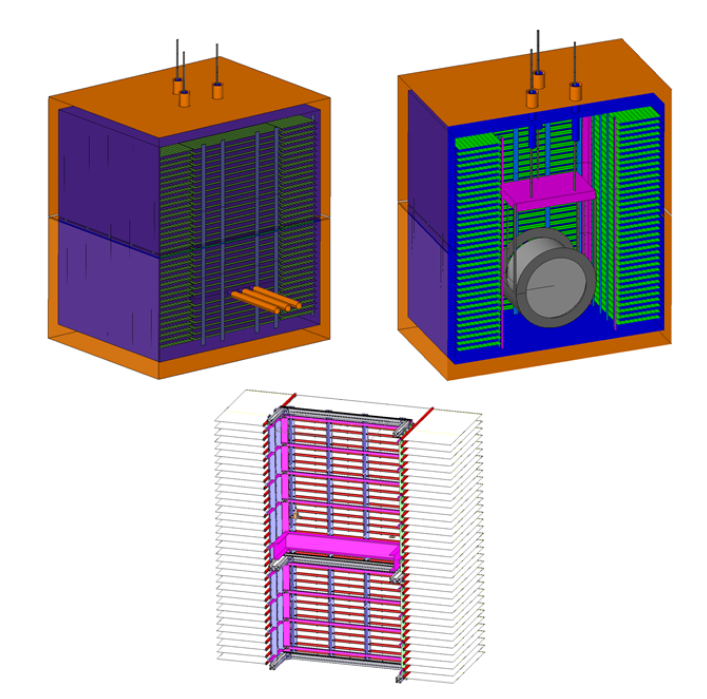}
    \caption{CAD and GMM design views of the cryostat}
    \label{fig:CAO_proto}
\end{figure}

With a worst-case approach, the external cryotat assumes:

\begin{itemize}
    \item The LN2 (liquid nitrogen) panels (orange in Figure~\ref{fig:CAO_proto}) is supporting the GHe (gaseous Helium) panels supporting themselves the external fins.
    \item The three apertures in front of the mirror are assumed to be with 30mm diameter through the LN2 shield and GHe panels and are aligned such that only one internal fin is removed (keeping two consecutive external fins). In addition, these three holes are equipped with LN2 shield and GHe tubes in order to minimize the incoming heat load in the system, see Figure~\ref{fig:CAO_proto}.
    \item The three suspension apertures are assumed with 50mm diameter on the top of the cryostat for the Titanium wired and cold harness, also equipped with LN2 shield + GHe tubes protruding inside the cryostat to radiative enhance thermalisation of the wires \& harnesses.
\end{itemize}

For the inner/suspended cryostat, the assumptions are the following:

\begin{itemize}
    \item Three Titanium wires coming from the marionette are assumed in Ti6Al4V with 4mm in diameter.
    \item 20 AWG24 phosphore bronze wires for sensors.
    \item 10 AWG32 phosphore bronze wires for temperature sensors (cooling down monitoring).
    \item  4 sapphire rods with 10mm diameter but with thinner ends with 3mm in diameter, supporting the mirror from the cold platform.
    \item  The interfaces between these sapphire rods and the CuCrZr blades of the cold platform are improved with indium sheets.
    \item Flexible thermal straps linked the fin frames (T profile) directly to the cold platform.
    \item No continuous power dissipation.
\end{itemize}

\begin{figure}[H]
    \centering
    \includegraphics[width = 0.75\linewidth]{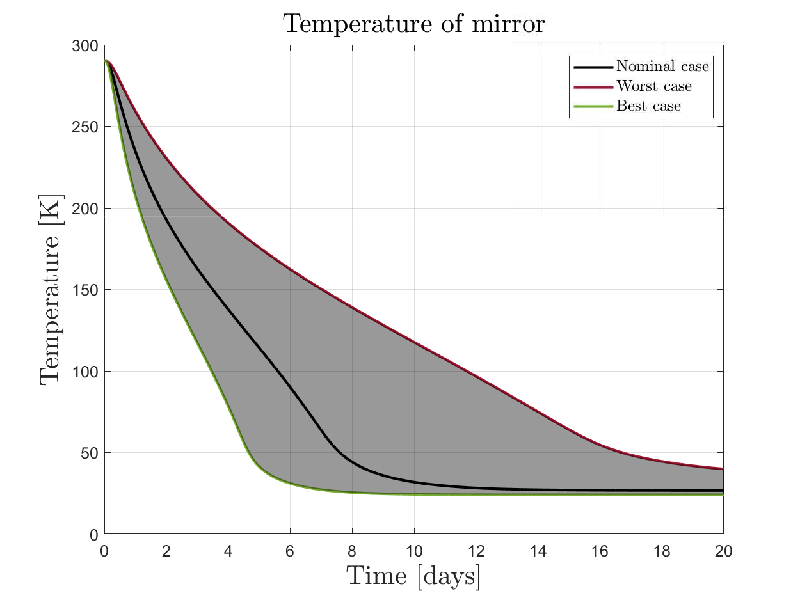}
    \caption{Temperature profiles of the mirror compared with worst case and best case assumptions.}
    \label{fig:proto_cooling_uncrt}
\end{figure}

To these numbers should be added some uncertainties due to the presence of large uncertainties in various parameters such as the emissivity of the paint, contact conductances through the screws and straps, thermal properties of materials at cryogenic temperatures, etc. Therefore, Figure~\ref{fig:proto_cooling_uncrt} depicts a range of possible temperature profiles for the cooling of the mirror, using a worst case approach.

\subsection{Future work} 
\begin{figure}[H]
    \centering
    \includegraphics[width = 1\linewidth]{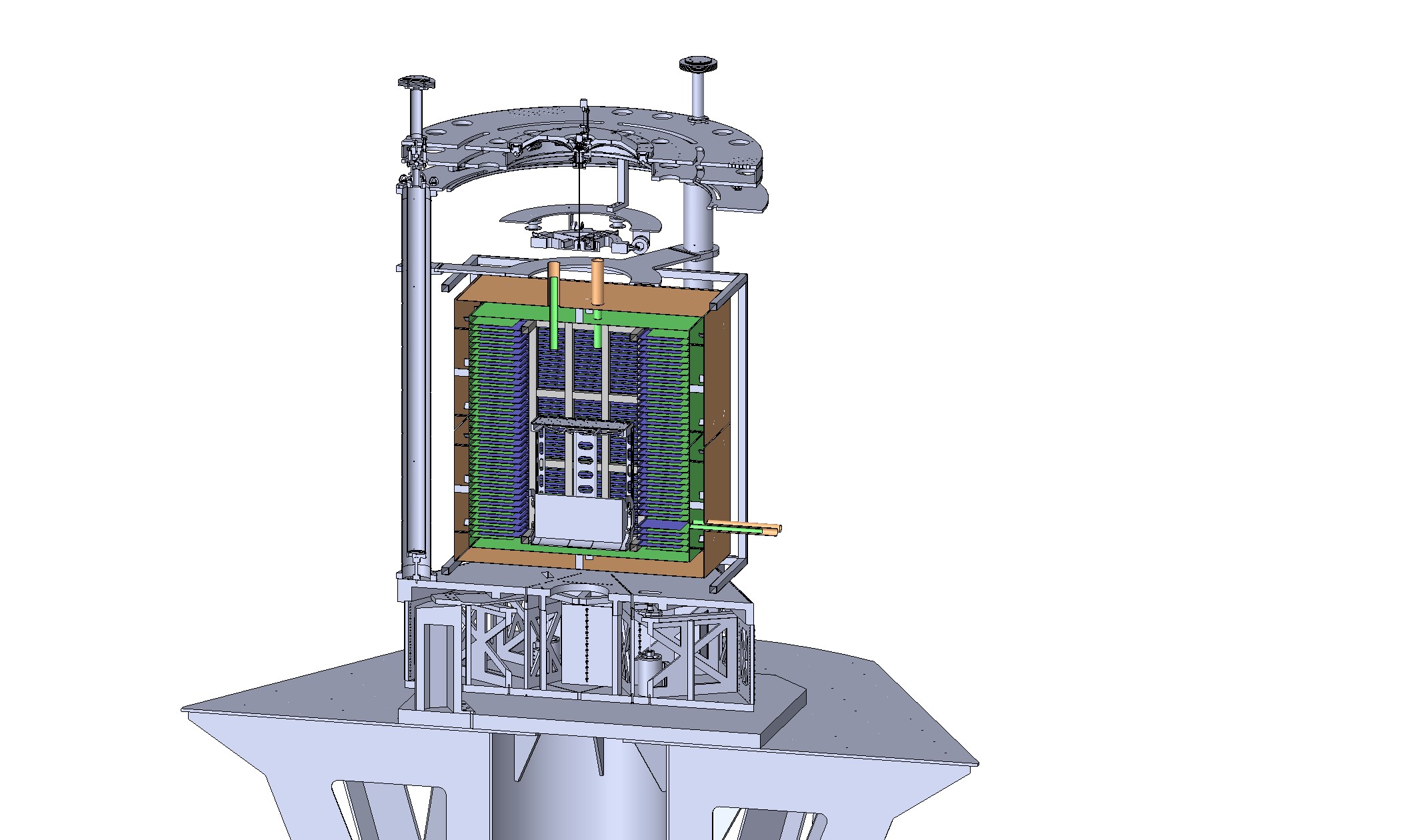}
    \caption{Preliminary design of the prototype on the optical bench of Focal 6.5 vacuum chamber (sectional view)}
    \label{fig:proto_design}
\end{figure}

In the next few months, specifically in close collaboration with the PML Lab (ULiège) and Nikhef, we will produce the final design of the prototype which will take place in the vacuum chamber, see Figure~\ref{fig:proto_design}. This design includes the pendulum, the cryostat, but also all of the external devices necessary for the experiments (thermal tents, interfaces, sensors,…). 
In parallel to the mechanical design, CSL will finalize the thermal simulations of the thermal tent/cryostat/mirror system. These simulations will provide the best and worst case based on realistic assumptions and taking into account some uncertainties base on our experience in such problem (mainly thermal contact, emissivity dependence with temperature). The only way to reduce the uncertainties would be to carry out specific cryogenic tests on representative elements in smaller chamber. Unfortunately, due to schedule and cost constraints, such characterizations cannot be foreseen in this project. Consequently, it is expected that large uncertainties will remain on the projected final temperature after a constrained cooling time of about 20 days. The E-test project will be an experimental concept validation for the cooling down method. 

\section{Cryogenic test bench}
\label{cryo_test}

\newcommand{\minSI}{\,\text{min}}
\newcommand{\kelvinSI}{\,\text{K}}
\newcommand{\mbarSI}{\,\text{mbar}}
\newcommand{\wattSI}{\,\text{W}}
\newcommand{\hourSI}{\,\text{h}}
\newcommand{\hzSI}{\,\text{Hz}}
\newcommand{\mumeterSI}{\, \mu\text{m}}
\newcommand{\mmSI}{\,\text{mm}}

\subsection{Design and use case of the test bench}

Within the E-Test Project a variety of cryogenic components are developed. The main E-TEST prototype will take multiple days to reach its vacuum and temperature goals. Quickly adjusting or repairing components inside the chamber is thus not possible. Therefore, testing them prior to using them inside of the E-TEST prototype is necessary. That is one of the use cases of this cryogenic test bench. The design parameters such as temperature and pressure level are chosen within the specification of the ET-LF configuration displayed in Tab. \ref{tab:Parameters_Cryostat}.
The cooling power is provided by a two-stage Gifford-McMahon (GM) cryocooler. This closed-cycle cryostat is driven by compressed helium provided by a water cooled compressor. 

\begin{table}[htb]
    \centering
    \begin{tabular}{|c|c|c|}
        \hline
        Parameter & ET-LF design goal & Cryogenic Test Bench \\
        \hline
        Minimum temperature & $ 10 - 20 \kelvinSI $ & $ 10 \kelvinSI $ \\
        Achievable vacuum level & $ > 10^{-10} \mbarSI $ & $ > 10^{-10} \mbarSI $\\
        Maximum power load & $ \text{several } \wattSI \text{ at } 10 \kelvinSI $ & $ \sim 1 \wattSI \text{ at } 10 \kelvinSI $ \\
        Cooldown time & several months & $ < 2 \hourSI $ \\
        \hline
    \end{tabular}
    \caption{Comparison of the chosen parameters for the Cryogenic Test Bench to ET-LF from \cite{ET20202} }
    \label{tab:Parameters_Cryostat}
\end{table}

\begin{figure}[htb]
    \centering
    \includegraphics[width = 0.75\linewidth]{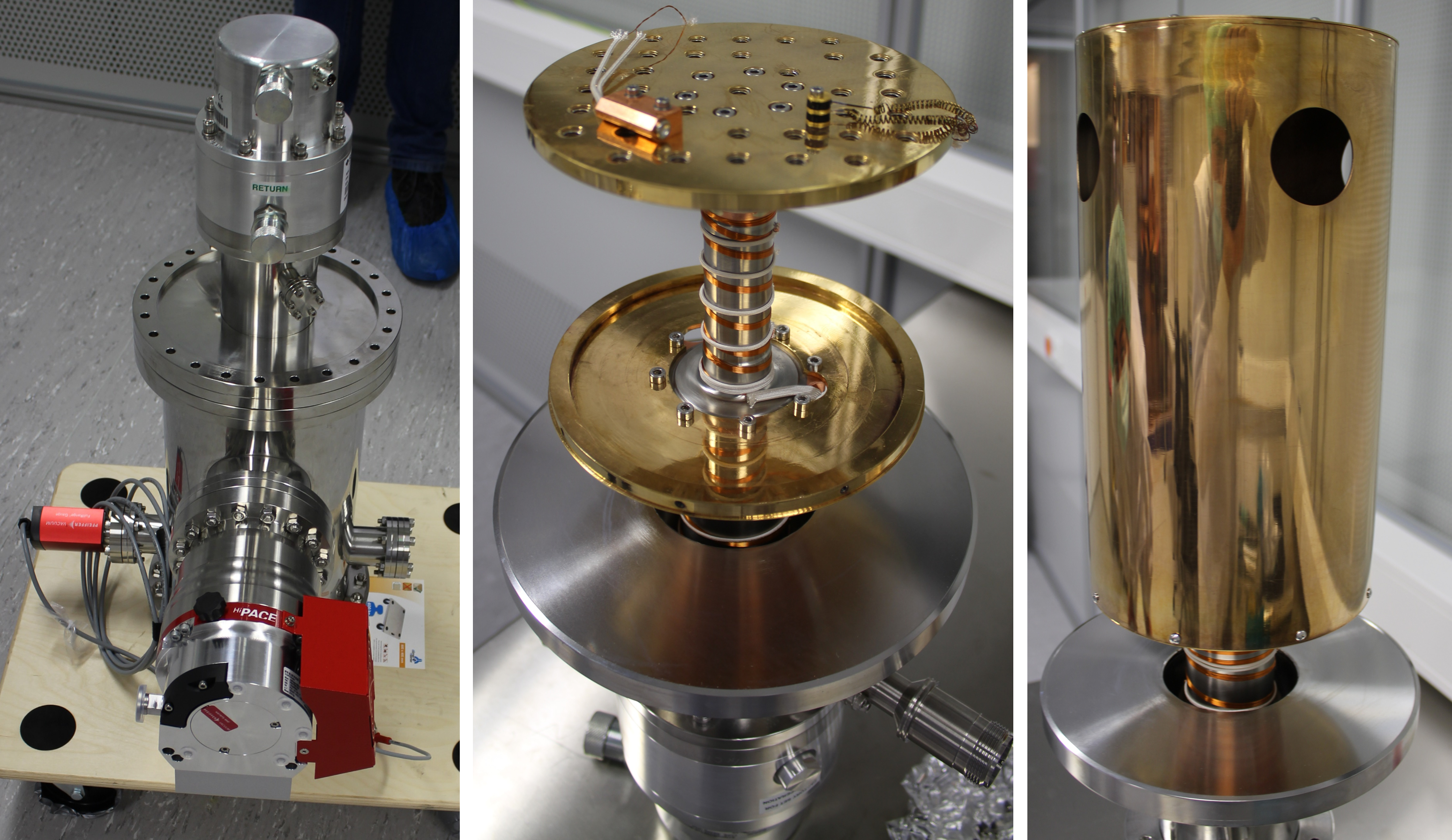}
    \caption{Left: Cryostat with turbomolecular pump attached; Middle: Open cryostat without heat shield, on the second stage, a heater cartridge and two temperature sensors (LS-DT-670B-CU) are located on the sample plate; Right: Open cryostat with heat shield}
    \label{fig:Cryostat}
\end{figure}

The cryocooler is placed inside an UHV vessel with 4 window ports. A heat shield is mounted on the first stage of the cryocooler. This reduces the power introduced in the cold environment by thermal radiation. The usable volume is cylindrical and has a diameter of $150 \mmSI$ and a height of $150 \mmSI$. The sample plate, which provides the mounting points for the tested components, is fastened to the second stage, seen in Figure \ref{fig:Cryostat}.

\begin{figure}[htb]
    \centering
    \includegraphics[width = 0.75\linewidth]{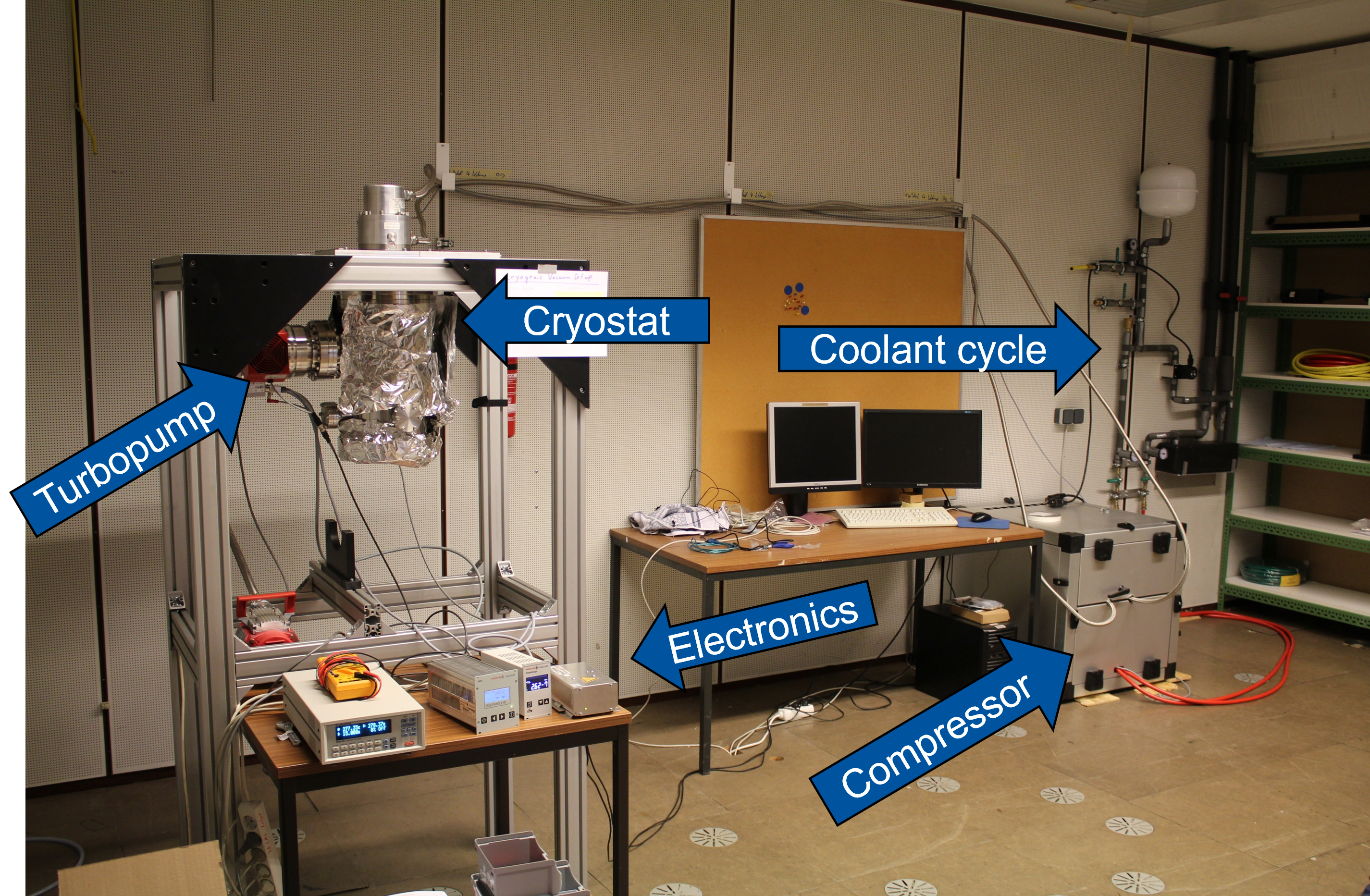}
    \caption{The cryostat is placed inside a holding structure mounted to the concrete floor. The aluminium extrusions are filled with sand to dampening higher frequency resonances. The control electronics are installed on a lab table next to the compressor and its coolant cycle.}
    \label{fig:Cryostat_setup}
\end{figure}

The whole cryogenic test bench is affixed inside a holding structure as can be seen in Figure \ref{fig:Cryostat_setup}. Among the improved seismic isolation, this allows an easier access to the insides of the cryostat. This structure is rigidly mounted onto the concrete floor of the room. By filling the hollow aluminium extrusions with sand, high frequency resonances are dampened. Next to the cryostat, the control electronics and the fore vacuum system are installed. The helium compressor is placed on a vibration dampening matt next to the water cooling system.

While the cryocooler is running, it induces vibrations to the sample plate. These vibrations are in an order of a few $ \mumeterSI $ in amplitude with a frequency of $ \sim 1 \hzSI $. To reduce these vibrations, the cryocooler can be turned off and  measurements can be performed while the structure is slowly warming up. Depending on the power output of the sample, these measurements can last up to $ 10 \minSI $ below $20 \kelvinSI $. To further reduce the vibration level, a valve can be closed and the turbomolecular pump shut down in addition.

\subsection{Commissioning of the test bench}

After the cryogenic test bench was successfully installed, different commissioning measurements were performed. Alongside with a vacuum leakage test, long-term vacuum tests over a few days and multiple temperature cycle measurements were performed. The minimum temperature achieved, $ T_{min}=6.7 \kelvinSI $, was reached at a minimal pressure level of $ P=6 \cdot 10^{-9} \mbarSI $. With a different pre-pump the setup was able to reach a minimal pressure level of $ P_{min}=4 \cdot 10^{-10} \mbarSI $ while cooling.

\begin{figure}[htb]
    \centering
    \includegraphics[width = 0.98\linewidth]{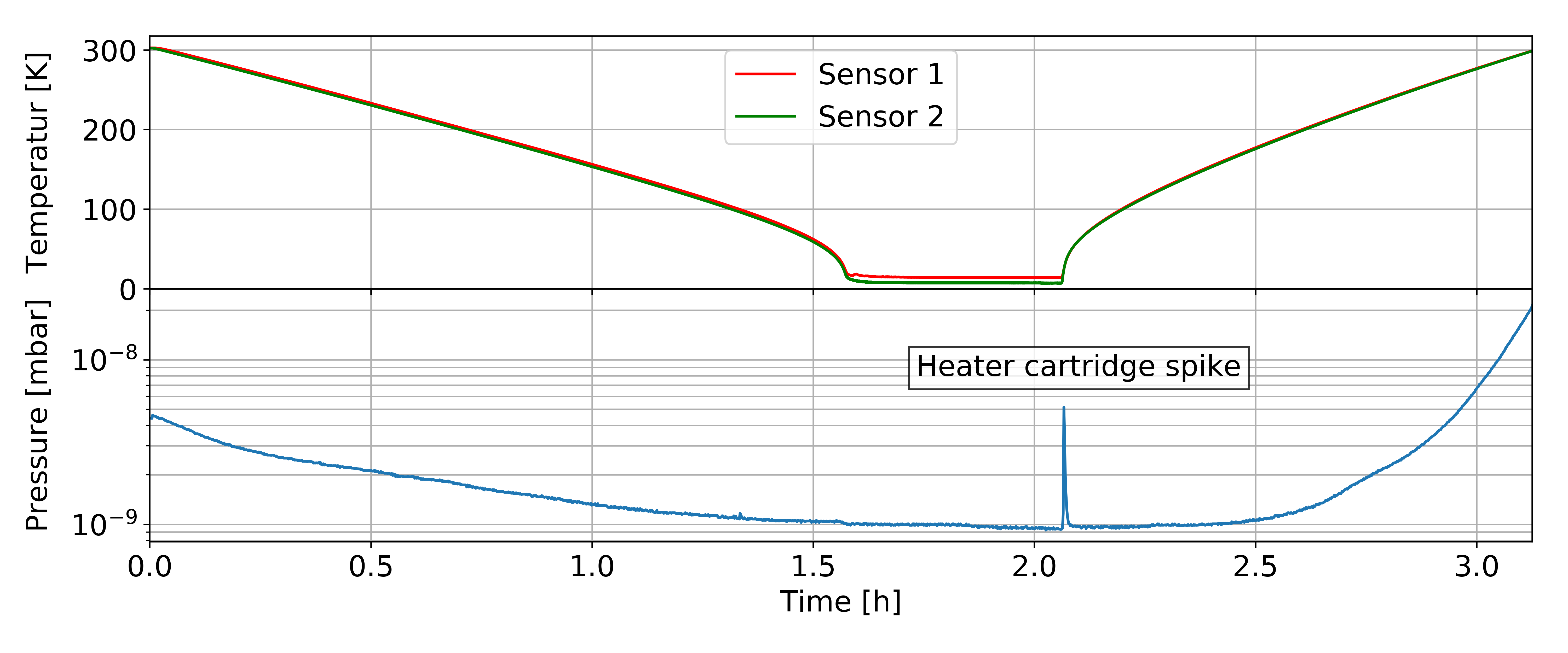}
    \caption{Complete Temperature cycle of cooling down to $ 7 \, \text{K} $ and heating back up. Sensor 2 is mounted on the sample plate, while sensor 1 is located next to the plate. The pressure spices in the beginning of the heating due to frozen out gases on the heater cartridge itself.}
    \label{fig:Typical_Temp_Curve_new}
\end{figure}

A typical cool down of the cryostat takes approximately $ 95 \minSI $ as can be seen in Figure \ref{fig:Typical_Temp_Curve_new}. The pressure level decreases with lower temperature due to the freeze-out of residual gases. Once cooled down, the temperature is stable within $ \pm 0.2 \kelvinSI $. The fluctuations are induced by the cyclic nature of the cooling process. Heating the setup to ambient temperatures takes approximately $ 60 \minSI$ using the heater at full power.

\begin{figure}[htb]
    \centering
    \includegraphics[width = 0.98\linewidth]{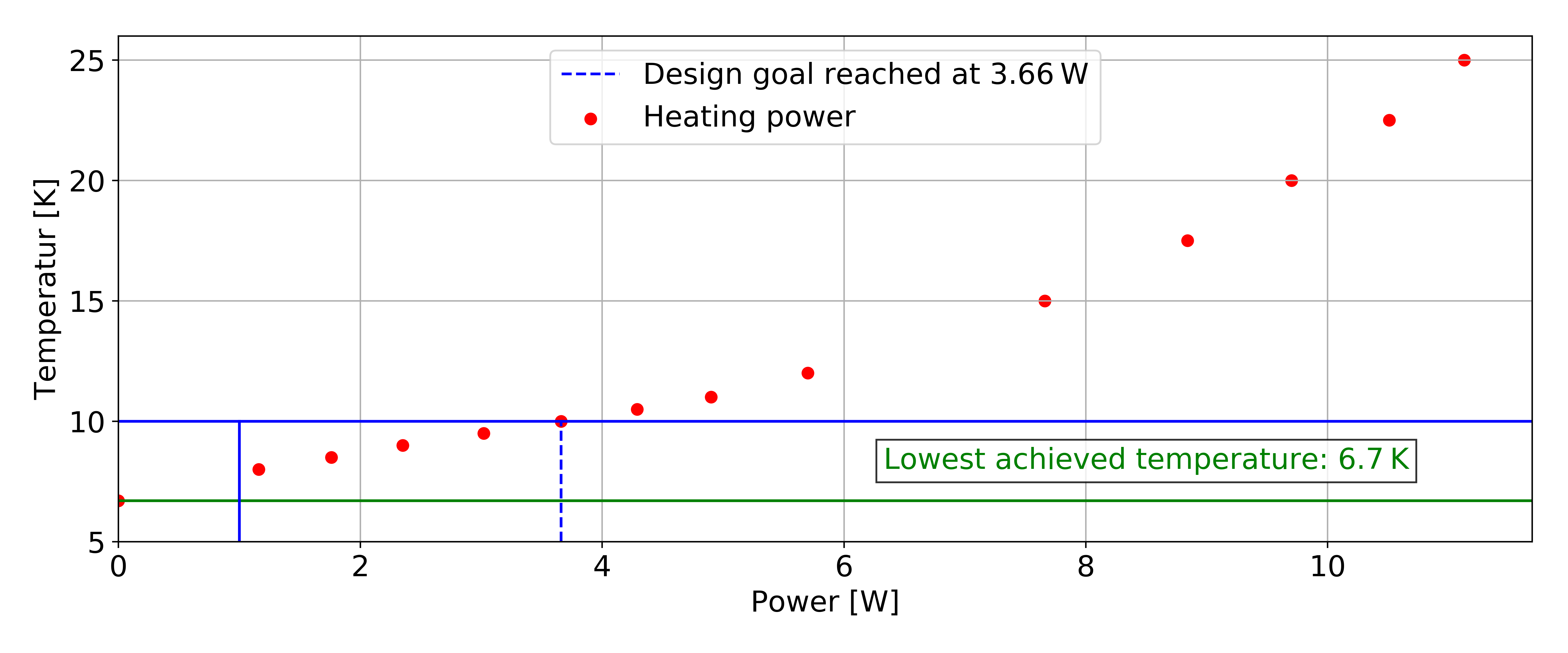}
    \caption{Comparison between the measured temperature of the sample plate with different heating power levels. The heater cartridge is operated by a Lake-shore control unit. At an additional $ 3.66 \wattSI $ our design goal of $ 10 \kelvinSI $ is reached. The lowest achieved temperature is $ 6.7 \kelvinSI $ with the cryostat.}
    \label{fig:Power_over_Temp}
\end{figure}

As an additional measurement, the reachable temperature for a specific heat load was measured. As seen in Figure \ref{fig:Power_over_Temp}, the setup is capable of reaching the target temperature of $ 10\kelvinSI$ at an additional heat load of $ 3.66 \wattSI $ which is far above the required $ 1 \wattSI $ design value. The used heater cartridge has a resistance of $ 40 \Omega$.

\section{Cryo-CMOS electronics}
\label{cryo_cmos}
Application Specific Integrated Circuit (ASIC) design has many advantages over designs entirely made with commercial discrete components, especially for applications with stringent requirements such as those imposed by a cryogenic cooling system. An ASIC, as the name implies, is tailored for the application. Considering the fact that research and development in Integrated Circuit (IC) design is on average 15 years ahead of the latest commercially available products, the advantage of ASICs is obvious. A single IC can integrate multiple functions while benefiting from extremely small size and significantly lower power consumption. This allows for sensor front-ends, Analog-to-Digital Converters (ADCs) and Digital-to-Analog Converters (DACs) to be realized in extremely small sizes and with reduced heat emission.
Even when the footprint of electronic circuits is not a limiting factor, there are disadvantages to using discrete components in extreme conditions such as those found it ET-LF. Commercially available discrete components, at temperatures close to those of liquid helium, have generally unpredictable behavior. Reduced performance \cite{Svindrych2008}\cite{LeGuevel2020} but also complete failure of integrated circuits can occur. A case in point are linear voltage regulators \cite{Homulle2018}, that often use a BJT (bipolar junction transistor) as the power element and are completely non-functional at low temperature due to the dramatic reduction in current gain in bipolar transistors \cite{Cressler1989}\cite{Cressler1993}\cite{Song2006}. Finally, a final key advantage of custom ICs over commercial components is the possibility to enclose the IC die in packages with reduced outgassing \cite{Sandor2000}, which is a key requirement to operate close to a delicate optical system such as ET-LF.
\vskip 5pt
Table \ref{tab:Cryo_IC_tech} shows the main semiconductor technologies used in the fabrication of integrated circuits for low temperature applications. Of all those listed, only CMOS technology can offer very large scale integration (VLSI) and rely on over 60 years of industrial advances and optimization \cite{Charbon2019}. This fundamental advantage, coupled with functionality down to temperatures of a few mK, was the reason behind the choice of CMOS technology for the development of custom ICs for ET-LF.

\begin{table}[htb]
    \centering
    \begin{tabular}{|c|c|}
        \hline
        Technology & Minimum operating temperature \\
        \hline
        Si BJT & 100 K \\
        Ge BJT & 20 K \\
        SiGe HBT & < 1 K \\
        GaAs MESFET & < 4 K \\
        CMOS & 30 mK and below \\
        \hline
    \end{tabular}
    \caption{Comparison of the different technologies employed for IC design at low temperatures. Minimum operating temperature shown is for reliable and still acceptable performance \cite{Song2006}\cite{Sebastiano2018}.}
    \label{tab:Cryo_IC_tech}
\end{table}
\vskip 5pt
The possibility to place the analog front-end, part of the signal processing and the analog-to-digital conversion directly in the cryogenic chamber and close to the sensor, allows to reduce the number of connections with the external environment to a minimum. This leads to a twofold advantage. On the one hand, the integrity of the signal is maintained and not compromised by the use of long cables often made by prioritizing thermal specifications over electrical ones. On the other hand, reducing the number of connections to the outside world allows to maintain a better thermal insulation and avoid the injection of unwanted thermal noise. 
\vskip 5pt
Even passive components such as resistors, capacitors and  inductors show significant variations with temperature and most of them are not suitable for cryogenic environment \cite{Homulle2018}.  Exemplary is the case of capacitors, whose behavior at low temperature is closely related to the type of dielectric used. Class I ceramic capacitors (e.g. NP0/C0G), PPS capacitors and tantalum polymer capacitors show a stable behavior down to liquid helium temperature. The rest of the commercially available capacitor technologies suffer from the reduced temperature (with expensive exceptions such as silicon dioxide capacitors, which are also generally hard to find). A brief summary of the cryogenic characteristics of the main commercially available dielectrics is given in Table \ref{tab:Caps_Cryo}. 

\begin{figure}[htb]
    \centering
    \includegraphics[width = 0.98\linewidth]{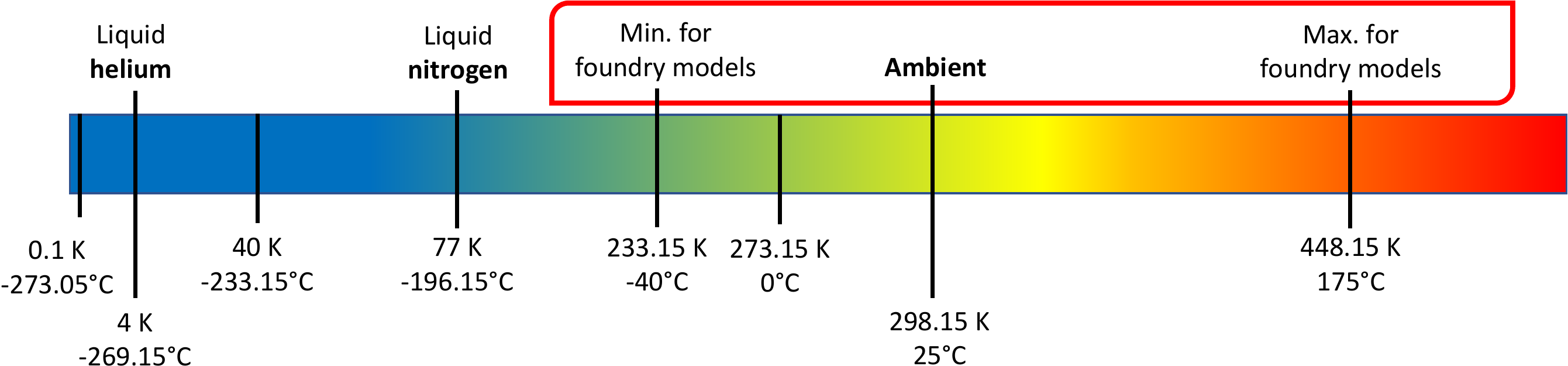}
    \caption{Standard temperature range for the electrical models (for SPICE-like circuit simulation software) distributed by semiconductor foundries. Outside of the highlighted interval, models are not reliable and can lead to errors beyond 100\% \cite{Akturk2010}.}
    \label{fig:cryoCMOS-temprange}
\end{figure}

\begin{figure}[htb]
    \centering
    \includegraphics[width = 0.65\linewidth]{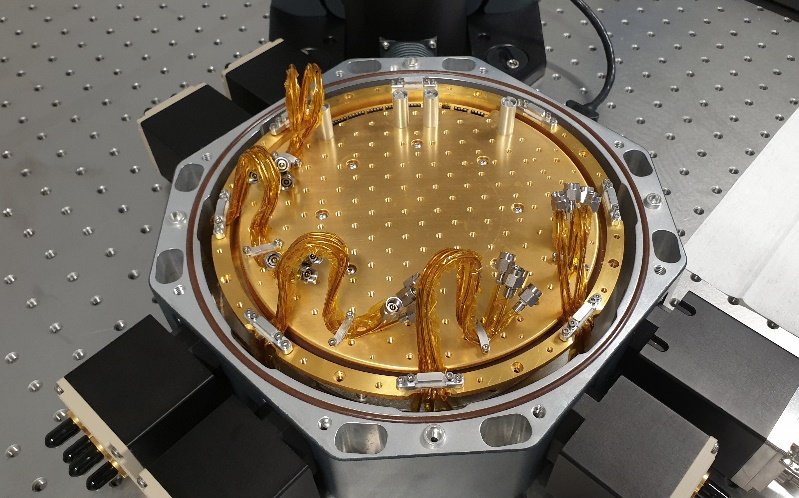}
    \caption{Retro-fitting of the cryostation with Kapton electrical insulation to enable ultra-low current measurements.}
    \label{fig:cryoCMOS-cryostatKUL}
\end{figure}

\begin{figure}[htb]
    \centering
    \includegraphics[width = 0.98\linewidth]{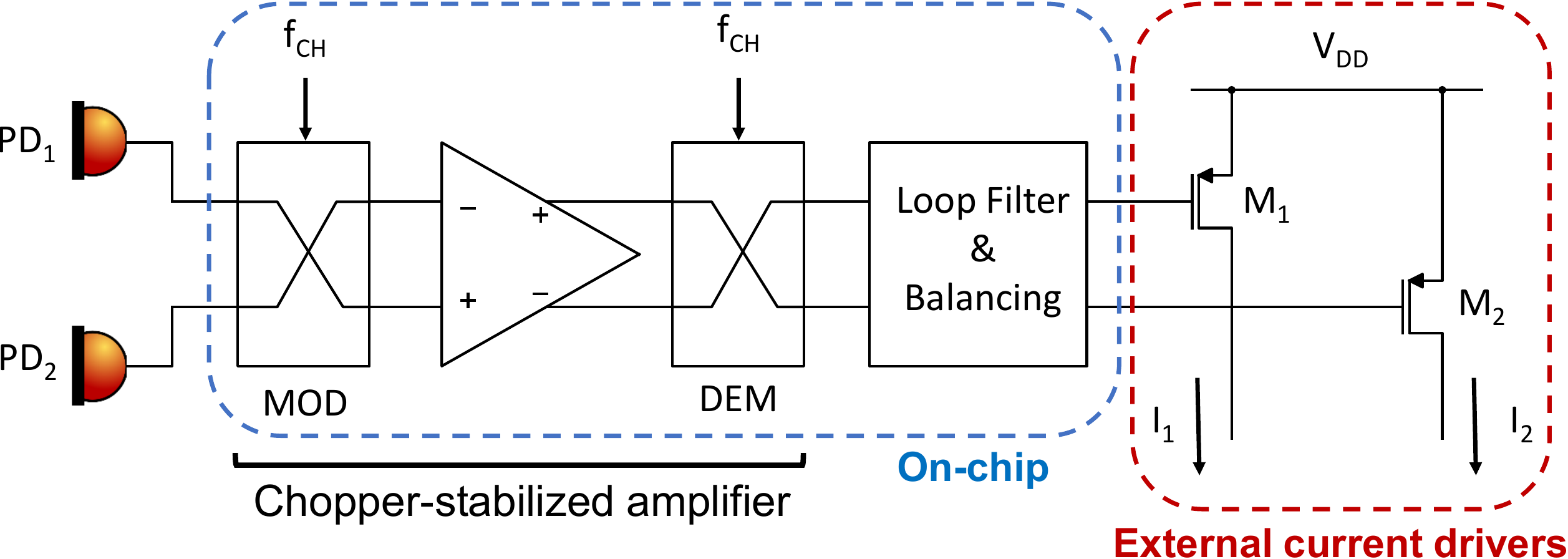}
    \caption{Conceptual schematic of the custom cryo-chip under design for the interferometric sensor developed at the UC Louvain group. Current drivers are kept off-chip for thermal reasons. Architecture is still provisional.}
    \label{fig:cryoCMOS-Acc}
\end{figure}

\begin{figure}[htb]
    \centering
    \includegraphics[width = 0.8\linewidth]{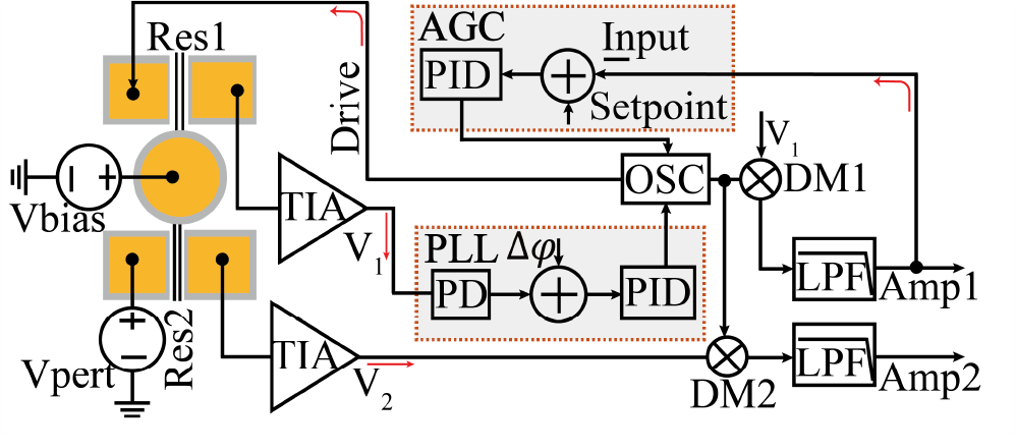}
    \caption{Test setup commonly used for weakly-coupled MEMS resonators. Adapted from \cite{Zhang2020}.}
    \label{fig:cryoCMOS-MEMS}
\end{figure}

\begin{table}[htb]
    \centering
    \begin{tabular}{|c|c|c|}
        \hline
        Dielectric & Capacitance variation & ESR variation \\
        \hline
        PPS & -0.63\%  & +374\% \\
        Ceramic X8L & -64\%  & +1271\% \\
        Ceramic C0G & -0.9\%  & +345\% \\
        Ceramic X6S & -30.5\%  & +1321\% \\
        Tantalum Solid & -31\%  & +200\% \\
        Niobium Oxide & -29\%  & +211\% \\
        $\mathrm{MnO_2 \ Tantalum}$ & -35\%  & +526\% \\
        Tantalum Polymer & -12\%  & -94\% \\
        \hline
    \end{tabular}
    \caption{Measured behavior of commercial capacitors at cryogenic temperature (6 K). Capacitance variations are referred to the low frequency value, while ESR variations are measured at the self-resonance frequency of the capacitor.}
    \label{tab:Caps_Cryo}
\end{table}
\vskip 5pt
CMOS technology, at temperatures close to those of liquid helium, shows several advantages \cite{Sebastiano2017,Incandela2018,Mehrpoo2019} including an increase in current drive capability \cite{Incandela2018,VanDijk2020} and maximum operating frequency \cite{Hong2008} (due to increased carrier mobility) \cite{Beckers2018} and a substantial reduction in leakage due to the very low carrier concentration in the absence of an electric field promoting dopant ionization \cite{Foty1990,Beckers2018model}. In addition, modern nanoscale technologies are normally free from non-idealities found in more mature technology nodes such as kink effect and hysteresis \cite{Incandela2018}.
\begin{figure}[htb]
    \centering
    \includegraphics[width = 0.49\linewidth]{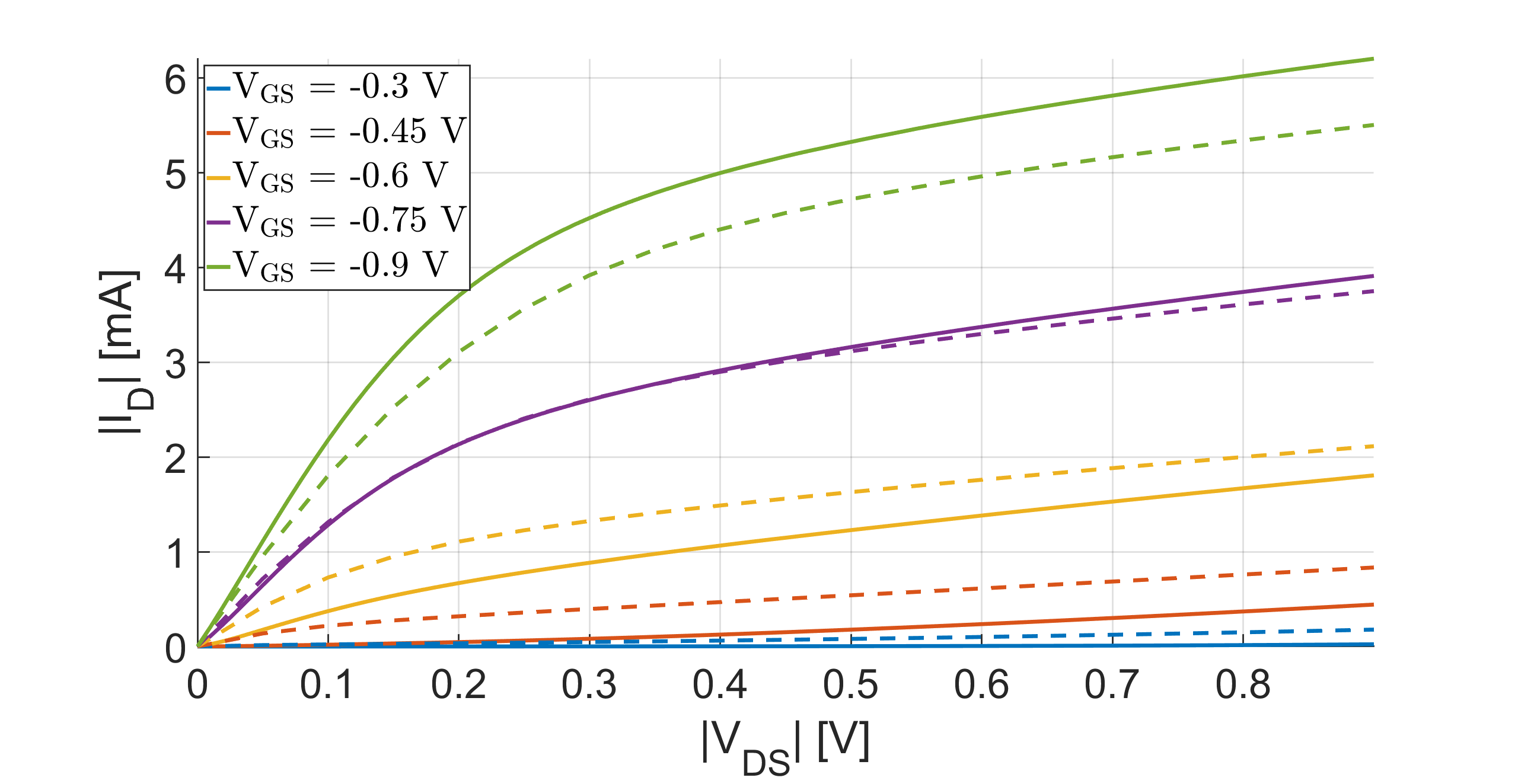}
    \includegraphics[width = 0.49\linewidth]{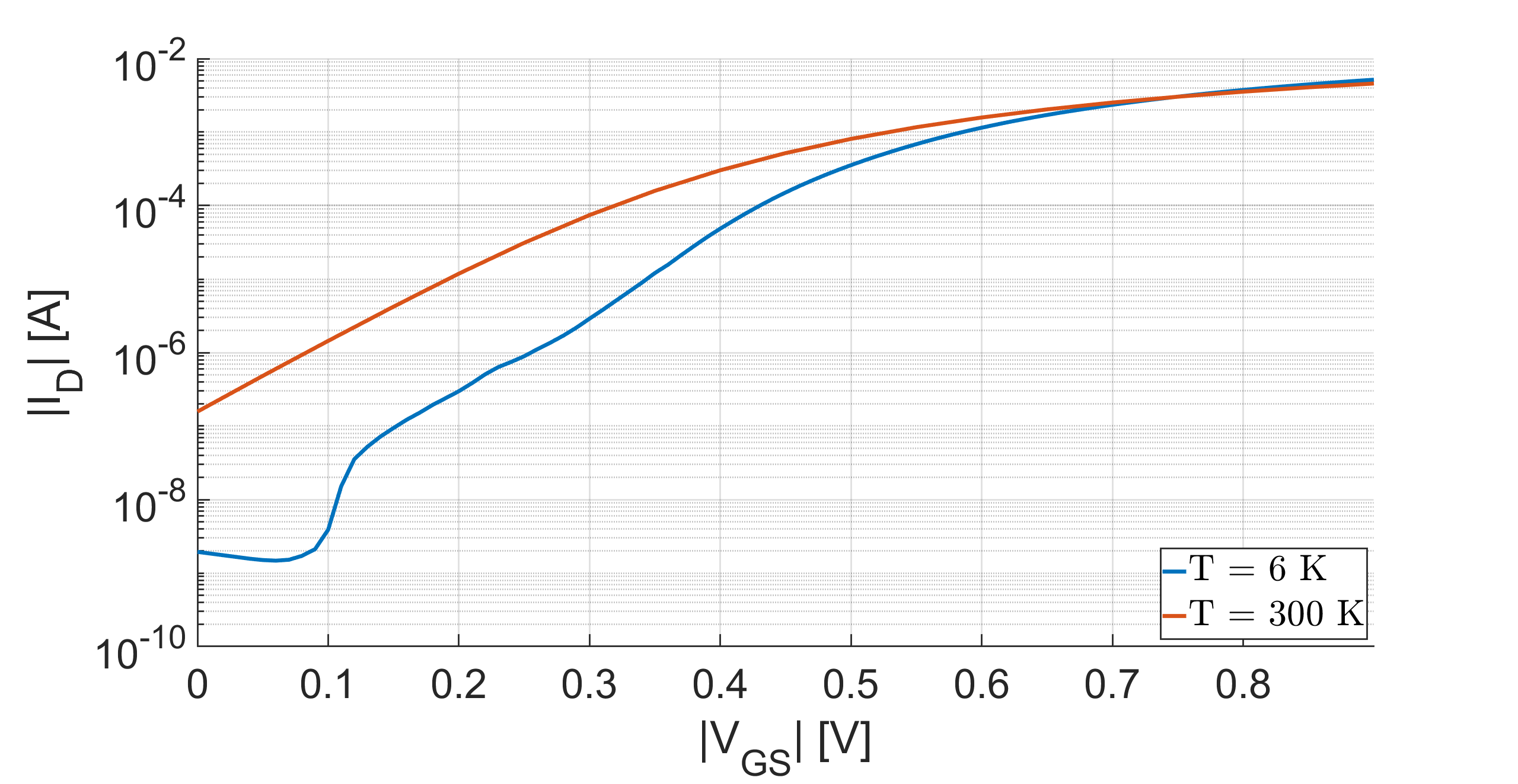}
    \caption{I-V characteristics of a pMOS transistor in a 40 nm bulk CMOS technology. Figure on the left hand side shows the increase in the threshold voltage (which lowers the curves for low values of the gate-source voltage $\mathrm{V_{GS}}$) and the current drive capability, i.e. the increase in the maximum current that the device is able to carry. In figure on the right hand side, the reduction of leakage when the device is turned off (low values of $\mathrm{V_{GS}}$) is clearly visible.}
    \label{fig:cryoCMOS-IV}
\end{figure}
However, the design of cryogenic (bulk) CMOS circuits also presents significant challenges. The shift in Fermi level at low temperature \cite{Beckers2018model} causes an increase in the threshold voltage \cite{Beckers2019}, as more band bending is required in the channel region to ensure the same carrier density \cite{Beckers2020}. The shift in the Fermi level is also believed to be the cause of the significant increase in flicker (1/f) noise at cryogenic temperatures, as it increases the energy range of the interface traps that contribute to the noise \cite{Hafez1990,Oka2020}. Broadband (white) noise, on the other hand, contrary to expectations, does not decrease proportionally with temperature since in nanoscale transistors a large part of it is shot-like, temperature-independent noise \cite{Smit2014}\cite{Chen2018}\cite{Chen2021}\cite{Peng2021}. Finally, differences between one device and another (also known as "mismatch") at low temperature are amplified \cite{THart2020}, leading to an increase in undesirable effects such as offset in precision amplifiers if appropriate design techniques are not employed to compensate for them \cite{Enz1996}. 
\vskip 5pt
The biggest challenge, however, is the absence of cryogenic models that can be used to simulate circuits in the design phase \cite{Incandela2018,Beckers2018model,Akturk2010}. The design and subsequent fabrication of an integrated circuit is in fact an extremely complex process that takes considerable time (several months) and, therefore, it is necessary to ensure a high level of confidence in the design phase to ensure the proper functioning of the fabricated circuit. For this reason, semiconductor foundries (such as TSMC and GlobalFoundries) release design tools called "Process Design Kits" or also "Product Design Kits" (PDKs) that include highly reliable electrical models to be used in complex circuit simulation software, able to reproduce also the statistical variations intrinsic to the manufacturing process and temperature behavior. Unfortunately, the models provided by foundries are only reliable within a relatively small temperature range (see Fig. \ref{fig:cryoCMOS-temprange}) and cannot reproduce the behavior of devices at cryogenic temperatures \cite{Incandela2018,Beckers2018model,Akturk2010,KeysightQuantum}. 
For this reason, we chose to edit the foundry models in order to account for the cryogenic variations and make them usable for cryogenic design too \cite{Incandela2018,Luo2019}. Test chips in different nanoscale technologies were fabricated to characterize active devices and integrated passive components within a high-vacuum, cryogen-free cryogenic chamber. The system was suitably modified (see Fig. \ref{fig:cryoCMOS-cryostatKUL}) to allow for high-precision current measurements \cite{KeithleyHandbook}, which are necessary to ensure model accuracy given the large reduction in leakage currents at cryogenic temperatures \cite{Incandela2018,Beckers2018,LakeshoreLowCurrents}. 
A parameter extraction procedure using state-of-the-art tools and algorithms has also been developed to allow direct modification of the foundry models, so as to considerably reduce the difficulties inherent in this type of process, which normally require years even with the resources of a foundry. Preliminary cryogenic models are currently in development and will be tested during 2022.  
\vskip 5pt
The ultimate goal is the design of two ASICs for the cryogenic mirror vibration control system in ET-LF. The first design concerns a custom chip to be inserted into the interferometric system developed by the research group at UC Louvain (see following sections). The fully-differential architecture in Fig. \ref{fig:cryoCMOS-Acc} is currently under development and a first prototype will be designed and fabricated in 2022. The design will include techniques for the compensation of cryogenic non-idealities such as increased flicker noise, in order to fully exploit the advantages discussed above.
The second design concerns a front-end for the readout and conditioning of a MEMS sensor based on weakly-coupled resonators, developed by the MNS research group at KU Leuven. The front-end for this type of sensors, in normal experimental setups, uses quite a complex system \cite{Zhang2020} including bulky laboratory equipment, as shown in Fig. \ref{fig:cryoCMOS-MEMS}. The aim of the joint effort of this project is to develop the MEMS accelerometer sensor with the highest resolution ever obtained, also integrating the necessary electronic circuitry in an extremely small space and with a negligible power consumption compared to traditional solutions.

\section{Cryogenic MEMS}
\label{cryo_iner}
Inserted into the continuous effort of suppressing Earthly vibrations that affect the suspended structure, a novel MEMS accelerometer based on the mode-localization effect is being proposed. This effect states that in a periodic system - two coupled identical resonators - the presence of a small perturbation in the original balanced system originates a vibration that gets spatially confined to a small region \cite{PhysRev.109.1492}.

Research has shown that the potential improvement in sensitivity is of 2 to 3 orders of magnitude when compared to 1 degree-of-freedom resonant sensors \cite{Spletzer2006,Spletzer2008}. It also has first order immunity to common mode changes such as temperature or pressure variations \cite{5168201}.

Weakly coupled resonators require that the stiffness of the coupler between two resonators is much smaller than the stiffness of the resonators themselves, typically an order of magnitude below. These two resonators must be identical in terms of mass and stiffness. For this symmetric system - the initial amplitude ratio between the first resonator and second resonator will be 1. When an acceleration force is acting on the device - if one resonator is driven to maintain its initial resonance at all times and the other experiences a change in stiffness due to the acting force, the symmetry is broken by a perturbation on of the resonators. Thus, the amplitude ration will no longer continue as 1, which means there is a spatially confined vibration on resonator two, and there is mode localization.

\begin{equation}\label{normalized1doffreq}
\lvert \Delta\omega \rvert \simeq \lvert \frac{\partial\omega_{o}}{\partial k}\rvert
\end{equation}

\begin{equation}\label{normalized1dof}
\lvert S_k^{\omega_o}\rvert = \lvert \frac{\partial\omega_{o}}{\delta k}\rvert \times \lvert \frac{k}{\omega_o}\rvert = \frac{1}{2}
\end{equation}

\begin{equation}\label{couplingratio}
K = \frac{k_c}{k}
\end{equation}

\begin{equation}\label{couplingchange}
\delta = \frac{\Delta k}{k}
\end{equation}

\begin{equation}\label{deltaml}
\Delta = \lvert \frac{u_i - u_{0i}}{u_{0i}} \rvert \simeq \lvert \frac{\delta}{4k_c}\rvert
\end{equation}

\begin{equation}\label{sensml}
\lvert S_k^{\Delta}\rvert = \lvert \frac{\partial \Delta}{\partial k}\rvert \times \frac{k}{\Delta} \simeq  \lvert \frac{k}{4k_c}\rvert
\end{equation}

For a single degree-of-freedom resonant device with change in frequency as output, the normalized sensitivity is stated in Equation~\ref{normalized1dof} \cite{Zhao2016z}, which yields $\frac{1}{2}$. In this equation, $k$ stands for the stiffness, and $\omega$ denotes the frequency. In a mode-localized device, this metric is the stiffness $4k$ over 4 times the coupling stiffness $k_c$, yielded by Equation~\ref{sensml} \cite{Zhao2016z}. This result means that the sensitivity is limited by how weak the coupling is and it is no longer limited at $\frac{1}{2}$.

On the other hand, the state-of-art for mode-localized MEMS accelerometers states that the best performer \cite{9354446} in noise performance and bias stability is limited at \(85 ng/\sqrt{Hz}\) and \(130 ng\), respectively. This can be partially explained by the lack of temperature-compensation systems or ASIC integration, as typically the theoretical thermo-mechanical noise of the proof-mass is at least one order of magnitude lower than the current reported state-of-art \cite{Pandit2019}.

In this project, the MEMS accelerometer will operate at the cryogenic temperature of 10 Kelvins and at a very high vacuum. This is an extremely controlled environment, in which environmental drifts are almost completely suppressed. Such an environment, in which a MEMS inertial sensor was never employed, requires a careful design process in which these operational conditions are taking into consideration.

To improve the performance of these devices, four major factors were identified: reducing the coupling factor to increase the sensitivity, suppressing temperature-related drifts, reduce electronic noise from circuitry, and reduce the thermal noise.

To reduce the coupling factor, an anchor coupler design was chosen - these types of devices achieved the lowest coupling factor in literature \cite{Zhang2020} and present a simplified structure. 
To suppress temperature-related drifts such as thermal stress created by different thermal expansion coefficients or variations in the Young's modulus \cite{s19071544}, an H-shape structure was employed - taking away thermal stress from the sensitive area \cite{s19071544}, specially present on the interface glass-glue-ceramic.
In order to reduce the electronic noise, a collaboration with Alberto Gatti from the MICAS group at KU Leuven is underway to develop a custom cryo-CMOS front-end for this device as replacing analog readout circuits with ASIC-level readouts is known to significantly improve MEMS sensors noise performance - it allows for the usage of techniques such as chopper stabilization which severely reduces flicker-noise.

\begin{figure}[htb]
    \centering
    \includegraphics[width = 0.65\linewidth]{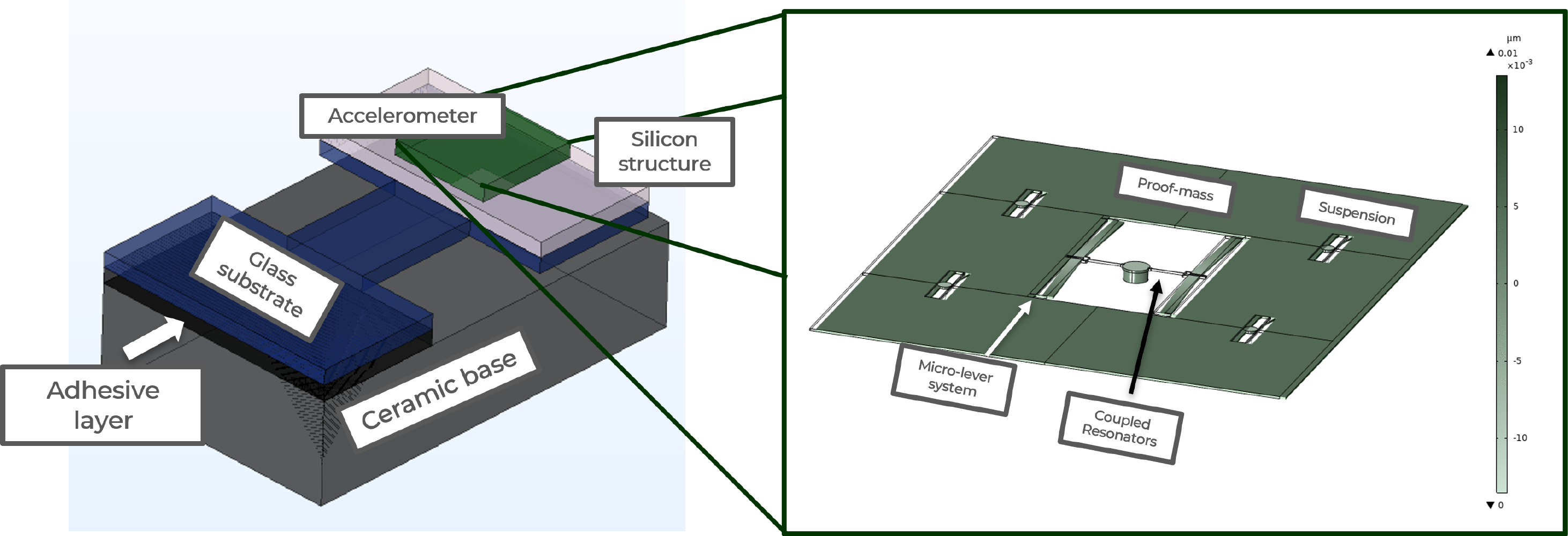}
    \caption{Complete preliminary model of MEMS mode-localized accelerometer and H-shape structure.}
    \label{fig:completemodelaccMNS}
\end{figure}

Taking all these considerations into account, a complete model was designed, illustrated in Figure~\ref{fig:completemodelaccMNS}. The sensor will be inserted in a H-shape structure composed by an Alumina base, an ultra-high vacuum epoxy glue, a borosilicate glass substrate and the Silicon structure. These materials have similar thermal expansion coefficients and can endure harsh temperatures - although there are no reports available for usage at 10 Kelvins.

The next steps include the simulation and optimization of the initial MEMS accelerometer model, the fabrication of a first complete design and the setup of a cryogenic characterization system.

\section{Development of cryogenic sensors}
\begin{figure}[htb]
    \centering
    \includegraphics[width = 0.98\linewidth]{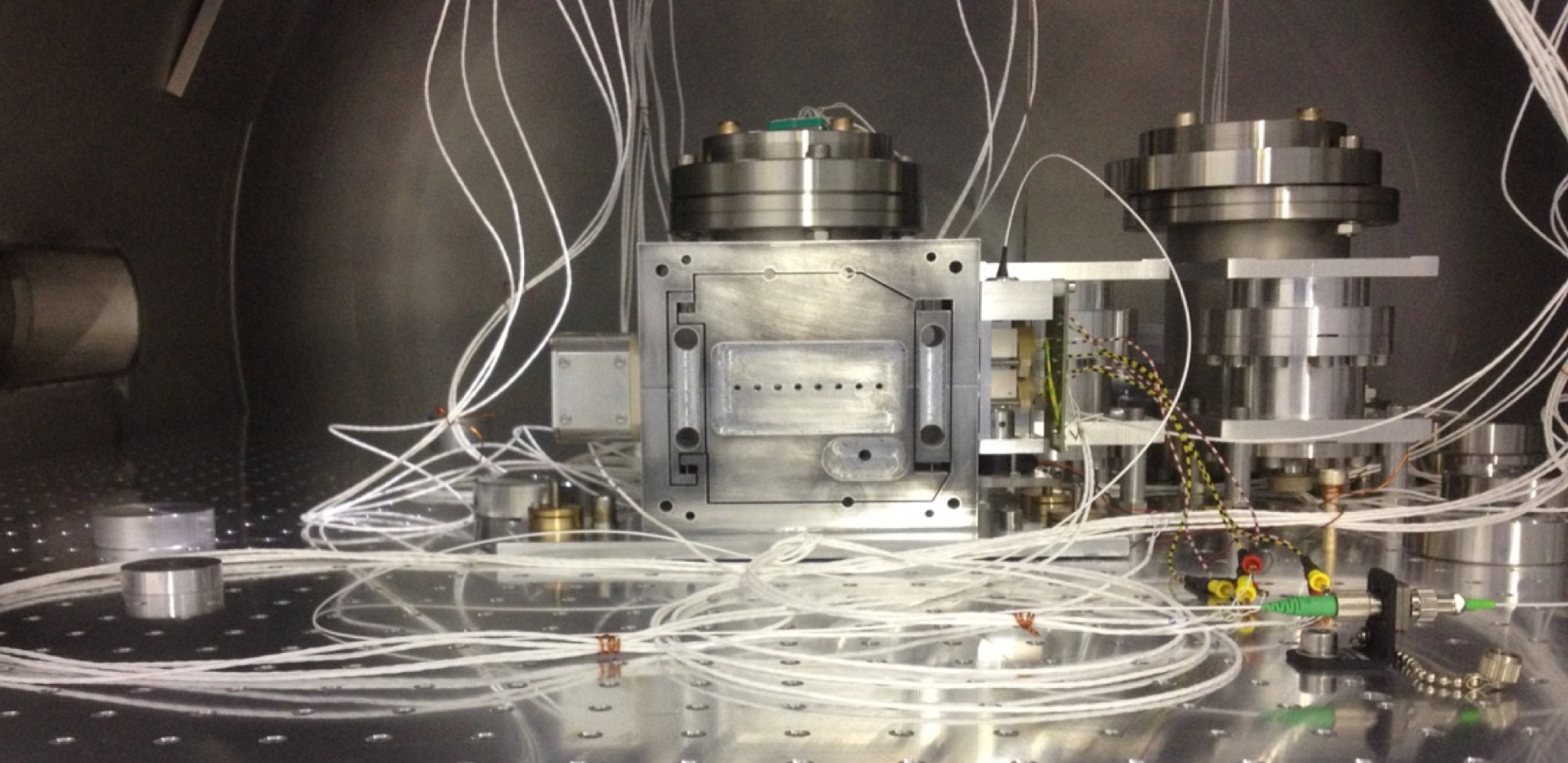}
    \caption{Photograph of the Nikhef room temperature monolithic accelerometer with an interferometric readout. The sensor is operated in force feedback mode using a coil-magnet actuator.}
    \label{fig:RTwlITFreadout}
\end{figure}

ET will extend the detection band down to 2 Hz and, at those frequencies, seismic noise and (mirror suspension) thermal noise are typically dominant. To detect the minuscule motion at the cryogenic penultimate stage, a new class of inertial sensors is needed. Shown in Figure~\ref{fig:RTwlITFreadout} is a monolithic inertial sensor with an interferometric readout which was developed at Nikhef between 2014-2018 \cite{vanHeijningen2018}. It achieved a displacement sensitivity of 8$\times10^{-15}$\;m/$\sqrt{\mathrm{Hz}}$ from 30 Hz onwards and a modelled shot noise limited 2$\times10^{-15}$\;m/$\sqrt{\mathrm{Hz}}$ above 10 Hz. 

\begin{figure}[htb]
    \centering
    \includegraphics[width = 0.98\linewidth]{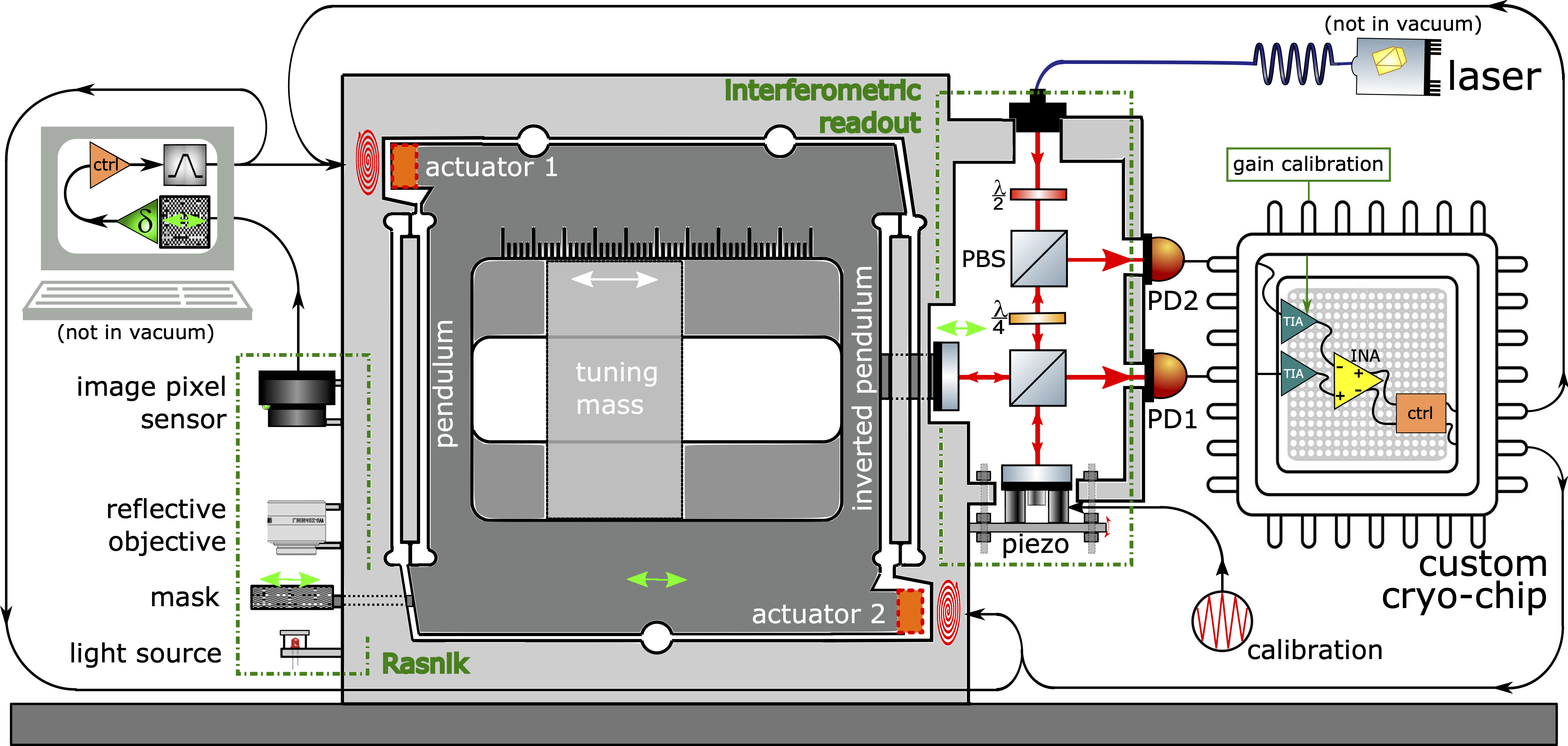}
    \caption{A cryogenic superconducting monolithic inertial sensor. An interferometric readout provides an error signal containing information on the position of the proof mass. The proof mass is inertially suspended from the frame by a regular pendulum and inverted pendulum. This is known as a Watt’s linkage and allows for an arbitrarily low natural frequency, increasing the mechanical sensitivity. The error signal is fed to the actuators to lock the mass with respect to the frame and is used as sensor output. Another loop, using a Rasnik \cite{Graaf2021} as long-range sensor, damps the resonance peak, reducing proof mass motion such that the other loop can work with high precision over a short range. The custom cryo-chip is under development using CMOS technology.}
    \label{fig:CSISoverview}
\end{figure}

A new proposal for a Cryogenic Superconducting Inertial Sensor (CSIS) \cite{Heijningen2020} aims to improve this lower bandwidth limit down to 1 Hz. A schematic overview of the sensor is shown in Figure~\ref{fig:CSISoverview}. The sensor self-noise features a fifty-fold reduction of thermal suspension noise over the previous design by increasing the mechanical $Q$ factor. This is achieved by not using magnets in the actuator. It vastly reduces eddy current damping which was a dominant (viscous) damping mechanism in the previous design.

\begin{figure}[H]
\centering
\includegraphics[width = 0.98\linewidth]{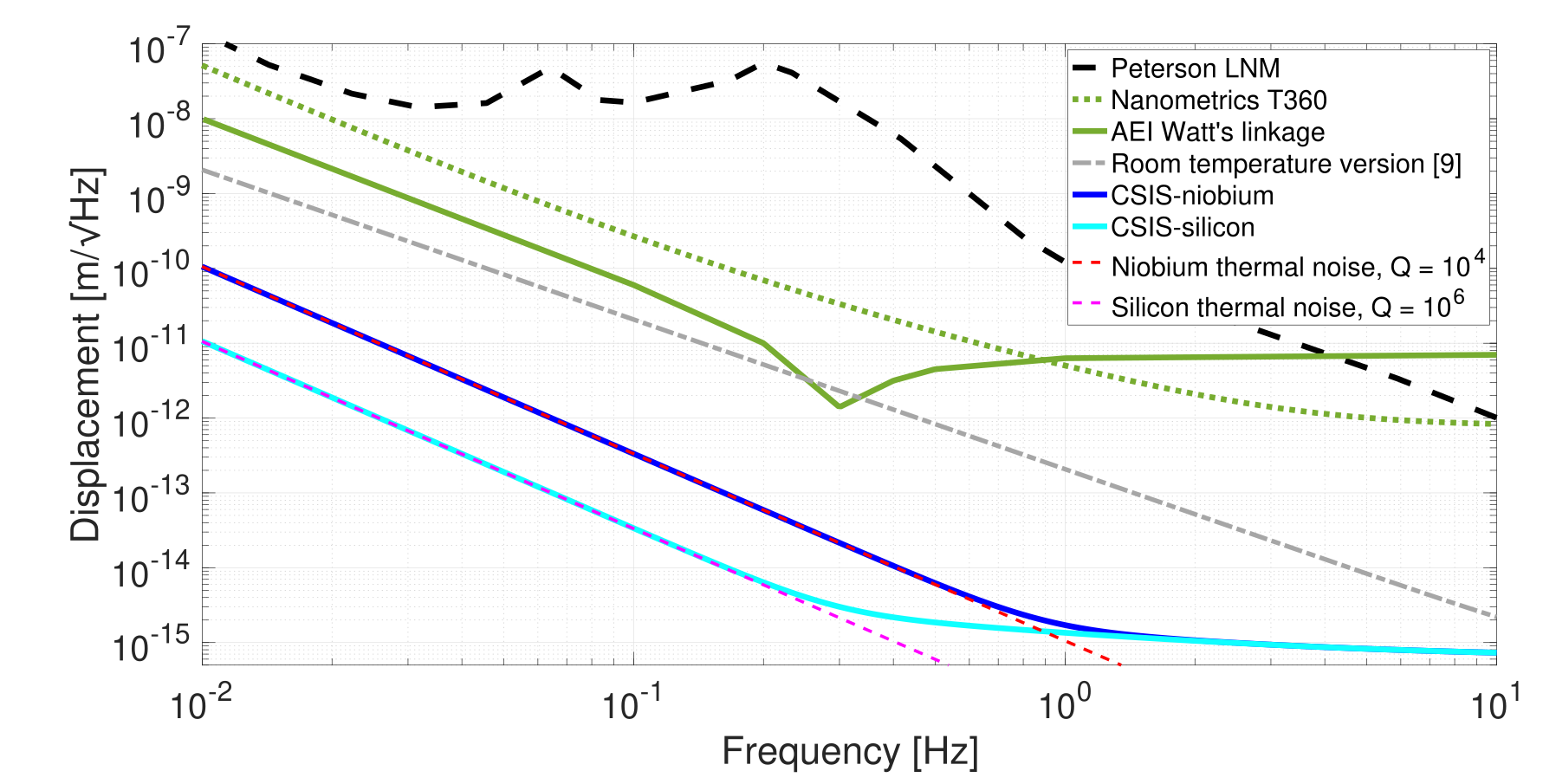}
\caption{Comparison between the Peterson low noise model (LNM), \cite{Pet1993} and minimum detectable inertial displacement for the state-of-the-art (the T360, \cite{T360GSN} and AEI’s accelerometer, \cite{Kirchhoff2017}) and the proposed CSIS. The currently investigated niobium and future plans for silicon ($Q$ $\approx 10^6$) show a thousand-fold improvement around 1 Hz.}
    \label{fig:NioSilCompToSOTA}
\end{figure}

A novel aspect is the thin film spiral coil actuators. The low-voltage superconducting coils are at cryogenic temperatures and exert a magnetic pressure on the proof mass surface by the Meissner effect. Because they are push actuators only, two per inertial sensor are needed as in Figure~\ref{fig:CSISoverview}. The absence of magnets ensures that the suspension is structurally damped and we assume a $Q$ factor of 10$^4$. At low frequencies we are thermal noise limited and with the fm/$\sqrt{\mathrm{Hz}}$ interferometric readout, inertial sensitivities shown in Figure~\ref{fig:NioSilCompToSOTA} are obtained.

 The penultimate mass of the ET mirror suspensions is conductively extracting the heat from the 10\;K mirror and is thus at around 5 K. The temperature necessary for niobium, the sensor mechanics material, to be in a full Meissner state is about 5 K. The penultimate mass is therefore the perfect place to use the sensor for monitoring the effect of the low-vibration cryogenic systems and possibly contribute to the suspension control. 
 
\begin{figure}[ht]
\centering
\includegraphics[width = 0.45\linewidth]{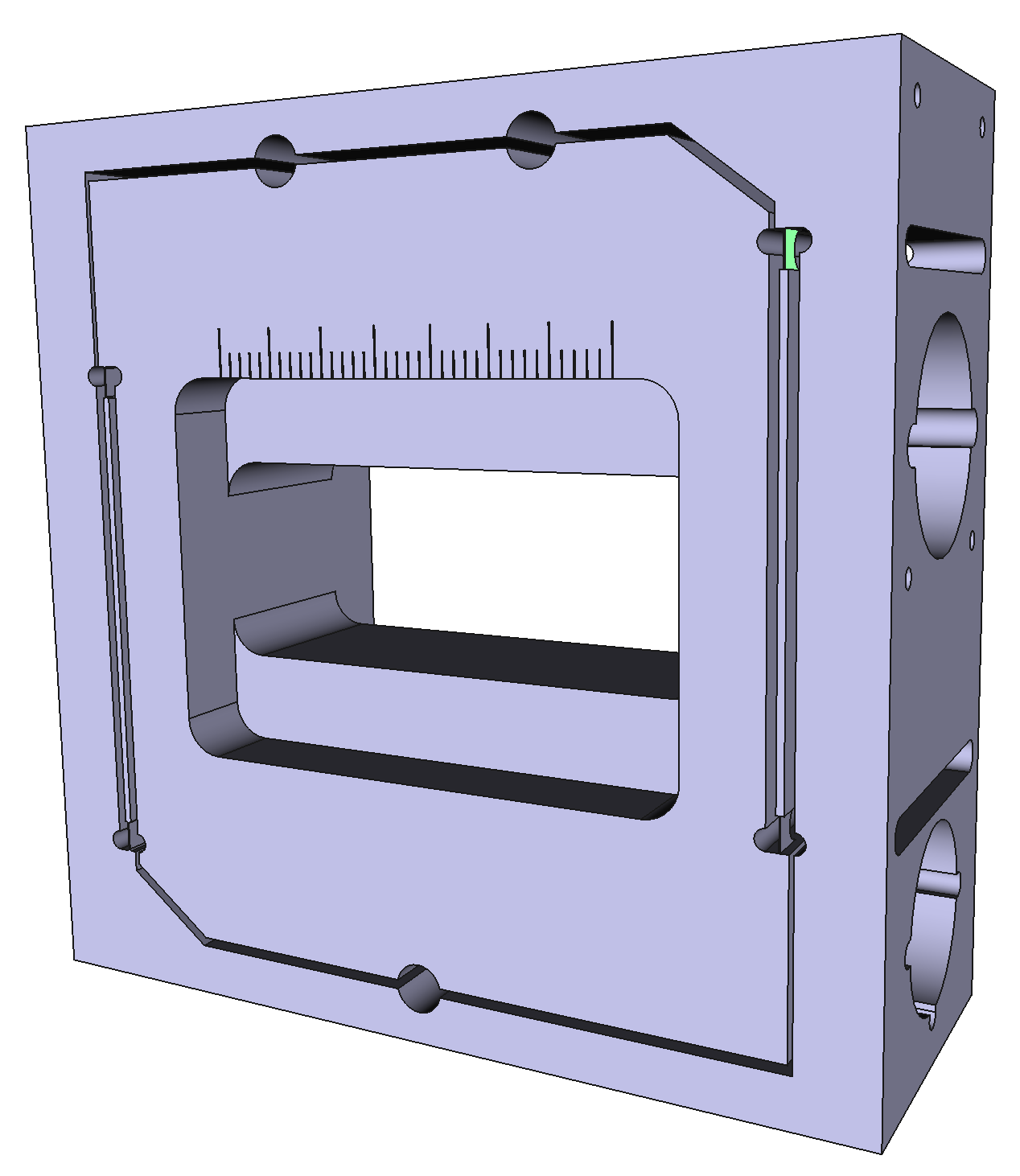}
\includegraphics[width = 0.2525\linewidth]{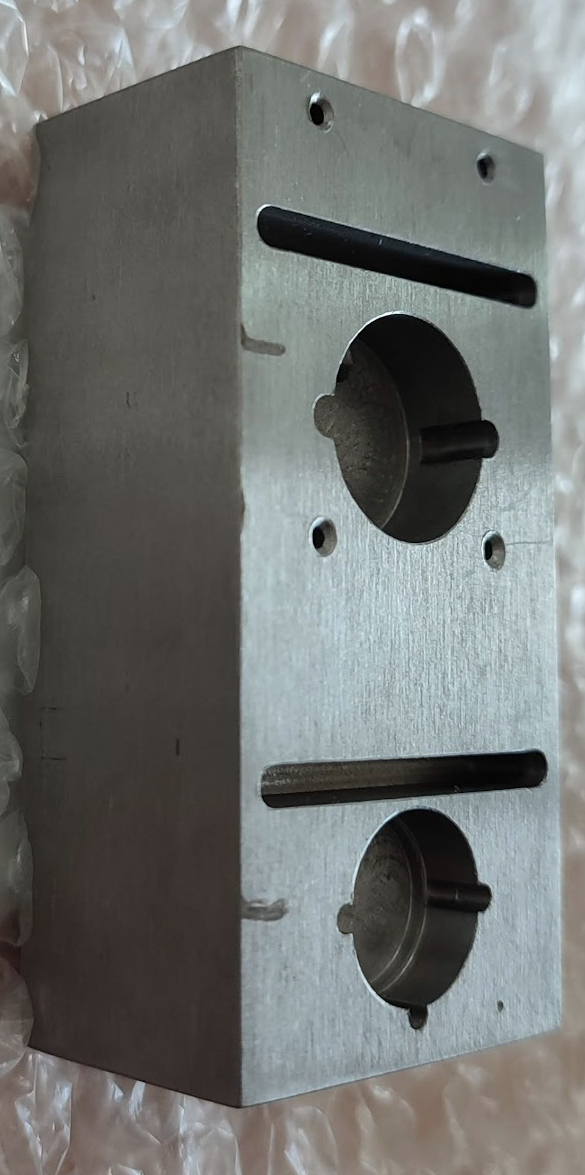}
\includegraphics[width = 0.26\linewidth]{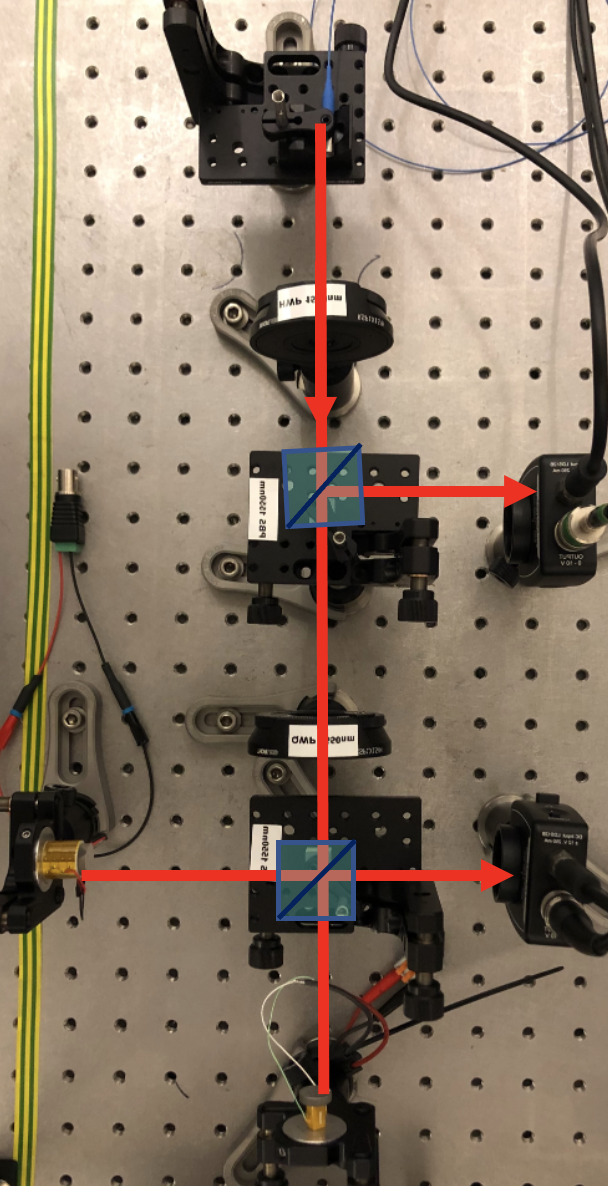}
\caption{Current status of CSIS: (left) CAD render of the monolithic sensor mechanics with one of the eight 100\;$\rm{\mu}$m thick flexures highlighted in green, (middle) a photograph of a niobium test block with the upper readout hole and lower actuator hole and (right) a top view of the interferometric readout.}
    \label{fig:CurrentCSIS}
\end{figure}

The sensor could operate at a temperature higher than 5 K. The cold platform temperature of the E-TEST prototype is certainly higher and the designed ET mirror temperature could be increased. Then we will abandon niobium in the actuator and use high critical temperature ($T_{\mathrm{c}}$) superconductors. Silicon could be used as Watt’s linkage material and high-Tc materials could be used for the coils and surfaces (see orange rectangles in Figure~\ref{fig:CSISoverview}). This will increase the complexity of the mechanics manufacturing, but it will improve sensor performance as shown in the modelled CSIS-silicon curve in Figure~\ref{fig:NioSilCompToSOTA}. Currently, the niobium sensor mechanics are being fabricated and a table-top version of the novel polarized interferometric readout is being characterised, as shown in Figure~\ref{fig:CurrentCSIS}. Progress with the superconducting actuators is summarised in ref.~\cite{ElvisTAUP}, where XRD measurements show the right crystallographic phase for superconducting Nb and NbN coils. For E-TEST, we plan to install 3 CSISs on the cold platform exploiting the low-vibration cryogenic environment to perform a self-noise measurement by huddle test. 

\label{cryo_sens}

\chapterimage{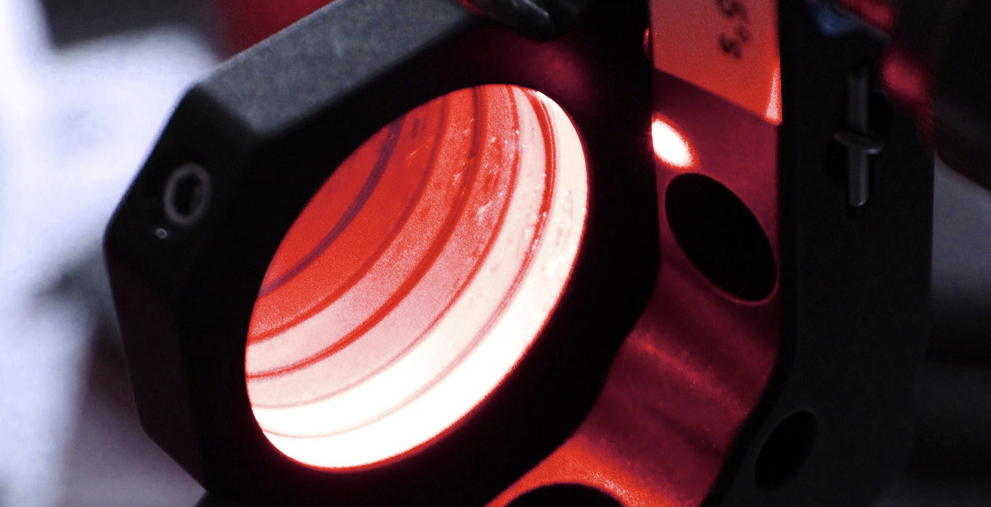} 
\chapter{Optical engineering}
\label{sec:optics}

\section{Overview}
\label{opt_over}
Besides the technology mentioned in the earlier chapters, also the necessary optical technology for ET-LF shall be developed and demonstrated within the E-TEST project. Since the hanging mirror from the suspensions in the cryogenic environment is a silicon mirror, which is opaque at the conventionally used wavelength of 1064 nm, new challenges arise: First of all, a new laser wavelength in the range of app. 1500-2000 nm has to be used, which will not be absorbed within the silicon mirror. Resulting from this, also new photodiodes and coatings have to be developed. Since the mirror shape might change when cooled down to the required 10 K, also a new interferometer to measure this behaviour is in development.  
Within E-TEST, a laser at a wavelength of 2090 nm, consisting of a crystal based Ho:YAG seeder and an ultra-stable holmium-doped fiber amplifier shall be developed. Therefore, e.g. the sensors and the coatings are also optimized to fit to the laser wavelength to enable a high level of synergy. In the following chapters, information on all the key technologies will be given. Afterwards the three big key experiments for the technology validation will be presented, which also covers the collaboration and shared frameworks of the WPT2 experiments.

\section{Laser at 2~$\mathrm{\mu}$m}
\label{sec:laser2um}
To achieve the very well-defined spectral parameters and in addition a high power, a MOPA (Master Oscillator, Power Amplifier) configuration is a well-established and typically used approach. In this approach, a seed laser at low power will be used to define the spectral parameters, e.g. the wavelength and the linewidth. Since such a seed laser typically has low power in the mW range, the radiation has to be amplified in e.g. a fiber-based power amplifier. By separating the power parameters from the spectral parameters, overall better parameters can be achieved, which is the reason why the MOPA concept will be used for the development of the E-TEST laser source. The development of such a system will be performed by Fraunhofer Institute for Laser Technology.
At one of the most common fiber laser wavelengths of 1064 nm, which is also used in most GWD, the lasers are a well-established technology. In ytterbium-doped fiber lasers, output power of up to 10 kW are achievable, solid state laser seed lasers with very well defined spectral properties in the kHz range are commercially available, the stabilization techniques of ultra-high stability lasers are demonstrated, and fiber laser components are well developed and at low prices affordable.
However, at a wavelength of 2090 nm, the development of lasers is not that advanced. Typically, holmium is used as an active dopant in crystals and fibers to enable lasing at a wavelength of 2090 nm. While at 1064 nm, e.g. NPROs (Non-Planar Ring Oscillators) are a commercially available low-bandwidth seed laser, there are no such lasers commercially available at ca. 2000 nm. For amplification in a to-be-used Holmium doped fiber laser even low output powers are a high challenge. The fiber laser components needed for the build-up of a fiber amplifier at 2090 nm are based on well-established technologies, but the unusual wavelength results in expensive, non off-the-shelf components, which have typically worse parameters (e.g. transmission loss or power capability) in comparison to 1064 nm fiber components.
For LF-ET a laser system with the following parameters is needed:

- Wavelength: 2090 nm

- Power: 5-10 W

- Spectral linewidth: 1 kHz

- Power Noise (RIN): < 1e-6 Hz/sqrt(Hz) @ 100 Hz

- Frequency Noise: 100 Hz/sqrt(Hz) @ 100 Hz (assumes 1/f spectrum)

- Linear polarization: > 98%

\begin{figure}[h]
    \centering
    \includegraphics[width = 0.7\linewidth]{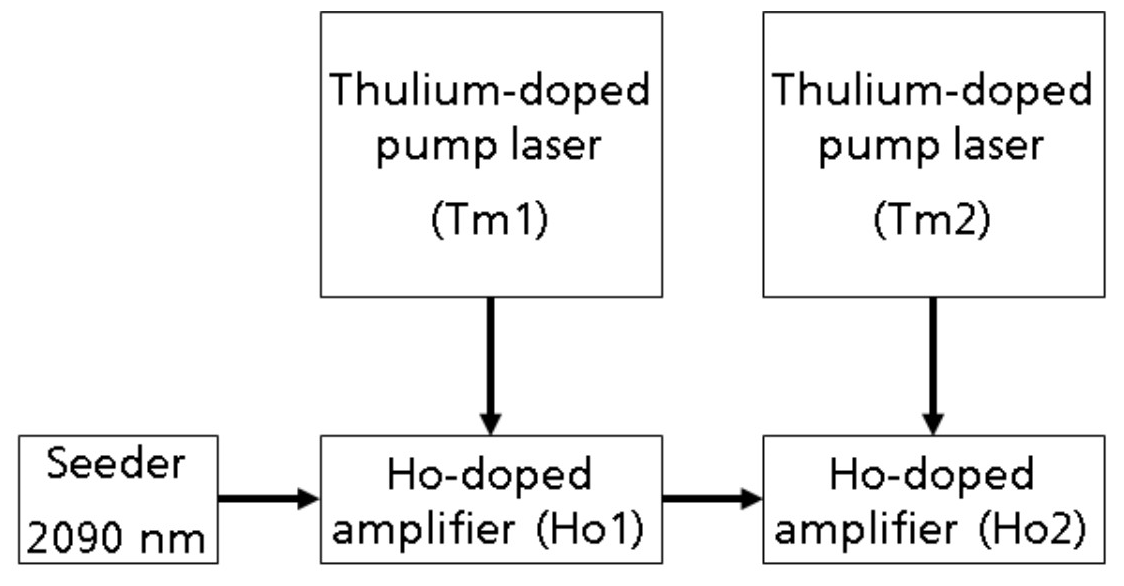}
    \caption{ Model of the laser concept. Within the Ho:YAG seeder, the spectral properties are defined. This will result in a low bandwidth and a wavelength of 2090 nm. Within the Ho-doped amplifier 1 $\&$  2 the power will be amplified}
    \label{fig:LaserConceptx}
\end{figure}

For such a system, a typical approach is the development of a MOPA system as described above. Therefore, it is planned to develop a solid-state laser based on Ho:YAG crystals as a seed laser, which defines the spectral properties. For the power amplification, a two-stage fiber amplifier based on holmium-doped fiber shall be developed. Even though the power noise is only defined at a single frequency, it is planned to measure it in different frequency bands to gather more information on the laser performance.
\ref{fig:LaserConceptx} shows the scheme of the full laser concept. In addition to the mentioned seed laser (Seeder) and the two amplifier stages (Ho1 $\&$ Ho2), pump lasers for Ho1 $\&$ Ho2 have to be used and developed (Tm1 $\&$ Tm2). These pump lasers are additional thulium-doped fiber lasers at a wavelength of 1950 nm. For the pumping of the Ho:YAG seeder, an ILT developed Tm:YLF laser will be used. Here, it will be tested whether the stability of the pump laser is high enough to achieve stable lasing of the seeder, or if another highly stable fiber based pump source has to be developed. Within the development, different concepts are tested and compared to find the most practical and satisfying concept.

\begin{figure}[h]
    \centering
    \includegraphics[width = 0.3\linewidth]{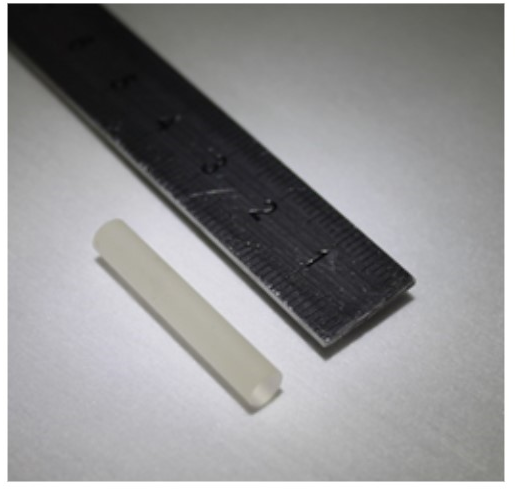}
    \caption{Ho:YAG crystals to be used for the ring oscillator}
    \label{fig:YAGCrystalx}
\end{figure}

While the goal is to develop an NPRO seed laser, the development is divided in two phases. Since the NPRO concept at 1064 nm shows satisfying results with respect to stability and linewidth, it is very hard to manufacture such a laser crystal and a lot of information on the crystal material itself are needed. Therefore, in the first phase, a ring oscillator will be developed. Such a concept has a significantly higher success probability and a lot of information on Ho:YAG and the power amplification within a fiber resonator can be gathered. For illustration, \ref{fig:YAGCrystalx}  shows the to-be-used Ho:YAG crystals for the ring oscillator. After the demonstration of the ring oscillator, an NPRO concept shall be tested. To enable active controlling of the laser parameters, it is planned to implement actuators to control these.

Since the seed laser is not available in the early stages of the project, a diode laser will be used as a seed source in the early stages of the fiber laser development, to prevent delay. With this diode laser, the power amplification can be demonstrated. Since the required relative intensity noise is very ambitious, the output power of the fiber amplifier will be actively controlled. This is done by adjusting the output power of the pump lasers. Based on the chosen concept, 4-7 individual fiber lasers are developed, which are all actively controlled by control algorithms. 

\begin{figure}[h]
    \centering
    \includegraphics[width = 0.5\linewidth]{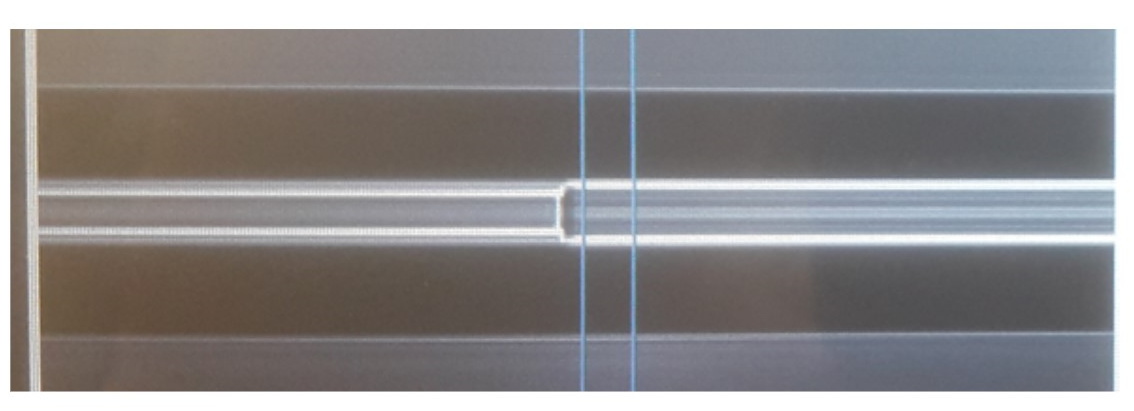}
    \caption{Ho:YAG crystals to be used for the ring oscillator}
    \label{fig:YAGCrystalxx}
\end{figure}

While a very useful characteristic of a fiber laser is the all-in-fiber concept, which results in e.g. no de-adjustment over time in comparison to other laser concepts, a process that needs to be perfected is the splicing of fibers. This way, two different (or same) types of fibers can be connected to enable the guidance of light within the fiber. A good splice is very important e.g. for a high efficiency and low noise laser. Especially for Thulium-doped fibers, the current state-of-the-art fibers are not as easy to splice as e.g. ytterbium-doped fibers, which results in higher losses and lower damage thresholds if the splice parameters are not optimized. Therefore, it is planned to perform studies to optimize the splices.
 \ref{fig:YAGCrystalxx}  shows an exemplary splice for illustration.

\section{Low noise coatings}

Many crystalline materials show lower optical absorption and mechanical loss than amorphous materials. However, the low and high refractive-index materials have to be lattice matched, which limits the number of possible materials significantly. Crystalline coatings made of AlGaAs and GaAs are under investigation and show promising performance. The requirement of a lattice matched substrate -- GaAs wafers in case of AlGaAs coatings -- for the growth procedure requires a transfer of the coating to a suitable substrate (e.g. SiO$_2$ or cSi) after production and also can limit the maximum coating size (to $\approx$ 20\,cm in case of GaAs). Defects in the bond for attaching the coating to the new substrate can lead to the coating detaching during the test-mass cooling procedure or to thermal noise from a change in bulk and shear loss. Alternative materials may solve some of these problems (e.g. AlGaP and GaP coatings can directly be grown on cSi). 

Part of the approved E-Test budget proposal is a contribution to the capital funding for the construction of a novel MBE coating facility at KU Leuven to investigate new crystalline coatings for application in future GW detectors. In the following sections a detailed motivation and description of this activity is given.  

Currently, the main methods to create state of the art oxide mirror coatings are based on Sputtering and Ion Beam (Assisted) Deposition (IBAD) techniques. There is already quite a large group of laboratories worldwide that focuses their efforts on using these techniques for oxide mirror coatings.

\subsection{Molecular Beam Epitaxy}

The Oxide Molecular Beam Epitaxy (MBE) technique is an interesting addition to the portfolio of techniques for mirror coatings and presents several unique and complimentary advantages. This technique has matured well over the past 30 years and the participants in the E-Test project have build 3 generations of small or large area oxide MBE systems. 

\underline{Overall principle}

\begin{figure}[h]
    \centering
    \includegraphics[height=8cm]{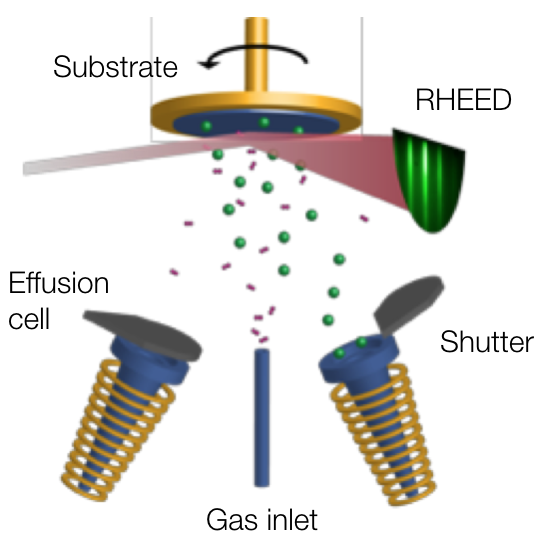}
    \caption{Schematic drawing of the main elements in a standard molecular beam epitaxy system.}
    \label{fig:MBE principle}
\end{figure}

The overall principle of MBE is illustrated in Figure~\ref{fig:MBE principle}. In short, a rotating and heated substrate is bombarded with a beam of atomic of molecular (precursors) species emanating from for instance an effusion cell. An effusion cell is essentially a crucible in which the material to be evaporated is inserted combined with a heating element wrapped around it and a shutter that opens and closes the path to the substrate. The deposited species on the substrate then react with oxygen species (atomic oxygen, molecular oxygen or ozone) emanation for instance from a gas inlet, to form on oxide film. The structure of the growing film is then monitored in real-time using an electron diffraction pattern created on a screen with a technique called Reflection High Energy Electron Diffraction (RHEED). 

Compared to other techniques, the main advantages of oxide MBE include: 

\begin{itemize}
    \item A low background (10$^{-11}$ Torr) and operational pressure (10$^{-8}$-10$^{-5}$ Torr). This enables an excellent control on (gaseous) contamination as well as oxygen content in the layers. For instance it allows to grow suboxides, i.e. oxides that do not have the highest oxidation degree; but many of which have a higher refractive index. In addition, it permits a good control on the oxidation modulation in a coating. A good example would be the case of combining aSi with an oxide as mentioned above.  Nevertheless if a higher oxidation state is required then atomic oxygen or ozone gases are added.
    \item A low energy deposition process in the sense that energetic species - as is typical in sputtering in IBAD - are not present during growth. Indeed MBE uses elemental thermal evaporation from sources operating at a maximum of 3000C or about 0.75 meV. Typically higher energetic growth processes are more prone to the appearance of 3D features and particles on the growing surface. In addition, diffusion reactions are also enhanced which can lead to a poorer control on interface sharpness and roughness.   
    \item An excellent uniformity control at or below 1 $\%$. The main sources - effusion cells and electron beam guns - are essentially point sources. This means that a simple geometric relation exist between the source and the evaporation profile. Hence, to improve uniformity over large areas  requires only to change the geometrical position of the source vis-a-vis the substrate. Interestingly, it does not need the uses of larger targets as might be required in large area IBAD or sputtering processes. 
    \item A directed "beam" deposition. The low pressure operational conditions ensure that the evaporating beam has molecular beam properties. This means that atoms evaporating from the hot source surface are not scattered on their path to the substrate. This has several advantages. First, it means that the beam can be directed only towards the substrate and that no atoms hit - and contaminate - the walls of the chamber. Accumulation of such "debris" deposits  on chamber walls can be a main source of subsequent particle contamination in the coatings which then necessitates frequent  cleaning of the chamber. Second, it enable to easily shape the coatings in a well defined manner, through the use of shutters and the substrate rotation. For instance composition-graded or thickness-graded profiles can be grown while conserving rotational symmetry. As an example, a rotational symmetric "curved" coating can be made whereby more substrate material is deposited at the edges. 
    \item In-situ thin film characterisation tools. Because of the low operational pressure several unique tools can be used to monitor the deposition process such as the beam fluxes and the growing film surface. The beam flux techniques include Quadruple Mass Spectrometry (QMS), Electron Impact Emission Spectroscopy (EIES) and and ionisation gauges. With these the composition of the evaporants to the substrate can be controlled with high precision ($\approx 1\%$). The film structure (roughness, crystallinity) can be measured in real-time using Reflection High Energy Electron Diffraction (RHEED) while the film composition can be measured using the x-rays emitted from the RHEED gun with a technique called Low Angle x-ray Spectroscopy (LAXS).  The RHEED technique becomes very important if the goal is to grow crystalline oxides. \\       
\end{itemize} 

The unique advantages of MBE apply to all types of coatings that can be made whether metals, semiconductors, oxides, nitrides, etc. Furthermore, they also apply to both amorphous as well as crystalline oxides. Nevertheless given the availability of in-situ characterisation tools, the biggest impact that can be made in this field with a new coating tool is related to exploring crystalline oxides.  

\underline{Crystalline coatings}

The motivation to explore crystalline oxides is similar to that used by the Cole et al.~\cite{Cole2016}, namely that the thermal noise of single crystals is currently still several orders of magnitude better than that of amorphous coatings. While single crystals can have very low mechanical loss angles of the order of $2 \times 10^{-9}$ (Al$_2$O$_3$) or $5\times 10^{-9}$ (SiO$_2$), the values for their amorphous coating counterparts is at best around $5 \times 10^{-5}$ for state of the art materials. Furthermore, there are a number of interesting features of oxide single crystals that may offer unique advantages in specific situations such as: \\

\begin{itemize}
    \item Explore tensor properties. The unique features of crystals are often tensor properties in contrast to those of amorphous materials. While the latter can typically be described by values averaged over the 3 crystal axes, this is often not the case for the crystalline materials.  This can be a blessing, for instance when the anisotropy of the thin film properties can be used to compensate those of the substrate. The relevant properties here are the elastic, thermo-optic and thermal expansion coefficients.  
    \item Tailoring thermal expansion. While most materials have positive thermal expansion coefficients along the three crystallographic axes, there are quite a number of materials that have anisotropic coefficients whereby the thermal expansion coefficient is negative (NTE) along one direction.  To date the highest NTE reported is $-1.3 \times 10^{-4}$/K which is about 200 times larger - in absolute value - than the PTE of fused quartz. That means that in principle such a 500 $\mu$m thick NTE coating may completely compensate the PTE of a 10 cm  thick fused quartz mirror. 
    \item Controlling the uniformity of elastic properties. While the origin of the mechanical losses in coatings are not fully understood or identified, it is reasonable to estimate that non-uniformity in structure and composition are some of the main culprits. Both have an immediate effect on the local elastic properties. Although on a macroscopic scale the elastic properties of thin films  are usually  considered uniform they are  most definitely not uniform on a microscopic /  atomic scale. Such local non-uniformity are for instance structural defects like dislocations, anti-phase grain boundaries, vacancies, interstitials, variations in local composition, density variations, etc. Hereby not only elasticity in terms of its static components - with properties such as Young's modulus, Poisson's ratio, etc.  - must be considered but also its dynamic components like the  velocities at which elastic waves propagate in solid media.
    \item Large family of interesting compounds. The amount of interesting oxide binary, ternary, quarternary, etc. materials is quite large compared to the crystalline AlGaAs and AlGaP compounds. Two families stand out namely the corundum type (R$_2$O$_3$) as well as the perovskite (ABO$_3$) type. The hexagonal corundum compounds include for instance Al$_2$O$_3$, Cr$_2$O$_3$, Fe$_2$O$_3$, Ti$_2$O$_3$, V$_2$O$_3$, Ga$_2$O$_3$ and In$_2$O$_3$. Within these materials it is also easily possible to create intermediary alloys, such as  (Cr$_x$Fe$_{1-x}$)$_2$O$_3$ or (Ga$_x$In$_{1-x}$)$_2$O$_3$. Here, the in-plane lattice parameters spans a broad range between 0.476 nm - 0.549 nm, while the refractive index (at 1064 nm) varies between 1.75 to 2.80. Epitaxial films of these materials can be  grown on sapphire or on Si(111) substrates. The perovskite family of compounds - like SrTiO$_3$ - is even much larger and can be well grown on Si(100) substrates. Note that there is also quite some structural variability within the perovskites that can be either cubic, tetragonal or orthorhombic. 
    \item Use of strain engineering. Thin crystal films can be quite easily deformed in a uniform and homogeneous manner through the use of epitaxial strain. This means that the lattice parameters of a substrate - or of a thick buffer layer - can be imposed on the subsequent films. In such a case, the film lattice parameters - and all its other properties -  can be tailored within a well defined range until optimal features are obtained. 
\end{itemize}

\subsection{Infrastructure components} 

Based on the above challenges, there are quite a few elements that need to be considered when defining the appropriate tooling to grow the thin film coatings. Both the individual elements as well as their combination will contain a number of technological innovations necessary to produce the high quality films needed for this application. 

\underline{Sources}

\begin{figure}[h]
    \centering
    \includegraphics[height=6cm]{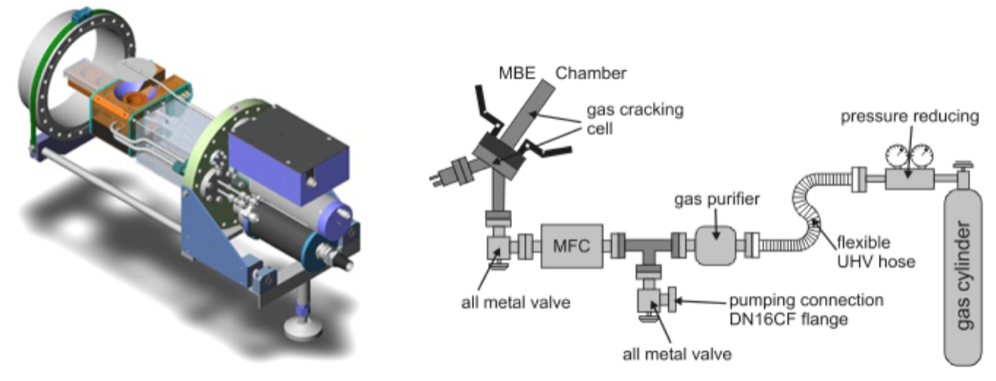}
    \caption{Schematic drawing of an electron beam evaporator (left) and a gas injection manifold (right).}
    \label{fig:Ebeam and gas manifold}
\end{figure}

Typically each element to be deposited requires its own source. Such a source can be an effusion cell - as illustrated above - an electron beam evaporator or a gas inlet system for gaseous (molecular/chemical) precursors. Each of these sources can come in different flavours. For instance under the umbrella of effusion cells, we have high temperature cells, low temperature cells, dual filament cells, cold-lip cells, valved crackers, etc. 
Among the different constructions of electron beam evaporators we have single crucible, multiple crucible as well as vertical evaporators. Depending on the amount of heat these sources produce, they are contained with a water cooled enclosure. A schematic drawing of an electron beam evaporator is shown in Figure~\ref{fig:Ebeam and gas manifold} (left). 

\begin{figure}[h]
    \centering
    \includegraphics[height=6cm]{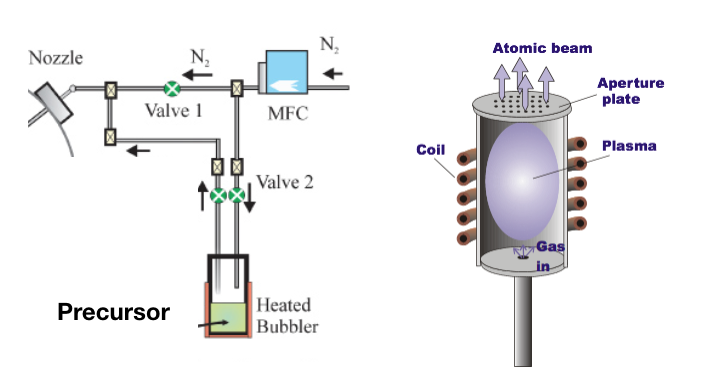}
    \caption{Schematic drawing of a precursor bubbler system (left) and an atomic beam source (right).}
    \label{fig:Bubbler and atomic}
\end{figure}

Also for the gaseous species injection there are a range of possible designs and options. One such gas manifold is illustrated in Figure~\ref{fig:Ebeam and gas manifold} (right). Depending on the type of gas to be used, instead of a gas cylinder a gas bubbler may be used and the gas lines might required to be heated. The bubblers are typically used when the gaseous precursors have a low vapor pressure. In that case the mode of operation is to use a (inert) carrier gas - such as nitrogen or argon - that is flown through the heated bubbler that contains the precursor. The overall principle of a bubbler system - with the Mass Flow Fontroller (MFC) that controls the gas flow rate - is illustrated in Figure~\ref{fig:Bubbler and atomic} (left). Note that in a typical system, several gas lines and manifolds will be put together in a rather complex system. 

Another important element is a source of reactive and/or highly energetic species. This can be performed by a range of different sources, mostly involving the creation of plasma's and/or high energy ion beams. One such radio frequency (RF) plasma source is depicted in Figure~\ref{fig:Bubbler and atomic} (right) where an oxygen plasma is generated inside a crucible leading to the generation of a relatively high amount of atomic oxygen (up to 30$\%$)  which then flows out through an aperture on top. Other implementations make use of a electron cyclotron resonance (ECR) to create a plasma. Furthermore when grids are added on top of these sources, then the application of a voltage allows to extract electrons or ions from the plasma leading to an ion source. 

\underline{Vacuum and pressure control}

\begin{figure}[h]
    \centering
    \includegraphics[height=5cm]{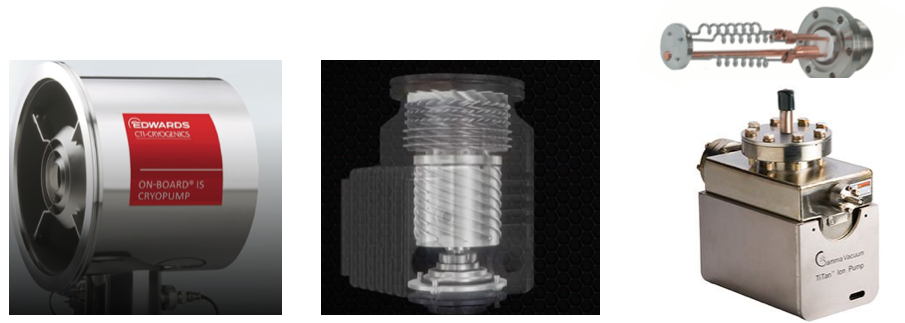}
    \caption{Pictures of a cryopump (left), a turbomolecular pump (middle) and an ion pump with titanium sublimation filaments (right).}
    \label{fig:Pumps}
\end{figure}

The base pressure inside an MBE system is typically of the order of 10$^{-11}$ Torr. This requires a range of pumping systems including cryopumps, turbomolecular pumps and their forepumps, ion pumps equipped with titanium sublimation filaments as well as liquid nitrogen cooled cryopanels. In standby operation, mostly the ion pumps with titanium sublimation and the cryopumps are active. However during film growth where the temperature inside the chamber increases and a gaseous oxidizing flow is necessary then the turbomolecular pumps are active while the others are valved off. Schematic pictures of the different systems are shown in Figure~\ref{fig:Pumps}.

\begin{figure}[h]
    \centering
    \includegraphics[height=5cm]{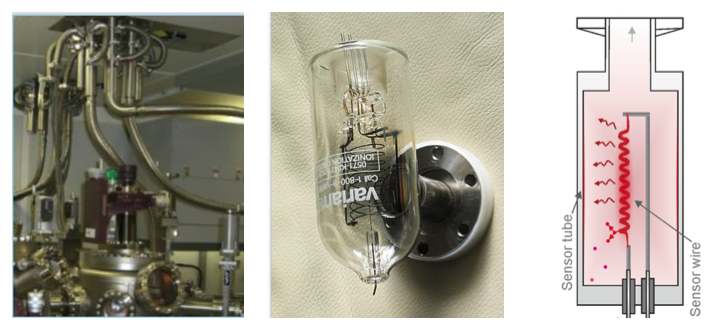}
    \caption{Pictures of a liquid nitrogen feed lines coming from a phase separator  (left), an ion gauge (middle) and a Pirani gauge filaments (right).}
    \label{fig:Liquid nitrogen pressure gauges}
\end{figure}

The liquid nitrogen that feeds the cryopanels of the system require a specific set of double walled vacuum tubings that transport the liquid nitrogen and a phase separator. The latter  recovers part of the liquid and separates it from the hotter gas that comes back after circulating through the panels. This concept is illustrated in Figure~\ref{fig:Liquid nitrogen pressure gauges} (left). 

The pressure in the different vacuum chambers is measured using a variety of instruments including ion gauges, Pirani gauges, baratrons, and residual gas analysers. The latter is used to monitor the composition of the base pressure as well as to detect small vacuum leaks and contaminations. The ion gauge and Pirani gauges are shown in Figure~\ref{fig:Liquid nitrogen pressure gauges} (middle and right). 

\underline{Flux control}

\begin{figure}[h]
    \centering
    \includegraphics[height=6cm]{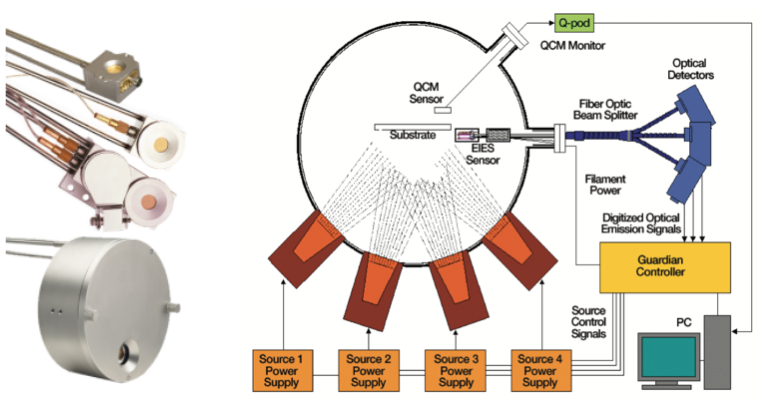}
    \caption{Examples of quartz crystal monitors (left) and an EIES system (right).}
    \label{fig:Quartz and EIES}
\end{figure}

In an MBE system, thin films of compound materials are formed by the combination of evaporants from different sources on the substrate. Hence it is quite essential to control the evaporation rate of each of the evaporants precisely. This is done by a combination of several techniques each with their advantages and disadvantages. 

Quartz crystal monitors use the measurement of a resonant frequency of a quartz crystal that shifts when material is added to its surface. Such monitors are used in many deposition processes and provide a good estimation of the flux rate. However since the conditions of deposition on the quartz surface are not the same as on the substrate - difference in temperature, sticking coefficient, oxidation coefficient etc. - it is not always the best technique. Furthermore its sensitivity is a strong function of temperature which can induce additional errors. Some examples of quartz crystal monitors are shown in Figure~\ref{fig:Quartz and EIES} (left).

Hence in most MBE system addition flux monitors are added for instance based on optical spectroscopy. In the case of Electron Impact Emission Spectroscopy (EIES), the flux of atoms or molecules is ionised under electron bombardment which results into the emission of light. In this case, different elements will give rise to a different emission spectrum. Using a combination of fiber optic beam splitters and optical detectors, the spectrum can be collected and used to control the deposition process as illustrated in Figure~\ref{fig:Quartz and EIES} (right). 

\begin{figure}[h]
    \centering
    \includegraphics[height=6cm]{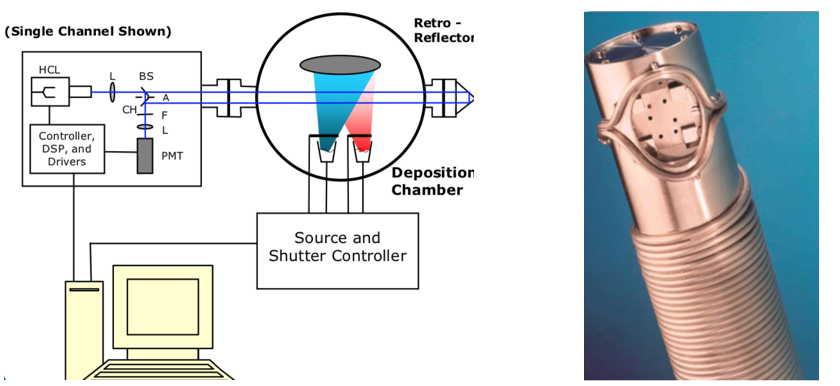}
    \caption{An implementation of AAS (left) and a picture of a cross-beam ion source as part of a QMS  system (right).}
    \label{fig:AAS and XBS}
\end{figure}

Other optical techniques are based on laser fluorescence or on Atomic Absorption Spectroscopy (AAS). Both of these methods have the advantage that no equipment - besides optical windows - needs to be added inside the deposition chamber since only interactions with light are important. An example of an implementation using AAs is shown in Figure~\ref{fig:AAS and XBS} (left). 

Although optical methods provide quite a flexible implementation, they are not the most sensitive instruments and for instance in the case of growing films with small doping levels, that sensitivity is not sufficient. An additional method is then to use Quadrupole Mass Spectrometry (QMS) systems equipped with specialized ion sources.  The ion sources have a cross-beam design to ensure that no evaporants are deposited on the inside of the source. Using special pulse counting techniques, such systems are the most sensitive tools for this application. 

\underline{In-situ thin film / crystal formation control} 

While the flux controls mentioned above enable to control the impinging amount of material on the substrate, this does not provide a characterisation of the growing films. Many factors play a role in determining the crystal phases, orientations, roughness, etc. These are a function of the temperature, oxidation coefficients, sticking coefficients, re-evaporation rates, etc. Hence it becomes important to use methods to monitor the growing material on top of the substrate.

There are multiple methods that are being used for this that are based on electron, photon and/or x-ray beams, impinging upon or emitted from the growing films. One technique using electron diffraction (RHEED) is already mentioned and illustrated above in Figure~\ref{fig:MBE principle}. This technique enables to control the crystal phases that are formed but does not give direct information about the composition of the films. 

\begin{figure}[h]
    \centering
    \includegraphics[height=4.8cm]{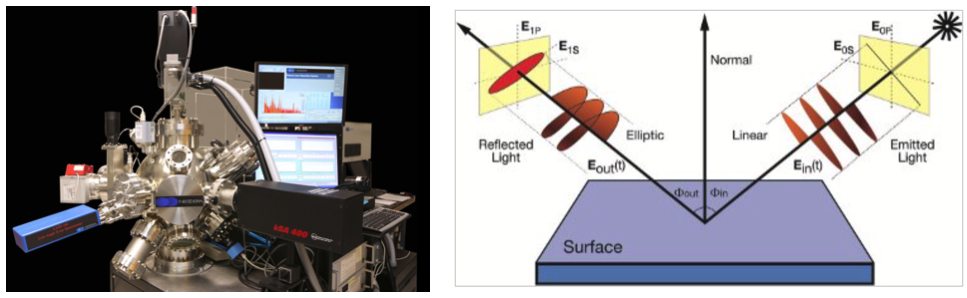}
    \caption{An implementation of LAXS (left) and a schematic picture of an ellipsometry measurement configuration. (right).}
    \label{fig:LAXS and SE}
\end{figure}

Another technique that uses the same electron source to bombard the surface and to collect the emitted element specific x-rays is LAXS. This technique is specifically designed to measure the surface chemical composition and to detect in-real time deviations from the overall stochiometry and is illustrated in Figure~\ref{fig:LAXS and SE} (left).  

The above techniques monitor what happens at or near the surface of the growing film. However during deposition process - certainly at high temperatures - a number of processes can take place below the growing surface such as reactions with the substrate or among the different layers or further oxidation processes, etc. To monitor those, one uses optical methods that have a much higher penetration depth than the electron beams from the preceding paragraph. 

One such method is Spectroscopic Ellipsometry (SE), a technique that measures the light polarization rotation as a function of the interaction with a thin film. The principle is illustrated in Figure~\ref{fig:LAXS and SE} (right) and this method allows to extract and monitor the refractive index and extinction coefficient as a function of the thin film thickness. Both of these quantities are essential parameters in the design of mirror coatings and can thus be accessed in real-time during the growth. 

\underline{Thin film temperature and strain control}

In most thin film deposition system, the substrate and thin film temperature is induced by light radiation for instance from a a hot filament or from a laser source. To measure the substrate temperature itself is not as straightforward since in most cases the substrate rotates which makes the use of thermocouples or resistive temperature sensors impossible. The most common method is to make use of light emission or pyrometry. This requires knowledge of the spectral emissivity of the substrate which is known or can be calibrated. One of the issues is that the filament radiation - as a function of wavelength - is partially absorbed and thus can reach the pyrometer and can interfere with the temperature measurement. 

\begin{figure}[h]
    \centering
    \includegraphics[height=4.3cm]{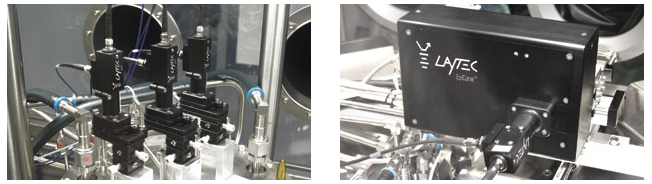}
    \caption{An implementation of a substrate/thin film temperature pyrometer (left) and a substrate bow measurement device (right).}
    \label{fig:Temperature and bow}
\end{figure}

As thin films grow, part of the radiation is now also absorbed in the film and the spectral emissivity becomes dominated by the thin film itself instead of the substrate. One method to then monitor the substrate temperature is  to measure the band gap of the film, which is also a function of the temperature. In principle this can be done by SE but there are also dedicated tools for this on the market such as the EpiTT from Laytec as depicted in Figure~\ref{fig:Temperature and bow} (left).  

Another critical parameter - certainly for thick films - is that of strain control. A large lattice and thermal expansion coefficient mismatch between film and substrate often leads to films which show cracks and a high density of other defects. To minimize the amount of strain being build up during growth and cooling it is important to be able to measure this in-situ. This is typically done by measuring how the entire wafer (substrate + film) curves up or downwards or in other words to determine the amount of bow. Such an instrument is shown in Figure~\ref{fig:Temperature and bow}. The amount of strain can then be tuned partially for instance through changes in composition. 

\underline{Beam of energetic particles}

An alternative method to control the strain and the overall density of thin films is to make use of ion bombardment. For instance in plasma assisted deposition, the properties of the plasma can be used to control the amount and even the sign of strain in the thin films. A part of this is related to the energy transfer from the energetic beam to the film and another part can be due to the implantation of extra species (oxygen, nitrogen, argon, etc.).

\begin{figure}[h]
    \centering
    \includegraphics[height=6cm]{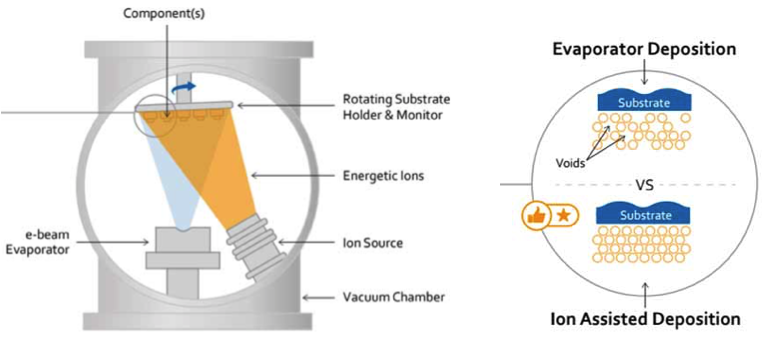}
    \caption{An implementation of an ion beam inside an evaporator (left) and the densification process that it induces (right).}
    \label{fig:Ion Beam}
\end{figure}

A further advantage of using energetic species is that they can improve the density of thin films (amorphous as well as crystalline). This  principle is illustrated in Figure~\ref{fig:Ion Beam} and is one of the main features of the Ion Beam Assisted Deposition (IBAD) technique, which is the standard technique used for GW mirror coatings.

\underline{In-situ annealing}

Crystal growth of thin films in general requires higher temperatures than their amorphous counterparts. This high temperature step can then either be applied during the growth or after the growth is finished. To minimize diffusion and reaction process either among the different layers or with the substrate, the goal is to expose the layers to a high temperature set for an as short a time as possible. Such approaches aiming to minimize the thermal budget to which the films are exposed are commonly called Rapid Thermal Annealing (RTA). 

There are a number of different annealing implementations available on the market. One approach uses for instance high intensity pulsed lasers and another uses flash lamps. This variety is necessary to control precisely the annealing duration as well as the photon wavelength that is used. Depending on the band gap and the optical properties of the materials to be annealed light of different wavelengths will be absorbed differently. 

\begin{figure}[h]
    \centering
    \includegraphics[height=6cm]{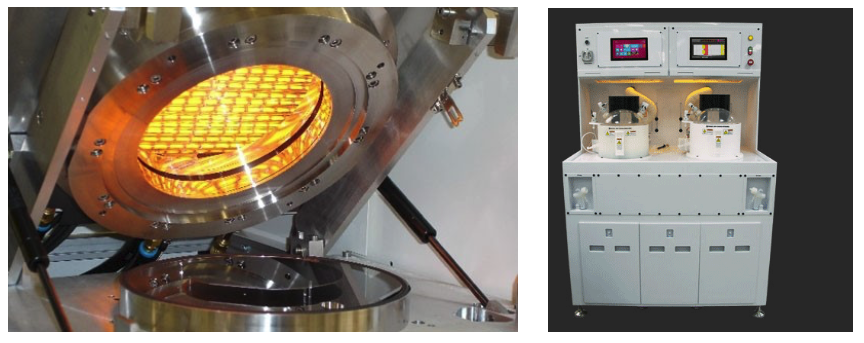}
    \caption{A rapid thermal annealing chamber with heating lamps (left) and a wet processing unit to clean wafers (right).}
    \label{fig:RTA and cleaning}
\end{figure}

The RTA process could in principle be done in the deposition chamber. However, the specifications of the heater and chamber environments for RTA are quite different from those of a typical substrate heater specially since a much higher amount of thermal power needs to be supplied and evacuated. Also the gaseous environment in which the annealing is performed plays  a crucial role since it may need to be performed in an oxygen, nitrogen, or an inert ambient pressure for which a deposition system is not designed. As a consequence in most systems, RTA takes place in a separate and optimized chamber. Such a chamber equipped with heating lamps is shown in Figure~\ref{fig:RTA and cleaning} (left). 

\underline{In-situ substrate cleaning}

Before the thin film growth process starts it is important to clean the silicon substrates in order to remove an amorphous silicon oxide surface layer, but also to eliminate organic residues, particles, etc. 
This is typically done using a series of wet chemistry processes such as dipping in cleaning solvents as well as in a buffered HF solution for a short time. After cleaning the wafer needs to be transferred to the vacuum system as soon as possible to avoid re-oxidation and contamination for instance due to the adsorption of water vapour. The Figure~\ref{fig:RTA and cleaning} (right) shows a wet chemistry processor that can be used for these purposes. Note that these are not optimal solutions since it usually takes quite some time before the wet solution is removed from the wafer and it is dried. Recently some in-situ dry cleaning processes have been developed and this will be the preferred solution if available. 

\underline{Load lock, Wafer Transfer and Storage} 

An ultrahigh vacuum deposition system can never be an isolated chamber otherwise any time a sample is transferred inside the chamber, it would need to be vented then pumped and backed out before the required base pressure is obtained again. 

Therefore, the main deposition chamber is connected with a set of ultrahigh vacuum chambers that enable to transfer a wafer from the ambient atmosphere through a series of glove boxes, load locks, pumping stages, transfer systems and wafer storage units. Depending on the final system configuration, the transfer systems also connect to the RTA chamber, the in-situ cleaning unit etc. 

Note that these wafer transfer requirements can turn out to be quite complicated and expensive. For instance in most RTA systems the wafer is facing upwards, while in an MBE system it faces downwards. To transfer the wafer from the RTA to the MBE then requires an 180 degrees flip operation under ultra high vacuum conditions. This can be performed either in a vacuum flipping station or as part of a standard atmospheric robot handler called a front opening universal pod (FOUP). 

The use of all this equipment is however critical for the final thin film  
quality. Atmospheric reactions (oxidation, water layer, carbonisation, etc. ...)and the presence of particles, dust can be avoided in this way, and therefore the contamination of the inside of the coaters can be avoided. Note that contamination has recently been identified within the Virgo coating community as one of the main issues to be addressed for the improvement also of the current generation of the amorphous mirror coatings.

There is no final layout available of the experimental setup yet as this will be part of the design process as well as depending on total cost.

\section{Photodiodes}
An important active part of the interferometer is the light detector. The photodetector detects photons exiting the interferometer at the dark fringe and thus it is susceptible to the photon number variations caused by the interaction of the GW with the interferometer. The photon detection has been developed to a perfection in previous GW collaborations ( LIGO, VIRGO, KAGRA), however these work at a different laser wavelength as planned for the Einstein telescope. In particular, the infrared wavelength $\sim$ 2 microns is one of the main innovation of the E-TEST development as it couples to requirements imposed by the low temperature operated Si mirrors and accompanying optics. This allows for a significant reduction of the vibrational noises at temperatures 5 – 20 K. To match the wavelength of the laser  ( 2 micron) a detector by optimal spectral dependence and as well the quantum efficiency needs to be chosen. The other GW collaborations use the laser wavelength at typically 1064 nm and these detectors don’t have therefore sufficient detectivity at 2 microns. Also, for novel future quantum-based methodologies using light squeezing, the quantum efficiency needs to be as high as possible, ideally close to 99 

Secondly, the low temperature operation allows for a significant reduction of the interferometer noise and thus improving the signal to noise (S/N) ratio as well as allowing to widen operation frequency for the GW detection. There are several components that enter into the play for reduction of the detection noise. In particular, in addition to vibrations and dealt with  in the WP1, there are two types of detection noises that need to be optimized, this in particular  the shot noise ( quantum noise) and the radiation pressure back action pressure noise, induced on the mirrors. These two noise inputs represent the fundamental physical limits of the interferometer sensitivity. The work of UHasselt contributes precisely to this main goal. e.g. the reduction of the detection noise. However, the physical limits of the detection can only be achieved if the all possible sources of the technical noises are reduced. This concerns for example the circuitry noise, amplifier noise, photodetector device noise  (a pin diode) which has been addressed by UHasselt work.

\begin{enumerate}
\item Optimization of the detection wavelength:

Based  on 3.2 developments of the highly stabilized laser at ILF, UHasselt was matching the laser wavelength by searching for a suitable detector devices. We have carried out set of measurements on different commercial devices, trying to find the best match with operation close to 2 micron range.  PbS and InGaAs diode have been tested and their absolute quantum efficiency has been measured ( see table 1). Figure \ref{fig:HamamatsuDiodes} summarized the available detectors. In collaboration with Hamamatsu several commercial detectors were identified and are beeing evaluted  in particular PbS detector and InGaAs G12181, G12182, G12183. From those IR-enhanced InGaAs ( series varies by different size and the dark current) is most interesting due to higher QE The peak sensitivity of the QE can also be slightly tuned  ($\sim$ 10 nm) by changing the temperature. The parameters are summarized in table below. Photocurrent measurement setup has been build and the detectors purchased for these measurements. We have built two different detection setups, a classical dc lock-in detection and homodyne detection. These systems are operational and being used for extensive detector testing. The testing involves the characterization of the dark current of the device that contributes to the noise level.

\begin{table}[htb]
    \centering
    \footnotesize{
    \begin{tabular}{|c|c|c|c|c|c|}
        \hline
        Material & Si & InGaAs & InAs & PbS & HgCdTe \\
        \hline
        Detectivity (cm $H^{1/2}$ $W^{-1}$) & & 5e12 & 5e11 & 6e10 & 7e10\\
        QE ($\lambda$) & 70 & 99 &&&\\
        Spectral range (nm) & UV-1000 &900-2600$^{*}$, *extended & 1200-3000 & 1000-3000 & 1800-6000  \\
        Maximum response & 1000 & 2000 & 3000 & 2200 & Tuanable by composition \\
        \hline
    \end{tabular}
    }
    \caption{The detectors tested in E-TEST project and their main specifications  }
    \label{tab:detectorstested }
\end{table}

\begin{figure}[h]
    \centering
    \includegraphics[width = 0.9\linewidth]{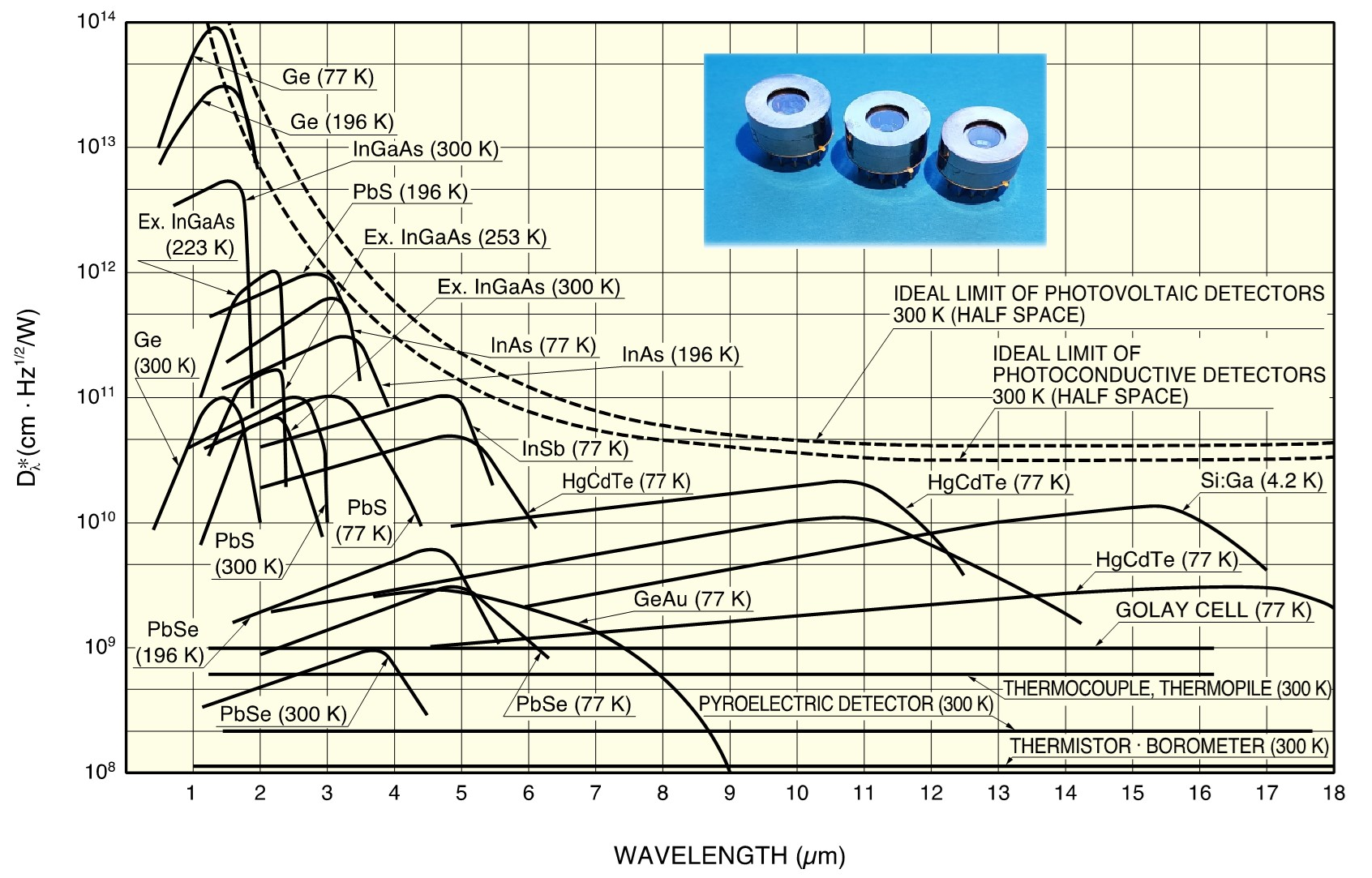}
    \caption{The spectral ranges of commercial Hamamatsu diodes relevant for the GW detection in IR 2 micron range}
    \label{fig:HamamatsuDiodes}
\end{figure}

\item The detector noise:

Extensive testing has been executed with the InGaAs detector, more specifically UHasselt optimized the detector circuitry. The tabulated pin device noise correspond 5 10$^{-13}$ W/sqrtHz. The most important part of the detection circuit is the current preamplifier. First tests started with the commercial preamplifier SRS 570. Further on, to improve the S/N ratio and the frequency range we and we have developed in collaboration with University of Stuttgart and tested novel trans-impedance preamplifier using multi-element pseudo resistors. In particular based on the University Stuttgart design we modified the preamplifier for our needs and constructed it on a PCB board. The developed pre-amplifier has a bandwidth of 10 kHz and the noise level of a 500 aW/sqrt(Hz). The preamplifier outperforms  SRS 570 when using test currents and these especially for higher bandwidth. This will give theoretical  > 10$^{11}$ noise suppression  in expected currents measured by the photodiode. The design of the PCB board allows a direct integration with PIN diode directly to minimise parasitic capacitance and limits the noise. This work will be carried out in the next period.

\begin{figure}[H]
    \centering
    \includegraphics[width = 0.9\linewidth]{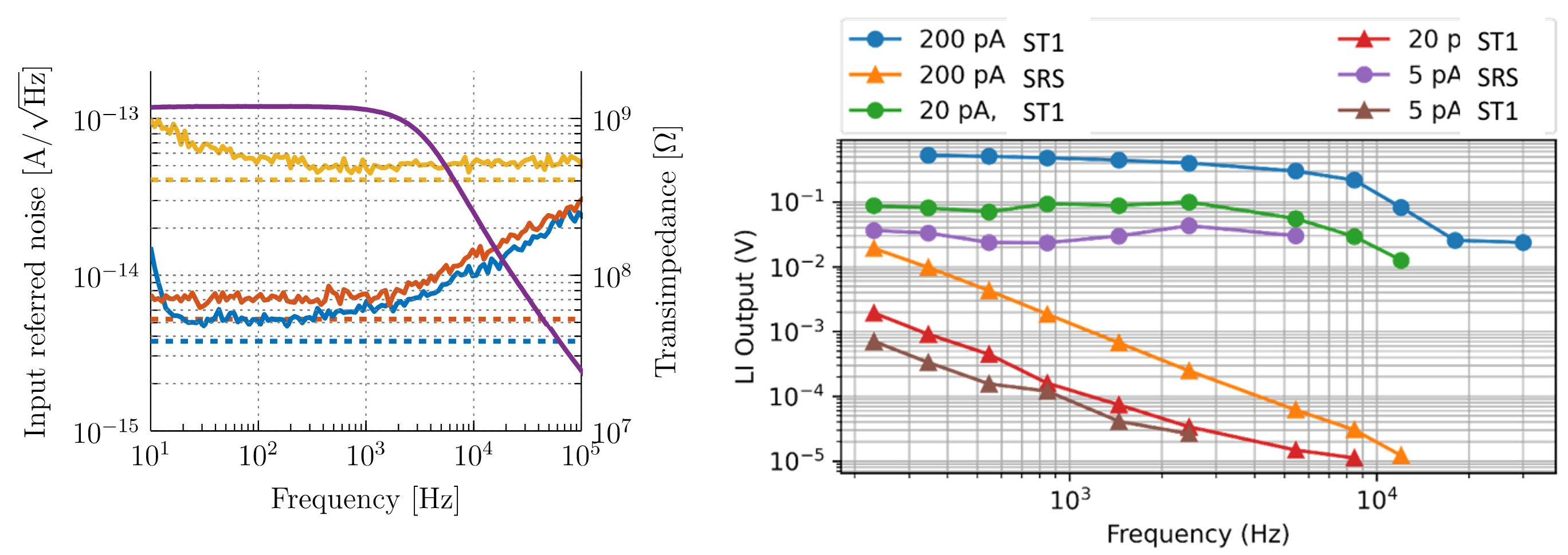}
    \caption{Left) Input referred noise for various dc feedback resistor values Rdc. Blue: Rdc = 150 G$\Omega$. Red: Rdc = 1.2 G$\Omega$. Yellow: Rdc = 10 M$\Omega$. For comparison, the dashed lines show the theoretical thermal noise of ideal resistors with noise- equivalent values of 1.2 G$\Omega$ (blue), 600 M$\Omega$ (red), and 10 M$\Omega$ (yellow). Right) Comparison of commercial SRS 570 and tailor made ST1 trans-impedance preamplifier performance expressed on detection voltage as a function of the frequency. The ST1 is still more sensitive to higher frequency and has  the current noise  $\sim 10^{-14}$ A/$\sqrt{}$Hz.}
    \label{fig:Inputreferrednoisefeedback}
\end{figure}

\item Further perspectives:

UHasselt was further involved in theoretical studies aimed at mathematical description of quantum detection, in particular squeezing and entanglement based method. This modeling allowed to calculate the theoretical limits of the detection taking into the account the shot noise and squeezing ratio and the back action noise. The article ( Entanglement Limits in Hybrid Spin-Mechanical Systems) has been submitted to the journal and it is available at \url{https://arxiv.org/abs/2108.13216}.

Based on these calculations UHasselt is designing the system for the squeezing. The squeezing setup will involve the tunable laser, that was procured, to find optimal squeezing configurations ( for at 2 micron wavelengths). We plan to apply the squeezed light configuration with heterodyne detection and devise the noise improvements. For this the performed calculations will be extended and used.

\end{enumerate}

\section{Technology validation and collaboration}
While all of the developed optical technologies are an important step towards the technological realization of ET-LF, it is also important that they are adapted to each other. One example for this is e.g. the sensitivity of the photo diodes, which has to be designed to fit to the laser wavelength to deliver the best results. While the earlier sections of this chapter described the technologies developed by the specific research facilities, this section will give an overview on their collaboration, and especially on the collaboration with the prototype developed in WPT1. To demonstrate this, three key experiments are planned, namely

- Measurement of the thermo-optic properties of the silicon test mass

- Mechanical quality factor measurement on the silicon substrate

\begin{figure}[H]
    \centering
    \includegraphics[width=.93\textwidth]{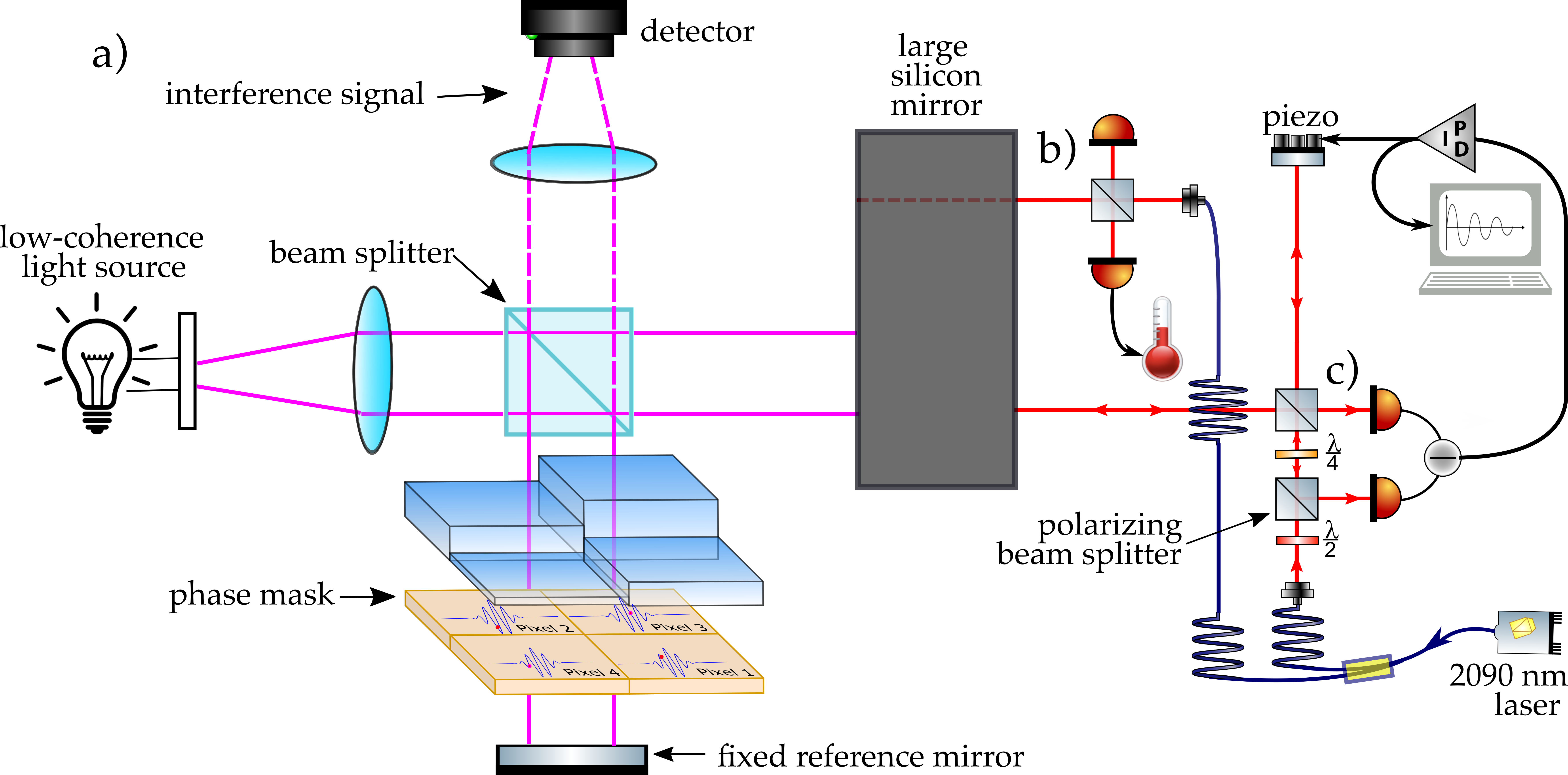}
    \caption{The 3 E-TEST prototype optical experiments: a) the white light interferometer, b) the temperature measurement and c) the quality factor measurement. Contrary to the suggestions of this figure, all experiments probe the same side of the mirror, having access through the 3 tubes into the cryostat visible in Figure~\ref{fig:E-TESTprtOvvw}. Experiment a) will also probe the full mirror surface, but on an optical bench outside the cryostat prior to cold suspension.}
    \label{fig:E-TESToptexps}
\end{figure}

- White light interferometry to characterize the mirror

\noindent
which will be described in detail in the following sections. A schematic representation of these three experiments is  shown in Figure \ref{fig:E-TESToptexps}. All of the experiments require a high collaboration between the different project partners, e.g. photo diodes are needed from UHasselt, the Laser is needed from Fraunhofer ILT and the white light interferometer from CSL. In addition, all experiments shall be performed on the prototype developed in WPT1.

\subsection{Measurement of the thermo-optic properties of the silicon test mass}

Due to its high refractive index of around 3.5 at a wavelength of 1550\,nm, the boundary between silicon and air reflects around 30\% of light. The two surfaces of the silicon test mass therefore form a Fabry-Pérot etalon, where light bounces back and forth. The individual light fields interfere constructively or destructively, depending on the optical path for one round-trip. As the silicon mirror gets cooled down, this optical path changes due to two effects: an expansion of the material due to temperature, described by the thermo-elastic coefficient $\alpha=dL/dT$, and a change in refractive index, described by the thermo-refractive coefficient $\beta=dn/dT$. Together, they form the thermo-optic coefficient of the silicon material. This thermo-optic coefficient can be measured by monitoring the interference fringes of the etalon, together with the substrate temperature \cite{Komma2012}. Conversely, knowing the thermo-optic coefficient, the observed interference fringes can be used to obtain the temperature evolution of the silicon substrate, which allows for a contactless measurement and monitoring of the silicon temperature at the location of the laser beam. This is especially relevant for E-TEST, where one is interested in the temperature gradient during cool-down, but at the same time cannot place conventional temperature sensors all over the test mass without impacting the performance of the seismic isolation, and more importantly, destroying the optical and mechanical quality of the mirror itself.

\begin{figure}[h]
    \begin{center}
        \includegraphics{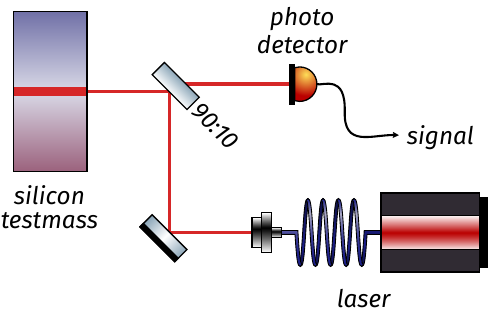}
    \end{center}
    \caption{Schematic layout of the experiment to measure thermo-optic properties of the silicon test mass.}
    \label{fig:to_experiment_layout}
\end{figure}

The experiment itself is straight-forward to setup, a schematic is given in Fig.~\ref{fig:to_experiment_layout}. A pick-off from the laser source at 2µm\footnote{alternatively, a different source at e.g. 1550nm can be used, as long as the wavelength sits within the transparency window of silicon}, see Section \ref{sec:laser2um}, will be sent orthogonally onto the silicon test mass. Thus, the beam will interfere with itself when reflected off the surfaces of the test mass. The reflected light is detected with a suitable photo detector and recorded with a data-acquisition system during the cool-down phase of the mirror.

\begin{figure}[h]
    \centering
    \includegraphics[width=9cm]{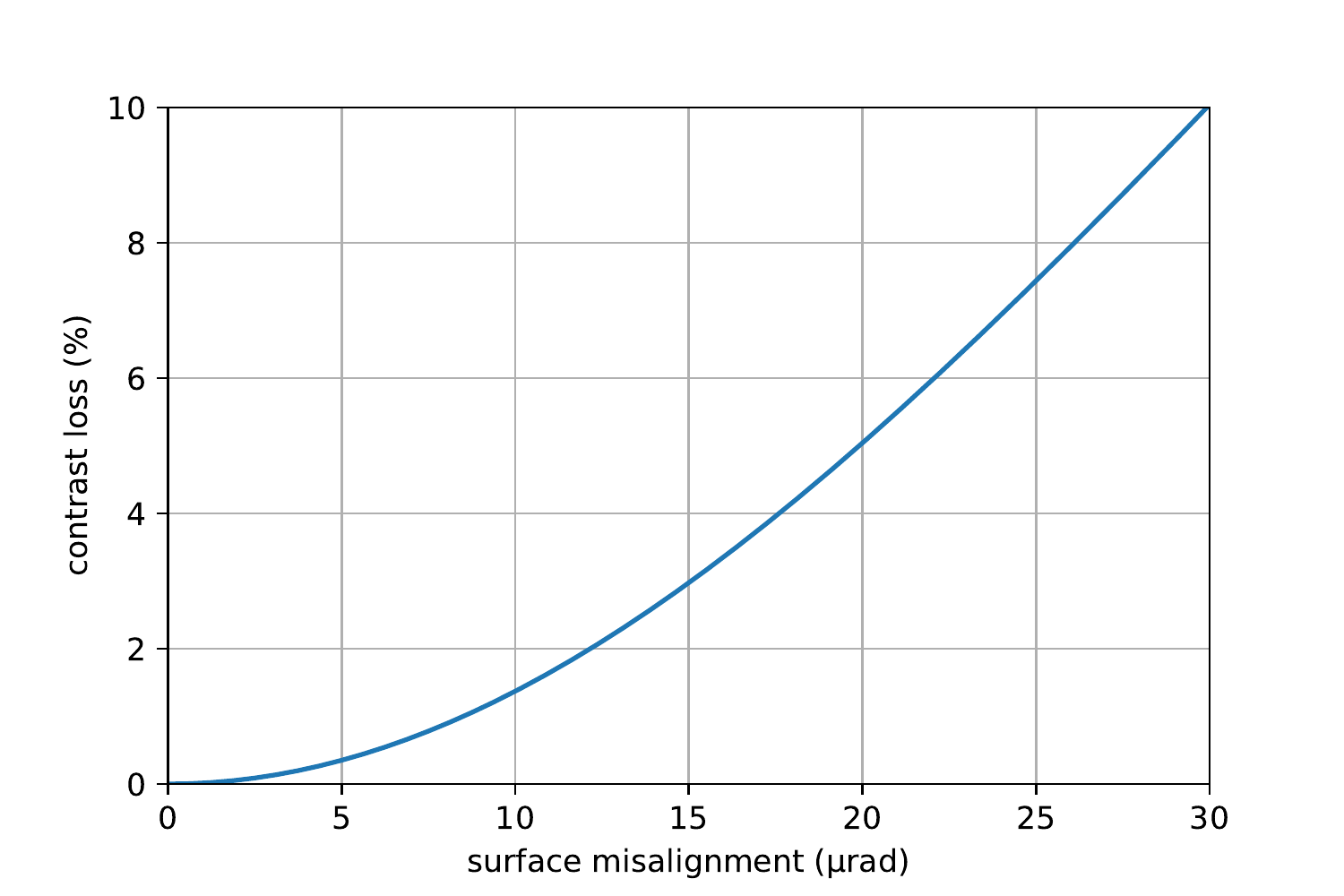}
    \caption{Loss of interference contrast for misalignment of the two surfaces of the silicon test mass. To keep loss to less than 10\%, any wedge of the test mass must remain less than around 30\,µrad (ca.\ 0.10\,arcmin).}
    \label{fig:surface_misalignment}
\end{figure}

The experiment requires the following components:
\begin{itemize}
    \item fibre-coupled laser source,
    \item fibre outcoupler/collimator,
    \item polarizing beam splitter and half-wave plate for power adjustment,
    \item pair of steering mirrors and mounts,
    \item beam splitting optic,
    \item photodetector,
    \item two-channel DAQ system,
    \item optical breadboard mounted at vacuum tank.
\end{itemize}
As shown in Figure~\ref{fig:surface_misalignment}, the two faces of the silicon test mass need to be parallel to within 30\,µrad, to achieve sufficient interference contrast.

\subsection{A mechanical quality factor measurement of the silicon substrate}
Another way of probing the quality of the mirror substrate is by exciting its mechanical internal modes and see how much energy they dissipate per cycle of the kHz mode. Using a tiny \textit{hammer}, \textit{i.e.} a small mass with a spring element that can be made with \textit{e.g.} a coil magnet actuator to tap the test mass ever so slightly, can excite all modes that do not have a node at the position of hammer impact. Once the modes are excited and ringing down it is up to an optical measurement to determine the envelop of the exponentially decaying ring-down.

The optical setup to do just that is depicted in Figure~\ref{fig:E-TESToptexps}b). An interferometer that uses polarizing optics to divert the light of both arms to two photodiodes~\cite{Badaracco2021}. The error signal is generated by subtracting the two out-of-phase photodiode output signals. This also ensures all common mode noises - typically laser noises such as amplitude fluctuation - are decreased to shot noise levels. The error signal is used to make the mirror attached to a piezo actuator follow to probed mirror point, such that the signal sent to the piezo actuator will show a ring down indicative of the mechanical losses of the mirror. Such interferometric readout systems can reach fm/$\sqrt{\rm{Hz}}$ sensitivity~\cite{JorisPhD2018}. 

The different modes can be isolated using a bandpass filter. By finite element analysis the frequency of each mode can be compared and the mode shape identified. This will also show which part of the mirror substrate (face) will move with a large enough amplitude for our sensitivity and an estimation of the expected signal strength can be made. The laser probe position on the mirror front surface can be varied by using the three different access tubes through the cryostat. 

\subsection{White Light Interferometry to characterize the mirror}
In the following paragraphs, we describe one of the optical experiments for the E-TEST project and the interest in implementing it on the Einstein Telescope \cite{jesus}.

\subsubsection{Purpose of the optical experiment}
We develop a metrology instrument to characterize the induced vibrations and wavefront or topology change of the silicon mirror at cryogenic temperatures.

\subsubsection{The added value of the system}
The instrument measures relative values of vibration and deformation of different points of the mirror surface instead of a single absolute value. We target a high precision on the measurements in the order of sub-nanometers. 

\subsubsection{Implementation in E-TEST and the future Einstein Telescope }
For the E-TEST project, we aim to build a proof of concept of the metrology instrument to test its innovative features. The final implementation of our optical experiment on the vacuum chamber for E-TEST depends on the state of the research and the results of the laboratory validation. In the framework of the Einstein Telescope, the mirror characterization done by the metrology instrument will be essential to ensure the accuracy of the results on the detection of gravitational waves.

\subsubsection{Working principle and optical setup of the instrument}
The induced vibrations and wavefront or topology change are two issues that may appear consequence of cooling down the suspended mirror to 10 K - 20 K. Reference \cite{Kaneda2004} already detected the permanent deformations of a mirror after the cooling down. As a result, we need to study these two side effects in detail on the Einstein Telescope framework.  

\begin{figure}[h]
  \centering
  \begin{minipage}[b]{0.57\textwidth}
    \includegraphics[width=\textwidth]{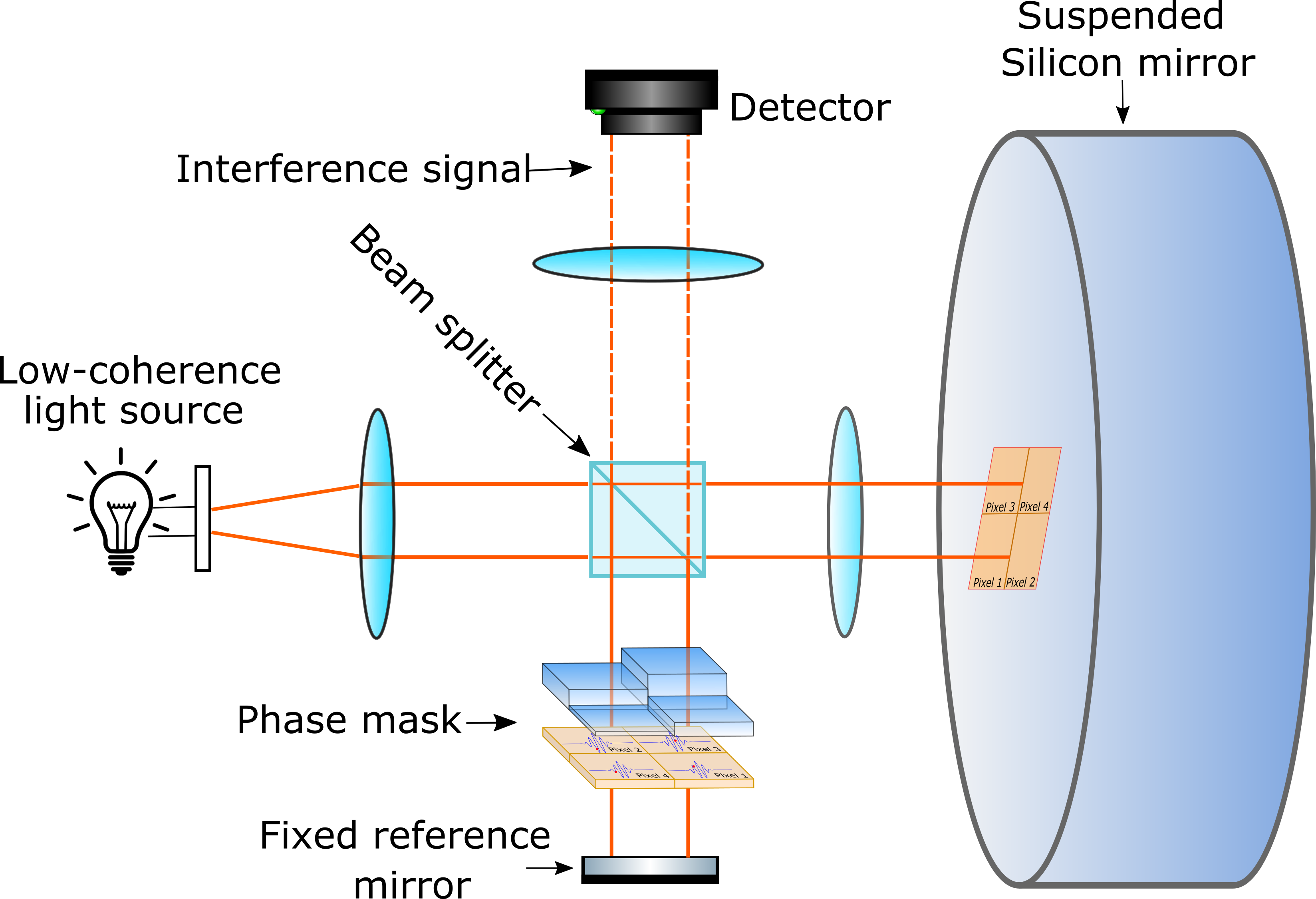}
  \end{minipage}
  \hfill
  \begin{minipage}[b]{0.40\textwidth}
    \includegraphics[width=\textwidth]{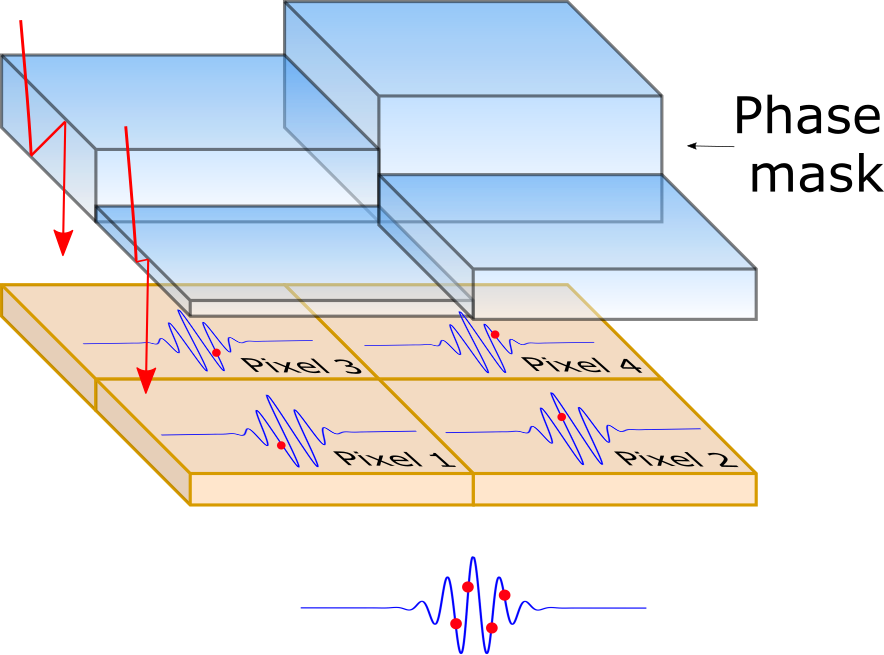}
  \end{minipage}
  \caption{a) Concept view of the phase mask designed to have single frame white light interferometry. b) Concept optical layout of the metrology instrument.}
  \label{fig:WLI_interferometer}
\end{figure}

To fulfil the accuracy requirements needed for the characterization, the metrology instrument is based on an interferometer optical layout. It is, therefore, a non-contact measurement solution. In addition, it does not implement movable components to avoid additional unwanted vibrations. In specific, we design a Michelson Interferometer that implements two sources of light. The primary light source for the mirror surface characterization is a white light source of short coherence. The second light source is a laser of high coherence that works as a complementary source during this first research phase. Figure \ref{fig:WLI_interferometer} a) represents a schematic representation of the metrology instrument setup.

The metrology instrument we are developing is a dynamic interferometer that captures a whole interference pattern in a single frame. To achieve this milestone, we design an innovative phase mask. The phase mask has a step design, where each step introduces a different Optical Path Difference (OPD) on the interferometer. The OPD is the inequality in the path followed by the light rays after the beamsplitter in each of the two arms of the interferometer. Therefore the phase mask simulates the changes in OPD introduced by the displacement of one of the interferometer mirrors by a PZT device in conventional interferometers. As a result, we also avoid additional unwanted vibrations from movable components. A schematic representation of the phase mask appears in Fig. \ref{fig:WLI_interferometer} b). The phase mask allows capturing Fields of View, which corresponds to different surface points, at different OPD’s in a single camera frame. Each of the interferograms measured contains the information of the relative displacement and deformation of this specific point. 

\begin{figure}[h!]
  \centering
  \begin{minipage}[b]{0.47\textwidth}
    \includegraphics[width=\textwidth]{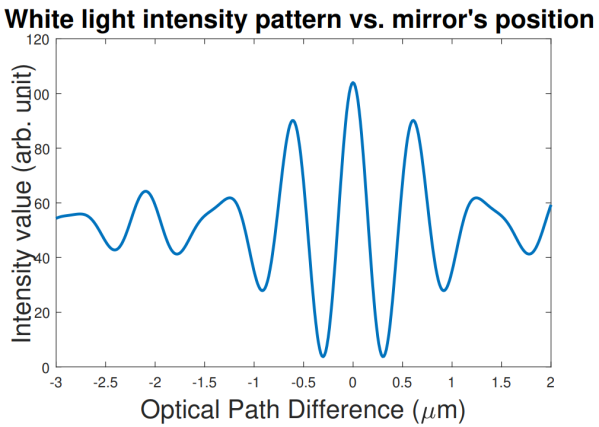}
  \end{minipage}
  \begin{minipage}[b]{0.45\textwidth}
    \includegraphics[width=\textwidth]{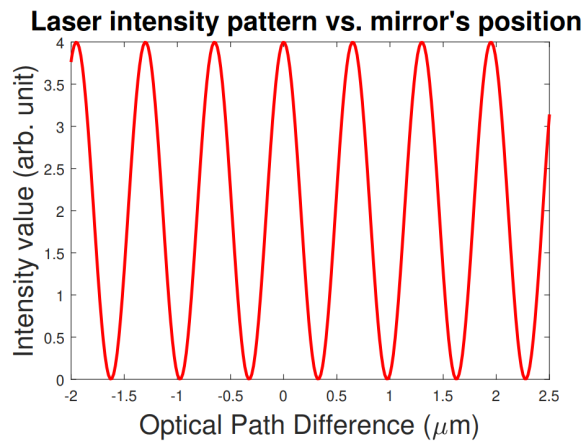}
  \end{minipage}
  \caption{a)Low coherence interferometry pattern vs b) high coherence interferometry pattern}
  \label{fig:interference_patterns}
\end{figure}

The polychromatic source that we use has a central wavelength of around 550 nm and at least a 100 nm bandwidth. The characteristic feature of low coherence interferometry interferograms is a well-differentiated central fringe. This feature allows an unambiguous determination of the position of the mirror since this zero-order fringe is visible only when the two arms of the interferometer have an equal path length \cite{Taylor}. Moreover, low coherence interferometry avoids spurious unwanted interferometry signals from scattered rays, dust, imperfections on the optical surfaces that may interfere with the measurement beam. Compared with monochromatic laser-based systems, white light interferometry creates a lower coherent noise and lower phase noise \cite{Deng2017}. The white light interference pattern is represented in Fig. \ref{fig:interference_patterns} a), which plots the intensity value captured for a specific measurement point at different OPD. Figure \ref{fig:interference_patterns} b) is the sinusoidal pattern created by a source with a high coherence like a laser for comparison.

The next step is to proceed with the interferogram analysis to determine the peak of the coherence envelope of the pattern. For this, we study the phase of the white light interference pattern. Besides, we implement new approaches for the interferograms analysis to avoid limitations from conventional Phase Shifting Interferometry algorithms. The number and height of the mask steps depend on the smallest vibration and deformation that we would like to measure. The mask material and the tolerance in manufacturing also influence the mask design. Finally, we perform the pattern analysis at the simultaneous interferograms captured at successive camera frames. As a result, we can determine the relative displacement of the white light interference pattern associated with the relative value of the mirror induced vibration and wavefront deformation. 

\subsubsection{Performance of the metrology device} 
The lateral resolution of the instrument will be between 10 $\mu$m to 20 $\mu$m, according to the detector pixel size. The phase resolution and, therefore, the height difference between consecutive steps of the phase mask is intended to be maximum $\lambda$/8, where $\lambda$ is 550 nm. Since the optical experiments are located outside the vacuum chamber, the relative motion between the metrology instrument and the mirror will be in the order of micrometres. The vibration of the mirror is expected to be in the order of a few nanometers. There will certainly be aberrations from the fact that one of the arms of the interferometers is inside the vacuum chamber and the other one outside. We are looking for solutions to minimize this effect.

\subsubsection{Development of the instrument} 
\begin{itemize}
    \item \textbf{Planned case:} we are building a prototype of the metrology instrument at the laboratory. The aim is to test the performance. In particular, to test the innovative concept of the phase mask for dynamic or single frame interferometry with a polychromatic source.

    \item \textbf{Ideal case:} to perform measurements on the suspended mirror used for E-TEST. We consider two options. The first option is to locate the metrology instrument outside the vacuum chamber, according to Fig. \ref{fig:option1_CSL_WP2}. It uses a 50 cm x 3 m space on the optical table. Nevertheless, we can compact the instrument by modifying the optical path of the light on the 3 m arm of the Michelson Interferometer. The second option is to place the instrument inside the vacuum chamber (and outside the cooling architecture of the mirror) over the optical table. That reduces an unstable environment (outside the vacuum chamber) and reduces induced aberrations from the optical window, Ref \cite{Vandenrijt2016}.
        \begin{figure}[h]
        \centering
        \includegraphics[scale = 0.39]{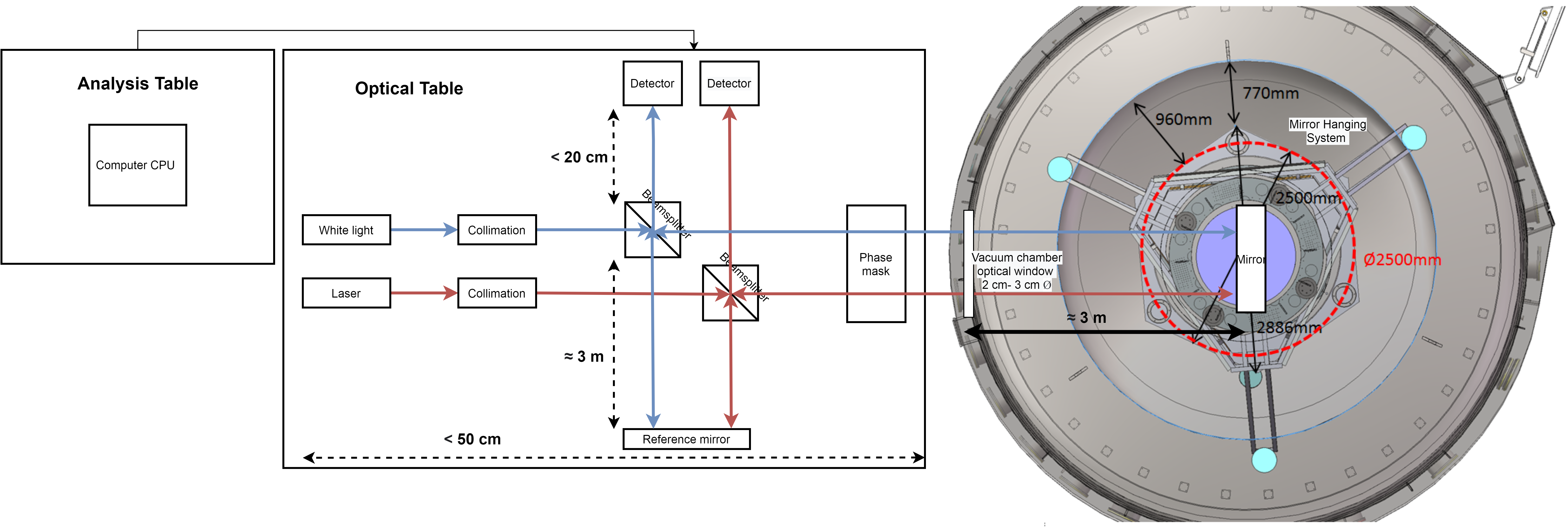}
        \caption{Concept view of the phase mask designed to have single frame white light interferometry.}
        \label{fig:option1_CSL_WP2}
        \end{figure}
\end{itemize}

\subsubsection{Component List}
\begin{itemize}
    \item White Light Interferometry
    \begin{itemize}
        \item 1xHalogen low coherence light source (in the visible and with 200 nm bandpass).
        \item Optical lenses for collimating the source.
        \item Optical filters to limit the source bandpass.
        \item 1xBeamsplitter (50/50).
        \item 1xPhase mask (on design).
        \item 1xPixelated photodetector.
        \item 1xAnalysis unit (CPU).
        \item 1xPZT if the instrument is located inside the vacuum chamber.
    \end{itemize}
    \item Large coherence interferometry
    \begin{itemize}
        \item We use similar components to the white light experiment except for the source. We use the Laser source developed by ILT for the E-TEST project. 
    \end{itemize}
\end{itemize}

\chapterimage{Figures/books.jpg} 



\clearpage 
\phantomsection
\addcontentsline{toc}{section}{Bibliography}
\bibliographystyle{et_dsd} 
\bibliography{bibFiles/ET_CDR_2019_References}


\nomenclature{ET}{Einstein Telescope }

\nomenclature{CE}{Cosmic Explorer }

\nomenclature{HF}{High Frequency }

\nomenclature{LF}{Low Frequency }

\nomenclature{GW}{Gravitational Wave}

\nomenclature{E-TEST }{Einstein Telescope Euregio-Meuse-Rhin Site and Technology}

\nomenclature{WP}{Work Package}

\nomenclature{CSL}{Space Centre of Liège}

\nomenclature{IPP}{Inverted Pendulum Platform}

\nomenclature{IP}{Inverted Pendulum}

\nomenclature{GAS}{Geometric Anti-Spring}

\nomenclature{DOF}{Degree Of Freedom}

\nomenclature{AP}{Active Platform}

\nomenclature{CP}{Cold Platform}

\nomenclature{ASD}{Amplitude Spectral Densities}

\nomenclature{HINS}{Horizontal Inertial Sensor}

\nomenclature{VINS}{Vertical Inertial Sensor}

\nomenclature{PML}{Precision Mechatronics Laboratory}

\nomenclature{Mir}{Mirror}

\nomenclature{Mar}{Marionette}

\nomenclature{J}{Jacobian}

\nomenclature{K}{Kelvin}

\nomenclature{ESA}{European Space Agency}

\nomenclature{GM}{Gifford-McMahon}

\nomenclature{ASIC}{Application Specific Integrated Circuit}

\nomenclature{IC}{Integrated Circuit}

\nomenclature{ADC}{analog-to-digital converters}

\nomenclature{DAC}{Digital-to-analog converters}

\nomenclature{CSIS}{Cryogenic Superconducting Inertial Sensor}

\nomenclature{LNM}{low noise model}

\nomenclature{MOPA}{Master Oscillator Power Amplifier}

\nomenclature{NPRO}{Non-Planar Ring Oscillators}

\nomenclature{MBE}{Oxide Molecular Beam Epitaxy }

\nomenclature{RHEED}{Reflection High Energy Electron Diffraction}

\nomenclature{QMS}{Quadruple Mass Spectrometry}

\nomenclature{RHEED}{Reflection High Energy Electron Diffraction}

\nomenclature{LAXS}{Low Angle x-ray Spectroscopy}

\nomenclature{MFC}{Mass Flow Fontroller}

\nomenclature{EIES}{Electron Impact Emission Spectroscopy}

\nomenclature{AAS}{Atomic Absorption Spectroscopy}

\nomenclature{QMS}{Quadrupole Mass Spectrometry}

\nomenclature{SE}{Spectroscopic Ellipsometry}

\nomenclature{IBAD}{Ion Beam Assisted Deposition}

\nomenclature{RTA}{Rapid Thermal Annealing}

\nomenclature{OPD}{Optical Path Difference}

\printnomenclature[0.75in] 
\newpage
\newpage 
\FloatBarrier
\end{document}